\documentclass[]{JFM-FLM_Au}
\usepackage[perpage]{footmisc} 
\usepackage{mathrsfs} 
\usepackage{subfigure}
\usepackage{graphicx}
\usepackage{epstopdf}
\usepackage{newtxtext}
\usepackage{newtxmath}
\usepackage{natbib}
\usepackage{hyperref}
\hypersetup{
	colorlinks = true,
	urlcolor   = red,
	linkcolor=blue,
	citecolor  =blue,
	filecolor=black,
}
\usepackage{amsfonts}
\usepackage{amsmath}
\usepackage{amssymb,color,mathtools}
\usepackage{bm}
\newcommand{\RomanNumeralCaps}[1]
\linenumbers

\lefttitle{Tao Chen, Jie-Zhi Wu, Feng Mao and Tianshu Liu}
\righttitle{Journal of Fluid Mechanics}

\title{A general kinematic theory of fluid-element rotation and intrinsic vorticity decompositions}

\author{Tao Chen\aff{1}, Jie-Zhi Wu\aff{2}, Feng Mao\aff{3} and Tianshu Liu\aff{4}}

\affiliation{\aff{1}School of Physics, Nanjing University of Science and Technology, Nanjing 210094, China
\aff{2}Retired from Chinese Aeronautical Establishment, Beijing 100102, China
\aff{3}Shenzhen TenFong Technology Co., Ltd., Shenzhen 518055, China
\aff{4}Department of Mechanical and Aerospace Engineering, Western Michigan University, Kalamazoo, Michigan 49008, USA}

\corresau{Tao Chen, \email{chentao2023@njust.edu.cn}}

\begin{document}
\maketitle
\begin{abstract}
The present study proposes a general kinematic theory for fluid-element rotation and intrinsic vorticity decompositions within the context of vorticity and vortex dynamics.
Both the angular velocities of material line and surface elements comprise a classical contribution driven by volume-element rotation (equal to half the local vorticity), and a strain-rate-induced specific angular velocity. Then, two direction-dependent vorticity decompositions (DVDs) are constructed, revealing the rigid rotation and spin modes of vorticity.
We derive intrinsic coupling relations for orthogonal line-surface element pair, elucidating their complementary kinematic and geometric roles. Notably, we rigorously prove that the spin mode (in the surface-element-based DVD) is identical to the relative vorticity in the generalized Caswell formula, thereby faithfully accounting for surface shear stress in Newtonian fluids. 
Next, within a field-theoretic framework, vorticity decompositions are proposed based on streamline and streamsurface using differential geometry. The physical roles of the six rotational invariants in the characteristic algebraic description (including the the normal-nilpotent decomposition (NND) of the velocity gradient tensor (VGT) and the resulting invariant vorticity decomposition (IVD)) are clarified through unified analysis with the DVD, Caswell formula, and Helmholtz-Hodge decomposition.
For flows projected onto the invariant plane orthogonal to the swirling axis (i.e., the real eigenvector of the VGT in the region with positive discriminant), we establish explicit relationships between DVD and IVD vorticity modes.  Interestingly, under the Frenet-Serret frame attached to a streamline, the VGT exhibits an intrinsic structure as an irreducible real Schur form. It is found that physical admissible DVD vorticity modes must be bounded by IVD modes in phase space, whereas enforcing a minimization principle naturally yields the Liutex formula. Finally, the effectiveness of theory is validated across diverse flows, from simple to complex. Results show that a coupled IVD-DVD analysis could enhance physical understanding of complex vortical flows under both algebraic and field-theoretic frameworks.
\end{abstract}
\tableofcontents
\section{Introduction}
\subsection{Literature review and recent development}\label{sec1p1}
Fluids are distinguished from any other continuous media by their constitutive inability to resist shear, which inherently induces rotation of elementary fluid elements (encompassing material lines, surfaces, and volumes).
This distinctive characteristic is precisely what renders fluid motion exhibiting unparalleled complexity compared to solid continuum mechanics.
The symmetric-antisymmetric decomposition (SAD) affirms that the velocity gradient tensor (VGT), $\bm{A}\equiv\bm{\nabla}\bm{u}$, can be decomposed into the strain-rate tensor $\bm{D}$ and the rotation-rate tensor $\bm{\varOmega}$, such that $\bm{A}=\bm{D}+\bm{\varOmega}$~\citep{Batchelor1967,WuJZ2015book}. Then, the vorticity $(\bm{\omega}\equiv\bm{\nabla}\times\bm{u})$, defined as the curl of the velocity field, emerges naturally as the dual vector of $\bm{\varOmega}$, providing a kinematic measure for the shear-driven transverse processes in complex flows. In contrast to the momentum description employing the primitive hydrodynamic variables (velocity and pressure), vorticity-based formulations usually offer superior physical intuition and mathematical elegance for complex fluid motions. The vorticity paradigm has been proven to be powerful for analyzing the ubiquitous hydrodynamic phenomena like Rayleigh–Taylor (RT) and Richtmyer–Meshkov (RM) instabilities, with broad applications across geophysical flows to engineering systems~\citep{WuJZ2015book,ZhouYe2017}. 
Historically, the concept of fluid rotation traces its origins to the classical Cauchy-Stokes theorem\footnote{Established by Cauchy in 1841 and Stokes in 1845.}, which formalized the local kinematic decomposition of fluid motion via the SAD~\citep{Truesdell1954}.
By analogous to rigid-body kinematics, the theorem asserts that the vorticity is just twice the angular velocity of a fluid volume element~\citep{Truesdell1954,Batchelor1967,Corrsin1972}, which becomes a cornerstone of modern vorticity and vortex dynamics.
However, a lesser-recognized aspect is that the fundamental concept of fluid-element rotation has evolved along two distinct but deeply interconnected conceptual pathways, remaining subject to ongoing investigation and not yet finished even today.

\textcolor{red}{The first line of development focuses on three key aspects: the rotation of a fluid volume element, the intrinsic vorticity splitting, and developing the effective criteria for vortex identification.}
The concept of vorticity was realized insufficient for fully characterizing the vortex motion after the groundbreaking work of Helmholtz in 1858 [English translation version~\citet{Helmholtz1858}], where the word ``rotation'' (originally employing the German terminology ``Rotationsgeschwindigkeiten") was used for half of the vorticity\footnote{The concept of vorticity was first used by Lord Kelvin (William Thomson, 1875) to describe local rotational motion, though without a rigorous mathematical definition~\citep{Thomson1878}. Later,~\citet{Lamb1916} formally introduced the term ``vorticity'' for $\bm{\omega}$ in the fourth version of his monograph \textit{Hydrodynamics}.
The components of $\frac{1}{2}\bm{\omega}$ were called the \textit{angular velocities} by Stokes (1851), \textit{Rotationsgeschwindigkeiten} by Helmholtz (1858), the \textit{component rotations} by Kelvin (1869), the \textit{molecular rotations} by Basset (1888), and the \textit{spin} by Clifford (1858), see~\citet{Truesdell1954}.}. 
As noted by~\citet[\S29]{Truesdell1954}, Joseph Bertrand (a French mathematician and physicist) made an observation in 1868\footnote{Another way of expressing the rotational character of Bertrand's shearing motion was put forward by St. Venant (1869) [in a footnote of~\cite{Truesdell1954}]: the straight streamlines are the only lines in the $(x,y)$ plane which are not suffering motion.} that 
simple shear flow exhibits nonzero vorticity despite lacking any intuitive swirling motion of a typical vortex--flow particles move in rectilinear trajectories--clearly highlighting a critical limitation in equaling vorticity with vortex.
To some extent, this example underscores two crucial insights: (i) Swirling motion, a hallmark of vortices, generally contributes to part of vorticity and disappears entirely in simple shear flow; (ii) Vorticity is not an appropriate physical measure to represent vortex in shear-dominated region.
Standpoint (ii) also applies to turbulent boundary layers: the association between regions of strong vorticity and actual vortices can be rather weak, especially in the near-wall region~\citep{Robinson1991}.
As the controversy was finally settled, one has realized the necessity of rationally distinguishing between the two elementary vorticity modes, that is, the rigid rotation mode (swirling-dominant) and the spin mode (shear-dominant), even for a fluid volume element without preferred spatial directionality.
Indeed, Prandtl's boundary layer at high Reynolds numbers is the most brilliant example of the spin-dominated vortical flow. 
As profoundly articulated by Professor Shijia Lu (1911-1986), the pioneering Chinese aerodynamicist and the sole female student of Ludwig Prandtl (1875-1953) at G\"{o}ttingen, during her late career circa 1980: ``\textit{The essence of fluid is vortices. A fluid cannot stand rubbing; once you rub it there appear vortices.}"~\citep{Wu2006vorticity}
This rubbing process between fluid and wall creates the boundary layer mediated by viscosity and the no-slip boundary condition. Once an attached boundary layer leaves a solid wall to become a free shear layer (governed by the Biot-Savart law), it can automatically roll into an axial vortex dominated by the rigid rotation mode, or vice versa, and in such processes vorticity is transferring and redistributed between the two elementary modes~\citep{WuJZ2015book,MaoFeng2022}.

Vortices are the sinews and muscles of fluid motion~\citep{Kuchemann1965}, despite the challenges in their precise definition~\citep{Wu2006vorticity,Saffman1992,Lugt1979}.
Over the past few decades, numerous local vortex identification criteria have been proposed to visualize coherent structures in turbulent flows, with most methods relying on the VGT $\bm{A}$. Among them, the widely-used ones include the vorticity number $m$~\citep{Truesdell1954},
the $Q$-criterion~\citep{Hunt1988,Jeong1995}, $\Delta$-criterion~\citep{Chong1990}, $\lambda_{2}$-criterion~\citep{Jeong1995}, and $\lambda_{\rm ci}$-criterion~\citep{ZhouJ1999}, etc. The Okubo-Weiss (O-W) criterion~\citep{Okubo1970,Weiss1991}, prevalent in oceanography and atmospheric studies, can be regarded as a two-dimensional (2D) reduction of the classical $Q$-criterion. However, the flow structures identified by these methods are often highly sensitive to  arbitrary threshold selection. Determining an appropriate threshold is challenging, as different values yield significantly different vortical structures~\citep{LiuCQ2018,LiuCQ2020}. Furthermore, they cannot unambiguously determine the swirling axis of a vortex, nor can they elucidate the physical constituents of distinct vorticity modes from vorticity alone. In fluids with small viscosity, the rolling-up of thin shear layers is the only known mechanism for the rapid formation of axial vortices. This physical mechanism arises from extremely strong coupling of longitudinal and transverse fields, which we refer to as the Klein-Kaden-Betz mechanism~\citep{Wu2006vorticity,MaoFeng2022}. These criteria exclude sheet vortices, preventing the KKB mechanism from being fully captured.

The rational characterization of these two distinct vorticity modes (i.e., rigid rotation and spin) had long remained ambiguous until~\citet{Kolar2004}, in looking for advanced axial-vortex criterion, recalled for one’s attention to this fundamental issue. He insisted that ``solely vorticity cannot distinguish between swirling motions and shearing motions'', leading to the triple decomposition of motion (TDM) for the VGT $\bm{A}$~\citep{Kolar2004,Kolar2007,Kolar2007IJHMF}. Building on the Schur transform~\citep{Schur1909}, the first version of the normal-nilpotent decomposition (NND-I) of $\bm{A}$ was proposed within a pure algebraic framework~\citep{LiZhen2010,LiZhen2014,LiuCQ2018,GaoLiu2019}. Importantly, NND-I implies~\textit{the Liutex-shear decomposition} for vorticity, which cleanly extracts the characteristic rigid rotation (Liutex/Rortex) to represent vortex, with the residual shear being separated simultaneously~\citep{LiuCQ2018,LiuCQ2020}. To avoid calculating the coordinate rotation matrices, an explicit method for determining the Liutex was proposed by~\citet{XuWenqian2019} and~\citet{WangYiqian2019}. While NND-I (the Liutex-vector method) has been successfully applied to identify coherent vortex structures in turbulence~\citep{LiuCQ2020}, it is observed that the rigid rotation mode is limited to regions where $\Delta>0$ which disappears for $\Delta<0$ (with $\Delta$ being the discriminant of the characteristic equation of $\bm{A}$). 
Recent work by~\citet{LiZhen2024} has clarified several confusing points about Schur forms and NND variants, which include the conceptual distinction between NND and TDM approaches, the intrinsic gap between complex and real NND formulations.
An improved real Schur form (denoted as NND-II) extends the definition of rigid-rotation mode to all flow regions regardless of $\Delta$'s sign, with unique matrix representation under proper conditions of uniqueness.
However, discontinuity issues and sign ambiguities associated with the two vorticity modes are found when applying NND-II to global flow diagnosis~\citep{ChenLiu2025POF}, while its physical interpretation remains incomplete under the pure algebraic framework. 
Furthermore, as noted by~\citet{LiZhen2024}, additional real NNDs beyond NND-I/II are also possible, though their physical significance requires further investigation in different situations.
In this work, we classify all NND-related matrices collectively as the characteristic representation, with the corresponding theory constituting the characteristic (algebraic) description, to distinguish it from the material/field descriptions discussed below.

\textcolor{red}{The second line of investigation adopts material and field descriptions with well-defined physical carriers, focusing on the kinematics of directed material line and surface elements.} The material description examines the angular velocities of infinitesimal directed material elements, while the field description physically captures unit-vector rotation through, for example, the streamline tangent vector or streamsurface normal vector. 
Historically, the theory of fluid mechanics has been predominantly formulated in terms of volume element (per unit volume or mass), while the complete rotational kinematics of directed material elements, involving more degrees of freedom beyond the coverage of vorticity, remained largely unexplored in the past except a few recent studies. The gap becomes particularly apparent as one's attention shifts to those organized coherent structures in turbulence, where the most familiar examples of directed material elements like vortex filaments and vortex sheets play fundamental roles, serving as the asymptotic models for axial vortices and shear layers. 

The foundational work of~\citet{Corrsin1972} first revealed this complexity through hydrogen-bubble tracer to study the angular dispersion of a material line element in isotropic turbulence. They demonstrated for the first time that a line element $\delta\bm{r}\equiv\delta{r}\bm{e}$ is rotated under the combined influence of the vorticity $\bm{\omega}$ (or the rotation-rate tensor $\bm{\varOmega}$) and the strain-rate tensor $\bm{D}$.
The evolution equation for the rate of change of the unit direction vector $\bm{e}$ was derived and expressed in Cartesian coordinates (see also~\citet{Lumley1987}), by which they successfully explained their experimental observations.
Recently,~\citet{MaoFeng2022} reformulated this equation into a compact vectorial form in a review paper on vortex dynamics, and explicitly identified the strain-rate-induced specific angular velocity $\bm{W}_{D}(\bm{e})$ of a line element. However, they did not explore the connection between $\bm{W}_{D}(\bm{e})$ and the fundamental vorticity modes, nor did they consider the rotation of directed surface element $\delta\bm{\Sigma}=\delta\Sigma\bm{n}_{\Sigma}$ (where $\bm{n}_{\Sigma}$ is the unit surface normal vector).
From a field-theoretic perspective,~\citet{ChenLiu2025POF} employed 
differential geometry to establish a curvilinear coordinate system in the neighborhood of a regular streamline segment for generic two-dimensional (2D) flow.
Working within the Frenet-Serret reference frame intrinsic to streamlines, they performed a first theoretical analysis on the rotational kinematics of directed line element pairs for all possible cases, and its fundamental relationships with the vorticity modes (rigid rotation vs. spin). Notwithstanding these advances, fundamental issues remain regarding the relationship among the rotational kinematics of these directed material elements, distinct vorticity modes, and the existing criteria for vortex identification (the first research line). A comprehensive understanding of these aspects is still lacking.

\subsection{Structure of this article}
The present study follows the second line of investigation. Section~\ref{Cauchy-Stokes theorem and concept of fluid rotation} begins with a critical re-examination of the Cauchy-Stokes theorem and historical perspectives on concept of fluid element rotation. Through rigorous analysis of the angular velocities of elementary fluid elements, we propose two direction-dependent vorticity decompositions (DVDs), formulated via directed material line and surface elements (\S\ref{Rotation of directed material line element and vorticity decomposition} and~\S\ref{Rotation of directed material surface element and vorticity decomposition}). 
In~\S\ref{intrinsic}, we establish the intrinsic relations for a pair of orthogonal line and surface elements on both kinematics and geometry. Section~\ref{spin_mode_interpretation} presents a rational interpretation on the spin mode via the generalized Caswell formula. Subsequently, followed by decompositions of spin and surface shear stress based on the triple decomposition of the strain-rate tensor. In~\S\ref{Field description based on streamline} and~\S\ref{Field description based on streamsurface}, we extend vorticity decomposition to the field-theoretic framework using streamline and streamsurface. Section~\ref{New physical insights on NND invariants} elucidates the physical roles of NND (normal-nilpotent decomposition) rotational invariants, suggesting a unified algebraic-field (or DVD-NND) perspective.
As illustrative examples, the proposed theory are applied to study Joseph Bertrand’s puzzle on the simple shear flow (\S\ref{Simple shear flow and Bertrand's puzzle}), potential flow outside a point vortex (\S\ref{point_vortex}), Burgers vortex (\S\ref{Burgers vortex}), vortex-sheet instability under small perturbation (\S\ref{Vortex-sheet instability sec}), Moffatt-Kida-Ohkitani (MKO) asymptotic vortex solution and associated line-element orientation instability (\S\ref{Moffatt-Kida-Ohkitani vortex solution}), and intriguing flow structures of Hexagon and North Polar Vortex on Saturn (\S\ref{Hexagon and North Polar Vortex on Saturn}). Finally, conclusions and discussions are documented in~\S\ref{Conclusions and discussions}.

\section{Revisiting Cauchy-Stokes theorem and concept of fluid rotation}\label{Cauchy-Stokes theorem and concept of fluid rotation}
The symmetric-antisymmetric decomposition (SAD) of the velocity gradient tensor (VGT) $\bm{A}\equiv\bm{\nabla}\bm{u}$ is given by~\citep{Batchelor1967,WuJZ2015book}
\begin{subequations}\label{eq1a1b}
	\begin{eqnarray}\label{eq1a}
		\bm{A}=\bm{D}+\bm{\varOmega},
	\end{eqnarray}
	\begin{eqnarray*}
		\bm{D}\equiv\mathscr{S}[\bm{A}]=\frac{1}{2}\left(\bm{A}+\bm{A}^{\rm T}\right),
	\end{eqnarray*}
	\begin{eqnarray}\label{eq1b}
		\bm{\varOmega}\equiv\mathscr{A}[\bm{A}]=\frac{1}{2}\left(\bm{A}-\bm{A}^{\rm T}\right),
	\end{eqnarray}
\end{subequations}
where $\bm{D}$ is the strain-rate tensor, and $\bm{\varOmega}$ is the rotation-rate tensor (or the vorticity tensor). The vorticity $\bm{\omega}\equiv\bm{\nabla}\times\bm{u}$ is deduced as the dual vector of $\bm{\varOmega}$ via $\bm{\omega}=\bm{\varepsilon}\bm{:}\bm{\varOmega}$, where $\bm{\varepsilon}$ is the permutation tensor.
For any vector $\bm{\xi}$, the identify $2\bm{\xi}\bm{\cdot}\bm{\varOmega}=\bm{\omega}\times\bm{\xi}$ holds exactly due to the vector-tensor duality. Here, the operators $\mathscr{S}$ and $\mathscr{A}$ represent evaluating the symmetric and antisymmetric parts of a second-rank tensor, respectively. ``${\rm T}$'' represents the transpose operation.

Applying the SAD to a point $\bm{x}$ and and its infinitesimal neighborhood in an instantaneous flow field yields the relative velocity $\delta\bm{u}$ between $\bm{x}$ and a neighboring point $\bm{x}^{\prime}$:
\begin{eqnarray}\label{CStheorem}
	\delta\bm{u}=\delta\bm{r}\bm{\cdot}\bm{A}(\bm{x})=\delta\bm{r}\bm{\cdot}\bm{D}(\bm{x})+\frac{1}{2}\bm{\omega}(\bm{x})\times\delta\bm{r},
\end{eqnarray}
where $\delta\bm{r}\equiv\bm{x}^{\prime}-\bm{x}=\delta{r}\bm{e}$ is the relative position vector (that corresponds to a material line element), $\delta{r}\equiv\lVert\delta\bm{r}\rVert$ is its magnitude, and $\bm{e}$ is a unit direction vector. Equation~\eqref{CStheorem} is the classical Cauchy-Stokes theorem (namely, \textit{the fundamental theorem of deformation kinematics}), which asserts that, to the first-order in the linear dimensions of a small region surrounding the position $\bm{x}$, the relative velocity $\delta\bm{u}$ (after excluding a uniform translation with velocity $\bm{u}(\bm{x})$) consists of the superposition of a pure straining motion characterized by the the strain-rate tensor $\bm{D}$ (which itself can be decomposed into an isotropic expansion and a straining motion without changing the volume), and a rigid rotation with the local angular velocity $\frac{1}{2}\bm{\omega}$~\citep{Batchelor1967}. Another interpretation of the last term can be found as ``the role of $\bm{\omega}$ is to rotate the line element $\delta\bm{r}$ around $\bm{x}$ with angular velocity $\frac{1}{2}\bm{\omega}$. Namely, the vorticity can be understood as twice of the angular velocity of a fluid element''. However, these interpretations contain two-fold ambiguities, as detailed below.

On one hand, a rigorous and logically consistent definition of the angular velocity of a fluid volume element must account for all the constituent motions in~\eqref{CStheorem}.
	As is well known, the vorticity inside a rotating rigid body must be twice its angular velocity at any instant. For general fluid motion, by comparing the last constituent term, $\frac{1}{2}\bm{\omega}\times\delta\bm{r}$, in~\eqref{CStheorem} with rigid-body rotation,~\citet{Truesdell1954} asserted that ``\textit{$\cdots$ the vorticity is no longer constant from point to point, but at each point it may be regarded as twice the angular velocity of a small element of the continuum}.'' The statement actually defines the local angular velocity of a fluid volume element $\bm{W}_{V}$ as half the local vorticity, that is, $\bm{W}_{V}\equiv\frac{1}{2}\bm{\omega}$.
	A similar but more specific description was provided by ~\citet{Corrsin1972}: ``\textit{the vorticity is just twice the angular velocity of a volume element of fluid}.''
	The underlying physical picture stems from the solidification of a fluid volume element suddenly into an isolated rigid body with all the surrounding material removed, which yields the formula of local velocity being the same as that of a rigid body. Similar interpretations were also suggested by Stokes (1845) for a small spherical element and generalized by Beltrami (1871). However, under the solidification assumption, the distinctive constituent motion $\delta\bm{r}\bm{\cdot}\bm{D}$ in~\eqref{CStheorem} caused by the strain-rate tensor of a fluid was not rationally considered in formulating a complete definition of $\bm{W}_{V}$.

On the other hand, the relationship between the rotations of material line and volume elements remains to be clarified. In our viewpoint, the Cauchy-Stokes theorem inherently describes the line-element rotation (rather than the volume-element rotation), which depends on the spatial orientation of the unit vector $\bm{e}$. 
	The angular velocity of a material line element generally incorporates the contribution from the strain-rate tensor $\bm{D}$ due to the constituent motion in~\eqref{CStheorem}, thereby making $\frac{1}{2}\bm{\omega}$ alone an incomplete physical measure of its rotation.
	According to~\citet{Truesdell1954}, Levy (1890) and Boussinesq (time unknown)\footnote{Truesdell (1954) did not provide the source of Boussinesq's explanation, but instead cited a reference from G. Jeff\'{e} (1921) published in Phys. Z. 22: 180-183.} claimed that $\bm{W}_{V}\equiv\frac{1}{2}\bm{\omega}$ was equal to the angular velocities of material line elements aligned with the principal axes of extension, in the ordinary sense of rigid rotations.
	Nevertheless, a rigorous definition of line-element angular velocity is still lacking, and no general mathematical proof has yet established the relationship between the angular velocities of material line and volume elements.

The above discussion mirrors the fundamental complexity of ``rigid rotation''
when taking into account the diversity of the elementary fluid elements (encompassing the material line, surface, and volume elements), and the associated physical carriers (including vortex filaments and sheets, as the asymptotic models of axial vortices and shear layers).
Moreover, the intrinsic relationship between the rigid rotation of these fluid elements and the swirling motion of an axial vortex remains poorly understood. The present study plans to provide a rigorous clarification of these concepts by defining precisely rigid rotation for each element type, disentangling the contributions from distinct modes of vorticity, and enhancing the theoretical basis for vorticity kinematics.

\section{Rotation of directed material line element and vorticity decomposition}\label{Rotation of directed material line element and vorticity decomposition}
From a Lagrangian perspective, the relative velocity $\delta\bm{u}$ is equal to the rate of change of a material line element $\delta\bm{r}=\delta{r}\bm{e}$, satisfying
\begin{eqnarray}\label{uequaldrdt}
	\delta\bm{u}=\frac{D\delta\bm{r}}{Dt}=\delta\bm{r}\bm{\cdot}\bm{A},
\end{eqnarray}
where $D(\cdot)/Dt$ is the material derivative. Then, by employing the chain rule for time derivative and the vector identity,~\eqref{uequaldrdt} is decomposed as
\begin{eqnarray*}\label{ttt1}
	\bm{e}\left(\frac{1}{\delta r}\frac{D\delta{r}}{Dt}\right)+\frac{D\bm{e}}{Dt}
	=(\bm{e}\bm{\cdot}\bm{A}\bm{\cdot}\bm{e})\bm{e}+\left[\bm{e}\times\left(\bm{e}\bm{\cdot}\bm{A}\right)\right]\times\bm{e}.
\end{eqnarray*}
This decomposition gives rise to two distinct physical processes governed by
\begin{subequations}\label{mm12}
	\begin{eqnarray}\label{mm1}
		\frac{1}{\delta r}\frac{D\delta{r}}{Dt}=\bm{e}\bm{\cdot}\bm{A}\bm{\cdot}\bm{e}=\bm{e}\bm{\cdot}\bm{D}\bm{\cdot}\bm{e},
	\end{eqnarray}
	\begin{eqnarray}\label{mm2}
		\frac{D\bm{e}}{Dt}=\left[\bm{e}\times\left(\bm{e}\bm{\cdot}\bm{A}\right)\right]\times\bm{e}.
	\end{eqnarray}
\end{subequations}
Equation~\eqref{mm1} describes the relative stretching rate of a material line element, solely determined by the strain-rate tensor $\bm{D}$ and the instantaneous orientation $\bm{e}$, where the SAD~\eqref{eq1a} has been used. Equation~\eqref{mm2} motivates defining \textit{the effective angular velocity of rigid rotation} of the unit vector $\bm{e}$ as\footnote{Equation~\eqref{pp2} and the second equality in~\eqref{pp6} share the same form as that of a line element in a rigid body.  The terminology ``rigid-like rotation'' or ``turning'' could be more precise than ``rigid rotation'' for emphasizing the object of research as a fluid line element. We use ``rigid rotation'' throughout this article for conciseness without causing any confusion.}
\begin{eqnarray}\label{WLeff}
	\bm{W}_{L}^{\rm{eff}}(\bm{e})\equiv\bm{e}\times\left(\bm{e}\bm{\cdot}\bm{A}\right),
\end{eqnarray}
where and below we use the notation $(\bm{e})$ occasionally to emphasize the generic dependence of a physical quantity not only on $\bm{x}$ but also the orientation of $\bm{e}$; the subscript ``L'' denotes the line element.
Note that $\bm{W}_{L}^{\rm{eff}}(\bm{e})$ must be perpendicular to $\bm{e}$, thereby representing the effective rigid rotation of $\bm{e}$ in a 2D plane normal to $\bm{W}_{L}^{\rm{eff}}(\bm{e})$.

Substituting directly the SAD~\eqref{eq1a} into~\eqref{mm2} and~\eqref{WLeff} yields
\begin{eqnarray}\label{pp1}
	\bm{W}_{L}^{\rm{eff}}(\bm{e})=\bm{W}_{D}(\bm{e})+\frac{1}{2}\bm{\omega}-\frac{1}{2}\left(\bm{\omega}\bm{\cdot}\bm{e}\right)\bm{e},
\end{eqnarray}
\begin{eqnarray}\label{pp2}
	\frac{D\bm{e}}{Dt}=\bm{W}_{L}^{\rm{eff}}(\bm{e})\times\bm{e},
\end{eqnarray}
where in~\eqref{pp1}, the $\bm{D}$-caused angular velocity component is introduced as
\begin{eqnarray}\label{pp3}
	\bm{W}_{D}(\bm{e})\equiv\bm{e}\times\left(\bm{e}\bm{\cdot}\bm{D}\right).
\end{eqnarray}

\begin{figure}[t]
	\centering
	\includegraphics[width=1.0\columnwidth,trim={0cm 6cm 4cm 3cm},clip]{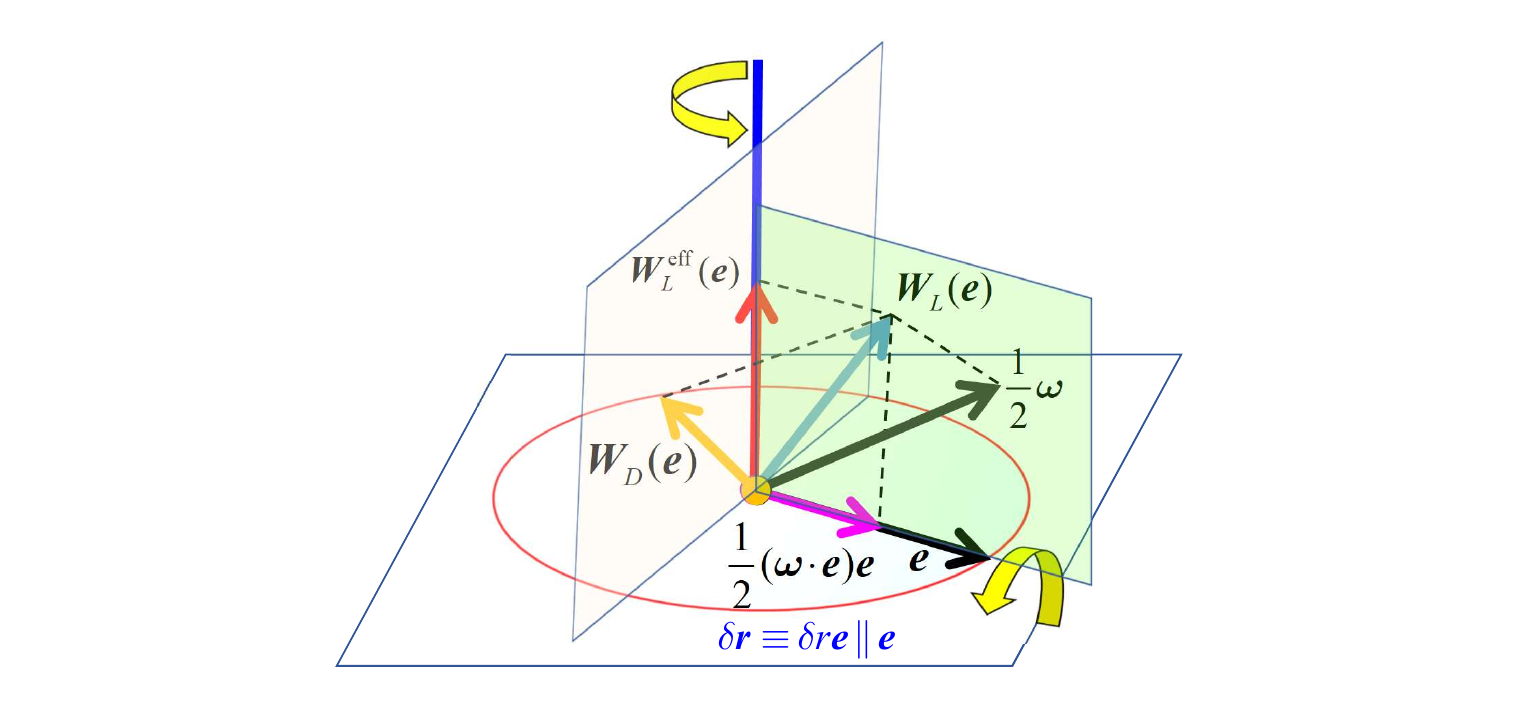}
	\caption{Rotation of a material line element $\delta\bm{r}\equiv\delta{r}\bm{e}$. $\bm{\omega}$ denotes the vorticity; $\bm{W}_{L}^{\rm{eff}}(\bm{e})$ and $\bm{W}_{L}(\bm{e})$ are the effective and total angular velocities of rigid rotation; $\bm{W}_{D}(\bm{e})$ is the specific angular velocity caused by the strain-rate tensor. The yellow arrows indicates the rotation directions of $\bm{W}_{L}^{\rm{eff}}(\bm{e})$ and the gauge term $\frac{1}{2}(\bm{\omega}\bm{\cdot}\bm{e})\bm{e}=\frac{1}{2}\bm{g}_{L}(\bm{e})$.} 
	\label{Line_element_rotation}
\end{figure}

A important observation from~\eqref{pp1} is that the term $\frac{1}{2}(\bm{\omega}\bm{\cdot}\bm{e})\bm{e}$, being parallel to $\bm{e}$, serves as a gauge term with two essential properties. It does not produce an effective rigid rotation of $\bm{e}$ in the plane orthogonal to $\bm{W}_{L}^{\rm{eff}}(\bm{e})$, yet it is mathematically indispensable for maintaining the vectorial equality.
Notably, this term makes no physical contribution to the material evolution rate described by~\eqref{pp2}. Since the rotation-rate tensor $\bm{\Omega}$ and vorticity $\bm{\omega}$ have been fully associated with the rigid rotation (as described by the last term in~\eqref{CStheorem}), decomposing vorticity along $\bm{e}$ becomes unnecessary for analyzing line-element rotation. We therefore incorporate it into $\bm{W}_{L}^{\rm{eff}}(\bm{e})$ to define~\textit{the angular velocity of rigid rotation} of $\bm{e}$:
\begin{eqnarray}\label{pp5}
	\bm{W}_{L}(\bm{e})\equiv\bm{W}_{L}^{\rm{eff}}(\bm{e})+\frac{1}{2}(\bm{\omega}\bm{\cdot}\bm{e})\bm{e},
\end{eqnarray}
leading to the simplification of~\eqref{pp1} and~\eqref{pp2} as
\begin{eqnarray*}
	\bm{W}_{L}(\bm{e})=\bm{W}_{D}(\bm{e})+\frac{1}{2}\bm{\omega},
\end{eqnarray*}
\begin{eqnarray}\label{pp6}
	\frac{D\bm{e}}{Dt}=\bm{W}_{L}(\bm{e})\times\bm{e}.
\end{eqnarray}
The physical quantities in~\eqref{pp1},~\eqref{pp5}, and~\eqref{pp6} are displayed in figure~\ref{Line_element_rotation}.

Equation~\eqref{pp6} supports the unverified assertion by~\citet{Corrsin1972} that~\textit{``a line element is rotated by both the vorticity and the strain rate"}. The term $\frac{1}{2}\bm{\omega}$ represents the classical rigid rotation driven by the vorticity, originating from the last term of~\eqref{CStheorem}.
The contribution of direction-dependent $\bm{W}_{D}(\bm{e})$ to line-element rotation rate was first identified by~\citet{Corrsin1972} in a study of the angular dispersion of a material line element in isotropic turbulence using hydrogen-bubble tracers, also in~\citet{Lumley1987}. As proved by~\citet{MaoFeng2022} in a recent review paper (with a new rigorous proof documented in Appendix~\ref{AP0}): 
\begin{eqnarray}\label{specific_WD}
	\langle\bm{W}_{D}(\bm{e})\rangle\equiv\frac{1}{4\pi}\oint_{\bm{\Sigma}(\bm{e})}\bm{W}_{D}(\bm{e})dS=\bm{0},
\end{eqnarray}
where $\langle\cdot\rangle$ denotes the surface average over all possible orientations
\begin{eqnarray*}
	\bm{e}=(\sin\phi\cos\theta,\sin\phi\sin\theta,\cos\phi)
\end{eqnarray*}
with the ending points located on the surface $\bm{\Sigma}(\bm{e})=\bm{\Sigma}(\theta,\phi)$ of unit sphere centered at the position $\bm{x}$. Here, $\phi$ and $\theta$ are the polar and azimuthal angles of the spherical coordinate system, and $dS=\sin\phi{d\phi}{d\theta}$. 

By~\eqref{specific_WD}, the first equality in~\eqref{pp6} suggests
$\langle\bm{W}_{L}(\bm{e})\rangle=\frac{1}{2}\bm{\omega}$. Since a fluid volume element (without preferred orientation) comprises material line elements spanning all spatial orientations, its rotational behavior fundamentally differs from that of any individual line element. Its angular velocity of rigid rotation $\bm{W}_{V}$ (``V'' denotes the volume element) can be reasonably defined as the mean angular velocity of its constituent line elements:
\begin{eqnarray}\label{pp7b}
	\bm{W}_{V}\equiv\langle\bm{W}_{L}(\bm{e})\rangle=\frac{1}{2}\bm{\omega}.
\end{eqnarray}
Equations~\eqref{specific_WD} and~\eqref{pp7b} indicate that $\bm{W}_{D}(\bm{e})$ induces no net contribution to the rigid rotation of a fluid volume element despite its existence for a fluid, thereby being referred to as~\textit{the specific angular velocity of a material line element}. Equation~\eqref{pp7b} clarifies why the long-unverified solidification hypothesis introduced by Stokes(1845) and~\citet{Truesdell1954}, along with the physical interpretation by~\citet{Corrsin1972} on the angular velocity of a fluid volume element (\S\ref{Cauchy-Stokes theorem and concept of fluid rotation}), remain valid even when only the final constituent motion in~\eqref{CStheorem} is considered, reinforcing the standard Cauchy-Stokes decomposition.

Additionally, when analyzed through~\eqref{pp6} while neglecting the axial stretching and shrinking motions, the three orthogonal material line elements, instantaneously aligned with the principal axes of the strain-rate tensor $\bm{D}$, exhibit identical rigid rotation characterized by the same angular velocity $\frac{1}{2}\bm{\omega}$ (where $\bm{W}_{D}=\bm{0}$ along these axes).
This result provides rigorous mathematical justification for Boussinesq's elegant reformation of Stokes's result [as cited in~\citet{Truesdell1954}]:``\textit{the local angular velocity is the angular velocity of the principal axes of extension in the ordinary sense of rigid rotations.''}

Rearranging the terms in~\eqref{pp1} yields the \textit{line-element-based triple decomposition of vorticity}:
\begin{eqnarray}\label{pp8}
	\bm{\omega}=\bm{R}_{L}(\bm{e})+\bm{s}_{L}(\bm{e})+\bm{g}_{L}(\bm{e}),
\end{eqnarray}
with the vorticity constituents given by
\begin{subequations}\label{eq14}
	\begin{eqnarray}\label{pp9}
		\bm{R}_{L}(\bm{e})\equiv2\bm{W}_{L}^{\rm{eff}}(\bm{e})=2\bm{e}\times\left(\bm{e}\bm{\cdot}\bm{A}\right),
	\end{eqnarray}
	\begin{eqnarray}\label{pp10}
		\bm{s}_{L}(\bm{e})\equiv-2\bm{W}_{D}(\bm{e})=-2\bm{e}\times\left(\bm{e}\bm{\cdot}\bm{D}\right),
	\end{eqnarray}
	\begin{eqnarray}\label{LLL1}
		\bm{g}_{L}(\bm{e})\equiv\left(\bm{\omega}\bm{\cdot}\bm{e}\right)\bm{e}.
	\end{eqnarray}
\end{subequations}
Each component depends on the orientation $\bm{e}$ of a material line element, while their sum yields the total vorticity describing the volume-element rotation.
$\bm{R}_{L}(\bm{e})$ in~\eqref{pp9} represents the effective rigid-rotation mode of a line element. 
If the unit vector $\bm{e}$ is instantaneously frozen to a sufficiently large imaginary rigid body (as an object of reference) rotating with the angular velocity $\bm{W}_{L}^{\rm{eff}}(\bm{e})$, then $\bm{R}_{L}(\bm{e})$ is the vorticity inside that body. 
$\bm{s}_{L}(\bm{e})$ in~\eqref{pp10} represents the spin mode of a material line element, which demonstrates how the strain-rate tensor $\bm{D}$ affects the specific rotation of a material line element. Obviously, it is implied from~\eqref{specific_WD} that $\langle\bm{s}_{L}(\bm{e})\rangle=\bm{0}$. $\bm{g}_{L}(\bm{e})$ accounts for the parallel component of vorticity along $\bm{e}$, originating from the gauge term in~\eqref{pp1}. The directional averages of $\bm{R}_{L}(\bm{e})$ and $\bm{g}_{L}(\bm{e})$ are evaluated to be
\begin{eqnarray*}
	\langle\bm{R}_{L}(\bm{e})\rangle=\frac{2}{3}\bm{\omega},~~\langle\bm{g}_{L}(\bm{e})\rangle=\frac{1}{3}\bm{\omega}.
\end{eqnarray*}
In addition, from~\eqref{pp5} and~\eqref{pp6}, we can write $\bm{R}_{L}(\bm{e})+\bm{g}_{L}(\bm{e})=2\bm{W}_{L}(\bm{e})$ from which an alternative binary vorticity decomposition can be obtained as 
\begin{eqnarray}
	\bm{\omega}=2\bm{W}_{L}(\bm{e})-2\bm{W}_{D}(\bm{e}).
\end{eqnarray}
This decomposition provides a refined understanding of vorticity based on a material line element,  and extends the classical Cauchy-Stokes decomposition by incorporating directional dependence, while still recovering the standard interpretation for volume element when averaged over all orientations.

\section{Rotation of directed material surface element and vorticity decomposition}\label{Rotation of directed material surface element and vorticity decomposition}
The shear/spin mode primarily manifests in two distinct scenarios: through viscous interaction between adjacent fluid surfaces, or at solid boundaries effects that become particularly significant at high Reynolds numbers. This physical understanding necessitates the rigorous treatment of the rotation of a material surface element, being represented by $\delta\bm{\Sigma}\equiv\delta\Sigma\bm{n}_{\Sigma}$, where $\delta\Sigma\equiv\lVert\delta\bm{\Sigma}\rVert$ quantifies the differential surface area, and $\bm{n}_{\Sigma}$ denotes the unit normal vector field. These elements may exhibit arbitrary curvature and can represent rigid or deformable boundaries, fluid-fluid interfaces, and embedded material surfaces within viscous flow domains.
For a vector field defined on the surface, we adopt the subscript convention which uses $\pi$ and $n$ to indicate its tangential and surface-normal components. The same convention are also applied for the tangential and normal derivative operators.

Following the surface deformation tensor introduced by~\citet{Dishington1965}
\begin{eqnarray}\label{ss1}
	\bm{B}=\vartheta\bm{I}-\bm{A}^{\rm T},
\end{eqnarray}
the material rate of change of $\delta\bm{\Sigma}$ is given by~\citep{Batchelor1967,WuJZ2005JFM}
\begin{eqnarray}\label{ss2}
	\frac{1}{\delta\Sigma}\frac{D\delta\bm{\Sigma}}{Dt}=\bm{n}_{\Sigma}\bm{\cdot}\bm{B}=\vartheta\bm{n}_{\Sigma}-\bm{A}\bm{\cdot}\bm{n}_{\Sigma},
\end{eqnarray}
where $\bm{I}$ is the identity tensor, and $\vartheta\equiv\bm{\nabla}\bm{\cdot}\bm{u}$ denotes the velocity divergence (dilatation).
It is noted that the right hand side of~\eqref{ss2} admits an equivalent vortical formulation $-(\bm{n}_{\Sigma}\times\bm{\nabla}_{\pi})\times\bm{u}=-(\bm{n}_{\Sigma}\times\bm{\nabla})\times\bm{u}$, where $\bm{\nabla}_{\pi}\equiv\bm{\nabla}-\bm{n}_{\Sigma}\partial_{n}$ is the surface gradient operator (acting on any tensor field well-defined on the surface), $\partial_{n}$ denotes the surface-normal derivative. Note that $\bm{\nabla}_{\pi}$ in the bracket can be replaced by the full gradient operator $\bm{\nabla}$, recovering the formulation used in~\citet{WuJZ2005JFM}. Since the tangential derivatives are involved alone, the material rate of change of a surface element is completely determined by the instantaneous surface geometry and the velocity distribution along the surface, while remaining independent of external flow conditions. 

Applying the chain rule for the left hand side of~\eqref{ss2} yields
\begin{eqnarray}\label{ss3}
	\frac{1}{\delta\Sigma}\frac{D\delta\bm{\Sigma}}{Dt}
	=\bm{n}_{\Sigma}\left(\frac{1}{\delta\Sigma}\frac{D\delta{\Sigma}}{Dt}\right)+\frac{D\bm{n}_{\Sigma}}{Dt},
\end{eqnarray}
while its right hand side can be equivalently decomposed as
\begin{eqnarray}\label{ss3RHS}
	\vartheta\bm{n}_{\Sigma}-\bm{A}\bm{\cdot}\bm{n}_{\Sigma}=(\bm{\nabla}_{\pi}\bm{\cdot}\bm{u})\bm{n}_{\Sigma}-[\bm{n}_{\Sigma}\times\left(\bm{A}\bm{\cdot}\bm{n}_{\Sigma}\right)]\times\bm{n}_{\Sigma}.
\end{eqnarray}
Equaling the right hand sides of~\eqref{ss3} and~\eqref{ss3RHS} yields the following orthogonal decomposition:
\begin{subequations}\label{ss45}
	\begin{eqnarray}\label{ss4}
		\frac{1}{\delta\Sigma}\frac{D\delta{\Sigma}}{Dt}=\bm{\nabla}_{\pi}\bm{\cdot}\bm{u},
	\end{eqnarray}
	\begin{eqnarray}\label{ss5}
		\frac{D\bm{n}_{\Sigma}}{Dt}=\bm{W}_{\Sigma}^{\rm{eff}}(\bm{n}_{\Sigma})\times\bm{n}_{\Sigma},
	\end{eqnarray}
\end{subequations}
where~\textit{the effective angular velocity of rigid rotation} of $\bm{n}_{\Sigma}$, being perpendicular to $\bm{n}_{\Sigma}$, is defined as
\begin{eqnarray}\label{ss7}
	\bm{W}_{\Sigma}^{\rm{eff}}(\bm{n}_{\Sigma})\equiv-\bm{n}_{\Sigma}\times\left(\bm{A}\bm{\cdot}\bm{n}_{\Sigma}\right).
\end{eqnarray}
Equation~\eqref{ss4} shows that the relative rate of change of the surface area $\delta\Sigma$ is solely determined by the surface velocity divergence $\bm{\nabla}_{\pi}\bm{\cdot}\bm{u}$. Equation~\eqref{ss5} describes the material evolution rate of the surface normal vector $\bm{n}_{\Sigma}$.
We observe that the tensor decomposition $\bm{A}\bm{\cdot}\bm{n}_{\Sigma}=(\partial_{n}u_{n})\bm{n}_{\Sigma}+\bm{\nabla}_{\pi}\bm{u}\bm{\cdot}\bm{n}_{\Sigma}$, where the normal relative stretching rate $\partial_{n}u_{n}=\bm{n}_{\Sigma}\bm{\cdot}\bm{A}\bm{\cdot}\bm{n}_{\Sigma}$, allows for an alternative representation of the right hand side of~\eqref{ss2} as $(\bm{\nabla}_{\pi}\bm{\cdot}\bm{u})\bm{n}_{\Sigma}-\bm{\nabla}_{\pi}\bm{u}\bm{\cdot}\bm{n}_{\Sigma}$. Consequently, in~\eqref{ss7}, $\bm{A}$ can be substituted with $\bm{\nabla}_{\pi}\bm{u}$.

\begin{figure}[t]
	\centering
	\includegraphics[width=1.0\columnwidth,trim={0cm 3.8cm 0cm 2.7cm},clip]{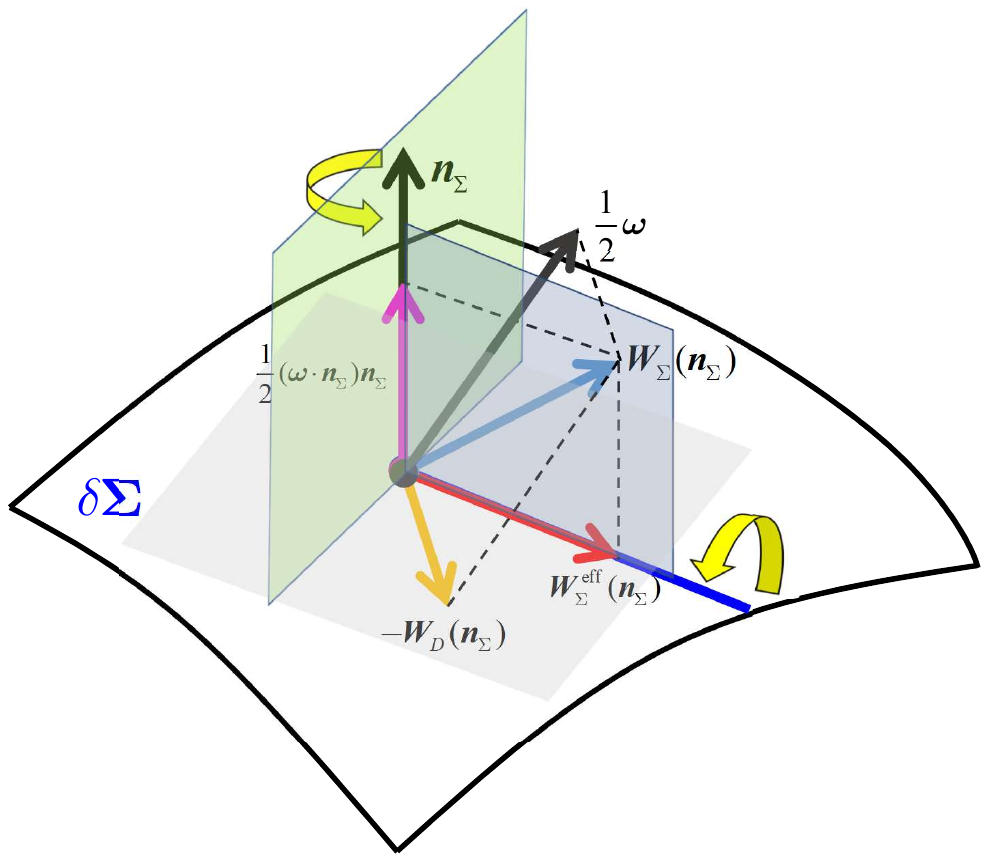}
	\caption{Rotation of a material surface element $\delta\bm{\Sigma}\equiv\delta{\Sigma}\bm{n}_{\Sigma}$. $\bm{\omega}$ is the vorticity; $\bm{W}_{\Sigma}^{\rm{eff}}(\bm{n}_{\Sigma})$ and $\bm{W}_{\Sigma}(\bm{n}_{\Sigma})$ are respectively the effective and total angular velocities of the line element; $\bm{W}_{D}(\bm{n}_{\Sigma})$ is the specific angular velocity caused by the strain-rate tensor. The yellow arrows indicate the rotation directions of $\bm{W}_{\Sigma}^{\rm{eff}}(\bm{n}_{\Sigma})$ and the gauge term $\frac{1}{2}(\bm{\omega}\bm{\cdot}\bm{n}_{\Sigma})\bm{n}_{\Sigma}=\frac{1}{2}\bm{g}_{\Sigma}(\bm{n}_{\Sigma})$.} 
	\label{Surface_element_rotation}
\end{figure}

Using the SAD in~\eqref{eq1a} to replace $\bm{A}$ in~\eqref{ss7}, we obtain
\begin{eqnarray}\label{ss8}
	\bm{W}_{\Sigma}^{\rm{eff}}(\bm{n}_{\Sigma})=-\bm{W}_{D}(\bm{n}_{\Sigma})+\frac{1}{2}\bm{\omega}-\frac{1}{2}\left(\bm{\omega}\bm{\cdot}\bm{n}_{\Sigma}\right)\bm{n}_{\Sigma}.
\end{eqnarray}
As observed from~\eqref{ss5}, the last term in~\eqref{ss8} makes no contribution to the material evolution rate of the normal vector $\bm{n}_{\Sigma}$. Therefore, we can define the angular velocity of rigid rotation of $\bm{n}_{\Sigma}$ as
\begin{eqnarray}\label{mmm1}
	\bm{W}_{\Sigma}(\bm{n}_{\Sigma})\equiv\bm{W}_{\Sigma}^{\rm{eff}}(\bm{n}_{\Sigma})+\frac{1}{2}\left(\bm{\omega}\bm{\cdot}\bm{n}_{\Sigma}\right)\bm{n}_{\Sigma}.
\end{eqnarray} 
From~\eqref{mmm1}, it follows that $\bm{\omega}\bm{\cdot}\bm{n}_{\Sigma}=2\bm{W}_{\Sigma}(\bm{n}_{\Sigma})\bm{\cdot}\bm{n}_{\Sigma}$, which holds only if the velocity remains continuous across $\delta\bm{\Sigma}$. This condition is consistent with the case of a rotating rigid body immersed in a viscous flow, where the no-slip boundary condition is enforced on the solid surface. Consequently, by combining~\eqref{ss5},~\eqref{ss8}, and~\eqref{mmm1}, we get
\begin{eqnarray*}
	\bm{W}_{\Sigma}(\bm{n}_{\Sigma})=-\bm{W}_{D}(\bm{n}_{\Sigma})+\frac{1}{2}\bm{\omega},
\end{eqnarray*}
\begin{eqnarray}\label{mmm2}
	\frac{D\bm{n}_{\Sigma}}{Dt}=\bm{W}_{\Sigma}(\bm{n}_{\Sigma})\times\bm{n}_{\Sigma}.
\end{eqnarray}
The physical quantities in~\eqref{mmm1} and~\eqref{mmm2} are illustrated in figure~\ref{Surface_element_rotation}.
Equation~\eqref{mmm2} indicates that a surface element undergoes the rigid rotation due to the combined effects of vorticity and strain rate, analogous to the interpretation of line element rotation insisted by~\citet{Corrsin1972}. However, the specific angular velocity of a surface element is identified as $-\bm{W}_{D}(\bm{n}_{\Sigma})$, differing from that of a line element by a sign change. Since a fluid element contains material surface elements oriented in all possible directions, the surface average of $\bm{W}_{D}(\bm{n}_{\Sigma})$ over all possible directions of $\bm{n}_{\Sigma}$ must vanish:
\begin{eqnarray}
	\langle\bm{W}_{D}(\bm{n}_{\Sigma})\rangle=\bm{0},
\end{eqnarray}
from which the first equality in~\eqref{mmm2} directly yields $\langle\bm{W}_{\Sigma}(\bm{n}_{\Sigma})\rangle=\frac{1}{2}\bm{\omega}$. 
Subsequently, a comparison with~\eqref{pp7b} gives
\begin{eqnarray}\label{pp7bsurface}
	\bm{W}_{V}=\langle\bm{W}_{\Sigma}(\bm{n}_{\Sigma})\rangle=\frac{1}{2}\bm{\omega},
\end{eqnarray}
demonstrating that the angular velocity of rigid rotation of a volume element can be equivalently interpreted as the mean angular velocity of its constituent surface elements averaged over all possible spatial orientations. In addition, since $\bm{W}_{D}(\bm{n}_{\Sigma})\perp\bm{n}_{\Sigma}$, the first equality in~\eqref{mmm2} implies that $\bm{W}_{\Sigma}(\bm{n}_{\Sigma})\bm{\cdot}\bm{n}_{\Sigma}=\frac{1}{2}\bm{\omega}\bm{\cdot}\bm{n}_{\Sigma}$ which just coincides with the surface-normal component of the ``mean angular velocity'' $\bm{W}_{\rm{KB}}$ introduced by~\citet{Batchelor1967} and Kelvin (1869) [see~\citet{Truesdell1954}]:
\begin{eqnarray}
\bm{W}_{\rm{KB}}\bm{\cdot}\bm{n}_{\Sigma}=\lim_{\epsilon\rightarrow 0}\frac{1}{2\pi\epsilon^2}\oint_{C}\bm{u}\bm{\cdot}d\bm{r}=\frac{1}{2}\bm{\omega}\bm{\cdot}\bm{n}_{\Sigma},
\end{eqnarray}
for any open disk (with the radius $\epsilon$) bounded by a circle $C$ centered at $\bm{x}$ and having the unit normal vector $\bm{n}_{\Sigma}$.

From~\eqref{ss8}, we obtain the~\textit{surface-element-based triple decomposition of vorticity}:
\begin{eqnarray}\label{ss9}
	\bm{\omega}=\bm{R}_{\Sigma}(\bm{n}_{\Sigma})+\bm{s}_{\Sigma}(\bm{n}_{\Sigma})+\bm{g}_{\Sigma}(\bm{n}_{\Sigma}),
\end{eqnarray}
with the $\bm{n}_{\Sigma}$-dependent vorticity constituents given by
\begin{subequations}\label{SEB}
	\begin{eqnarray}\label{ss10a}
		\bm{R}_{\Sigma}(\bm{n}_{\Sigma})\equiv2\bm{W}_{\Sigma}^{\rm{eff}}(\bm{n}_{\Sigma})=-2\bm{n}_{\Sigma}\times\left(\bm{A}\bm{\cdot}\bm{n}_{\Sigma}\right),
	\end{eqnarray}
	\begin{eqnarray}\label{ss10b}
		\bm{s}_{\Sigma}(\bm{n}_{\Sigma})\equiv2\bm{W}_{D}(\bm{n}_{\Sigma})=2\bm{n}_{\Sigma}\times\left(\bm{D}\bm{\cdot}\bm{n}_{\Sigma}\right),
	\end{eqnarray}
	\begin{eqnarray}\label{ss10c}
		\bm{g}_{\Sigma}(\bm{n}_{\Sigma})\equiv\left(\bm{\omega}\bm{\cdot}\bm{n}_{\Sigma}\right)\bm{n}_{\Sigma}.
	\end{eqnarray}
\end{subequations}
In~\eqref{SEB}, $\bm{R}_{\Sigma}(\bm{n}_{\Sigma})$ and $\bm{s}_{\Sigma}(\bm{n}_{\Sigma})$ characterize the rigid rotation and spin modes of a surface element, respectively. As demonstrated in~\S\ref{spin_mode_interpretation}, $\bm{s}_{\Sigma}(\bm{n}_{\Sigma})$ is directly linked to the surface shear stress acting on a surface element. $\bm{g}_{\Sigma}(\bm{n}_{\Sigma})$ represents the gauge contribution in determining $D\bm{n}_{\Sigma}/Dt$, which however, cannot be neglected in this decomposition. If the surface belongs to part of a stationary solid boundary, then the boundary vorticity is solely determined by the spin mode $\bm{s}_{\Sigma}(\bm{n}_{\Sigma})$ due to the velocity adherence. From~\eqref{mmm1} and~\eqref{mmm2}, we derive the relationship $\bm{R}_{\Sigma}(\bm{n}_{\Sigma})+\bm{g}_{\Sigma}(\bm{n}_{\Sigma})=2\bm{W}_{\Sigma}(\bm{n}_{\Sigma})$, and the alternative formulation of vorticity decomposition
\begin{eqnarray}
	\bm{\omega}=2\bm{W}_{\Sigma}(\bm{n}_{\Sigma})+2\bm{W}_{D}(\bm{n}_{\Sigma}).
\end{eqnarray}

\section{Intrinsic relations for orthogonal line and surface elements}\label{intrinsic}
Consider a pair of orthogonal line and surface elements $\left(\delta\bm{r},\delta\bm{\Sigma}\right)=(\delta{r}\bm{e},\delta\Sigma\bm{n}_{\Sigma})$ where $\bm{e}=\bm{n}_{\Sigma}$ at a time instant.
Since a volume element $\delta{V}$ is given by their product: $\delta{V}=\delta\bm{r}\bm{\cdot}\delta\bm{\Sigma}=\delta{r}\delta{\Sigma}$, there must exist a kinematic relationship between the angular velocities of rigid rotation of the line and surface elements. It is shown that the sum of~\eqref{mm1} and~\eqref{ss4} yields
\begin{eqnarray}\label{VVV1}
	\frac{1}{\delta r}\frac{D\delta r}{Dt}+	\frac{1}{\delta\Sigma}\frac{D\delta\Sigma}{Dt}=\vartheta.
\end{eqnarray}
Equation~\eqref{VVV1} can also be derived by evaluating the material derivatives of both sides of $\ln\delta{V}=\ln\delta{r}+\ln\delta{\Sigma}$ and using the relation
$\vartheta=\delta{V}^{-1}D\delta{V}/Dt$. It suggests that the dilatational motion of a volume element results from the superposition of surface stretching motion and normal-direction extension.

Due to the presence of strain rate for fluid, neither $\bm{W}_{\Sigma}(\bm{n}_{\Sigma})$ nor $\bm{W}_{L}(\bm{n}_{\Sigma})$ equals to $\frac{1}{2}\bm{\omega}$  individually (a condition satisfied only for rigid-body rotation), while each differs from $\frac{1}{2}\bm{\omega}$ by the specific angular velocity $\bm{W}_{D}(\bm{n}_{\Sigma})$.
From the first equalities in~\eqref{pp6} and~\eqref{mmm2}, the sum and difference of $\bm{W}_{\Sigma}(\bm{n}_{\Sigma})$ and $\bm{W}_{L}(\bm{n}_{\Sigma})$ are given by
\begin{subequations}\label{uuu810}
	\begin{eqnarray}\label{uuu8}
		\bm{W}_{\Sigma}(\bm{n}_{\Sigma})+\bm{W}_{L}(\bm{n}_{\Sigma})=\bm{\omega},
	\end{eqnarray}
	\begin{eqnarray}\label{uuu10}
		\bm{W}_{\Sigma}(\bm{n}_{\Sigma})-\bm{W}_{L}(\bm{n}_{\Sigma})=-\bm{s}_{\Sigma}(\bm{n}_{\Sigma})=-2\bm{W}_{D}(\bm{n}_{\Sigma}).
	\end{eqnarray}
\end{subequations}
Equivalently, in terms of vorticity constituents,~\eqref{uuu10} can be expressed as
\begin{eqnarray}\label{uuu11}
	\bm{R}_{L}(\bm{n}_{\Sigma})-\bm{R}_{\Sigma}(\bm{n}_{\Sigma})=2\bm{s}_{\Sigma}(\bm{n}_{\Sigma}).
\end{eqnarray}
Equations~\eqref{uuu8} and~\eqref{uuu10} establish the general relations between the angular velocities of $\left(\delta\bm{r},\delta\bm{\Sigma}\right)$, revealing their complementary nature in both kinematic and geometric aspects. 
First, equation~\eqref{VVV1} shows that the angular velocity of rigid rotation for a fluid volume element is the arithmetic mean of those for the orthogonal line and surface elements,
\begin{eqnarray}\label{VVV2}
	\bm{W}_{V}=\frac{1}{2}\left[\bm{W}_{\Sigma}(\bm{n}_{\Sigma})+\bm{W}_{L}(\bm{n}_{\Sigma})\right].
\end{eqnarray}
Second,~\eqref{uuu10} demonstrates that the surface element's specific angular velocity determines, by a factor of two, the difference in angular velocities between line and surface elements.
Taking the directional averages of both sides of~\eqref{VVV2} over all possible orientations of $\bm{n}_{\Sigma}$ recovers the previous result for volume element $\bm{W}_{V}=\frac{1}{2}\bm{\omega}$.

Taking the cross product of $\bm{n}_{\Sigma}$ with both sides of~\eqref{uuu8} yields
\begin{eqnarray}\label{VVV3}
	\frac{D\bm{e}}{Dt}+\frac{D\bm{n}_{\Sigma}}{Dt}=\bm{\omega}\times\bm{n}_{\Sigma}.
\end{eqnarray}
Then, the combination of~\eqref{VVV1} and~\eqref{VVV3} leads to
\begin{eqnarray}\label{VVV4}
	\frac{1}{\delta\Sigma}\frac{D\delta\bm{\Sigma}}{Dt}
	+\frac{1}{\delta{r}}\frac{D\delta\bm{r}}{Dt}=\vartheta\bm{n}_{\Sigma}+\bm{\omega}\times\bm{n}_{\Sigma},
\end{eqnarray}
where the explicit dependence on $\bm{e}$ in the angular velocities has been formally eliminated. Equation~\eqref{VVV4} can alternatively be derived by summing~\eqref{uequaldrdt} and~\eqref{ss2}, utilizing the relations $\bm{B}+\bm{A}=\vartheta\bm{I}+2\bm{\varOmega}$ and $2\bm{n}_{\Sigma}\bm{\cdot}\bm{\varOmega}=\bm{\omega}\times\bm{n}_{\Sigma}$.

\section{Physical interpretation on spin mode of vorticity}\label{spin_mode_interpretation}
Let us consider a material surface element $\delta\bm{\Sigma}\equiv\delta\Sigma\bm{n}_{\Sigma}$.
Building on the generalized decomposition of the VGT $\bm{A}\equiv\bm{\nabla}\bm{u}$ for arbitrarily moving and continuously deforming surfaces~\citep{WuJZ2005JFM}, we express the VGT in a more convenient form:
\begin{eqnarray}\label{uuu0}
	\bm{A}=\bm{n}_{\Sigma}\bm{n}_{\Sigma}\partial_{n}u_{n}+\bm{n}_{\Sigma}\left(\bm{\omega}_{r}\times\bm{n}_{\Sigma}\right)+\bm{n}_{\Sigma}\left(\bm{W}_{\Sigma}\times\bm{n}_{\Sigma}\right)
	-\left(\bm{W}_{\Sigma}\times\bm{n}_{\Sigma}\right)\bm{n}_{\Sigma}
	+{\left(\bm{\nabla}_{\pi}\bm{u}\right)_{\pi}},
\end{eqnarray}
where $\bm{\omega}_{r}\equiv\bm{\omega}-2\bm{W}_{\Sigma}(\bm{n}_{\Sigma})$ represents the relative vorticity with $-2\bm{W}_{\Sigma}(\bm{n}_{\Sigma})$ accounting for the additional vorticity from the angular velocity of rigid rotation of $\bm{n}_{\Sigma}$. The constraint in~\eqref{mmm1} requires $\bm{\omega}_{r}$ to remain within the tangent space of $\delta\bm{\Sigma}$. The decomposition in~\eqref{uuu0} reveals distinct physical interpretations for each component: the normal-normal (N-N) component is determined by $\partial_{n}u_{n}$, representing the relative stretching rate normal to the surface; the normal-tangential (N-T) and tangential-normal (T-N) components are coupled to both the relative vorticity $\bm{\omega}_{r}$ and the surface element's angular velocity $\bm{W}_{\Sigma}(\bm{n}_{\Sigma})$; the tangential-tangential (T-T) component is fully characterized by the surface velocity distribution, independent of off-surface velocity gradients.

The SAD of~\eqref{uuu0} yields the intrinsic formulations of the rate-of-strain tensor $\bm{D}$ and vorticity tensor $\bm{\varOmega}$:
\begin{subequations}
	\begin{equation}\label{uuu1}
		\bm{D}=\bm{n}_{\Sigma}\bm{n}_{\Sigma}\partial_{n}u_{n}+\frac{1}{2}\bm{n}_{\Sigma}(\bm{\omega}_{r}\times\bm{n}_{\Sigma})+\frac{1}{2}(\bm{\omega}_{r}\times\bm{n}_{\Sigma})\bm{n}_{\Sigma}+\mathscr{S}[{\left(\bm{\nabla}_{\pi}\bm{u}\right)_{\pi}}],
	\end{equation}
	\begin{equation}\label{uuuu1}
		\bm{\varOmega}=\frac{1}{2}\bm{n}_{\Sigma}(\bm{\omega}\times\bm{n}_{\Sigma})-\frac{1}{2}(\bm{\omega}\times\bm{n}_{\Sigma})\bm{n}_{\Sigma}+\mathscr{A}[{\left(\bm{\nabla}_{\pi}\bm{u}\right)_{\pi}}].
	\end{equation}
\end{subequations}
Equation~\eqref{uuu1} represents the generalized Caswell formula (alternatively known as the Caswell-Wu formula), while~\eqref{uuuu1} essentially realizes the orthogonal decomposition of vorticity in a tensorial form. 
Substituting~\eqref{uuu1} into~\eqref{ss10b} yields an elegant surface identity:
\begin{eqnarray}\label{uuu2}
	\bm{s}_{\Sigma}(\bm{n}_{\Sigma})=2\bm{W}_{D}(\bm{n}_{\Sigma})=\bm{\omega}_{r},
\end{eqnarray}
showing that the spin (in the surface-element-based vorticity decomposition) is identical to the relative vorticity.
To our knowledge, this is an important result that has never been reported in prior literature. An immediate consequence of~\eqref{uuu2} is the universal expression for surface shear stress (exerted by the fluid at the $\bm{n}_{\Sigma}$-side to the surface) in Newtonian fluids:
\begin{eqnarray}\label{uuu3}
	\bm{\tau}=\mu\bm{\omega}_{r}\times\bm{n}_{\Sigma}=\mu\bm{s}_{\Sigma}(\bm{n}_{\Sigma})\times\bm{n}_{\Sigma},
\end{eqnarray}
with $\mu$ denoting the dynamic viscosity. An intrinsic orthonormal surface reference frame consists of a conjugate orthogonal pair $(\bm{\tau},\bm{\omega}_{r})$ on the surface and the surface normal vector $\bm{n}_{\Sigma}$, analogous to the $\bm{\tau}$ frame~\citep{Wu2006vorticity} on a stationary solid boundary. For a material line element $\delta\bm{r}=\delta{r}\bm{e}$ with complementary surface element  $\delta\bm{\Sigma}$ (where $\bm{n}_{\Sigma}=\bm{e}$), we find the spin terms satisfy $\bm{s}_{L}(\bm{n}_{\Sigma})=-\bm{\omega}_{r}=-\bm{s}_{\Sigma}(\bm{n}_{\Sigma})$, and the surface shear stress is $\bm{\tau}=\mu\bm{n}_{\Sigma}\times\bm{s}_{L}(\bm{n}_{\Sigma})$. These relations fundamentally validate the spin mode representation through $\bm{s}_{\Sigma}(\bm{n}_{\Sigma})$ and $\bm{s}_{L}(\bm{n}_{\Sigma})$.

Applying the intrinsic triple decomposition of the strain-rate tensor, $\bm{D}=\vartheta\bm{I}+\bm{\varOmega}-\bm{B}$~\citep{WuJZ2015book}, yields a double decomposition of the spin mode $\bm{s}_{\Sigma}(\bm{n}_{\Sigma})$ as
\begin{subequations}
	\begin{eqnarray}\label{aaa2}
		\bm{s}_{\Sigma}(\bm{n}_{\Sigma})=2\bm{n}_{\Sigma}\times(\bm{n}_{\Sigma}\bm{\cdot}\bm{\varOmega})-2\bm{n}_{\Sigma}\times(\bm{n}_{\Sigma}\bm{\cdot}\bm{B}).
	\end{eqnarray}
	On the right hand side of~\eqref{aaa2}, the first term, induced by $\bm{\varOmega}$, is evaluated as
	\begin{eqnarray}\label{aaa3}
		\bm{s}_{\bm\varOmega}\equiv2\bm{n}_{\Sigma}\times(\bm{n}_{\Sigma}\bm{\cdot}\bm{\varOmega})=\bm{\omega}-(\bm{\omega}\bm{\cdot}\bm{n}_{\Sigma})\bm{n}_{\Sigma},
	\end{eqnarray}
	and the second term due to $\bm{B}$ is evaluated as
	\begin{eqnarray}\label{aaa4}
		\bm{s}_{\bm{B}}&\equiv&-2\bm{n}_{\Sigma}\times(\bm{n}_{\Sigma}\bm{\cdot}\bm{B})\nonumber\\
		&=&-2\bm{W}_{\Sigma}(\bm{n}_{\Sigma})+2(\bm{W}_{\Sigma}(\bm{n}_{\Sigma})\bm{\cdot}\bm{n}_{\Sigma})\bm{n}_{\Sigma}.
	\end{eqnarray}
\end{subequations}
By applying the relation $\bm{\omega} \cdot \bm{n}_{\Sigma} = 2\bm{W}_{\Sigma}(\bm{n}_{\Sigma}) \cdot \bm{n}_{\Sigma}$, the sum of~\eqref{aaa3} and~\eqref{aaa4} recovers the exact relative vorticity $\bm{\omega}_{r}$, and we find $\bm{s}_{\bm B}=-2\bm{W}_{\Sigma}^{\rm{eff}}(\bm{n}_{\Sigma})=-\bm{R}_{\Sigma}(\bm{n}_{\Sigma})$. The respective contributions to the surface shear stress are
\begin{subequations}\label{tau_decomp}
	\begin{eqnarray}\label{tau_decomp1}
		\bm{\tau}=\bm{\tau}_{\bm\varOmega}+\bm{\tau}_{\bm{B}},
	\end{eqnarray}
	\begin{eqnarray}\label{tau_decomp2}
		\bm{\tau}_{\bm\varOmega}\equiv\mu\bm{s}_{\bm\varOmega}\times\bm{n}_{\Sigma}=\mu\bm{\omega}\times\bm{n}_{\Sigma},
	\end{eqnarray}
	\begin{eqnarray}\label{tau_decomp3}
		\bm{\tau}_{\bm{B}}\equiv\mu\bm{s}_{\bm{B}}\times\bm{n}_{\Sigma}=-2\mu\bm{W}_{\Sigma}(\bm{n}_{\Sigma})\times\bm{n}_{\Sigma}=-2\mu\frac{D\bm{n}_{\Sigma}}{Dt},
	\end{eqnarray}
\end{subequations}
In~\eqref{tau_decomp2}, $\bm{\tau}_{\bm\varOmega}$ arises from the tangential component of the surface vorticity. The surface deformation tensor $\bm{B}$ does not appear in the momentum equation for a fluid with constant shear viscosity, since $\bm{\nabla}\bm{\cdot}\bm{B}=\bm{0}$~\citep{Wu1995POF}.
However, it plays a critical role in generating the local surface shear stress. Although the integral of $\bm{t}_{s}\equiv-2\mu\bm{n}_{\Sigma}\bm{\cdot}\bm{B}$ over a closed boundary surface always vanishes (as guaranteed by the Gauss theorem or the generalized Stokes theorem), its local contribution remains significant.
As the tangential component of $\bm{t}_{s}$, $\bm{\tau}_{\bm{B}}$ in~\eqref{tau_decomp3}, represents the shear stress contribution, governed by the material derivative of $\bm{n}_{\Sigma}$. Physically, $\bm{\tau}_{\bm{B}}$ can be interpreted as the tangential viscous resistance to changes in the orientation of $\delta\bm{\Sigma}$. 

Recall that the {Helmholtz-Hodge decomposition} provides a global decomposition of the velocity field~\citep{WuJZ2015book}:
\begin{eqnarray}\label{uuu6}
	\bm{u}=\bm{\nabla}\Phi+\bm{v},~~\bm{\nabla}\bm{\cdot}\bm{v}=0,~~\bm{\nabla}\times\bm{v}=\bm{\omega}.
\end{eqnarray}
This decomposition is unique in either an unbounded domain, or a bounded domain where $\bm{v}$ is parallel to the boundary.
Expanding~\eqref{ss7} gives
\begin{eqnarray}\label{uuu5}
	\bm{W}_{\Sigma}^{\rm{eff}}(\bm{n}_{\Sigma})=-\bm{n}_{\Sigma}\times\left(\bm{\nabla}_{\pi}u_{n}+\bm{u}\bm{\cdot}\bm{K}\right),
\end{eqnarray}
where $\bm{K}\equiv-\bm{\nabla}_{\pi}\bm{n}_{\Sigma}$ is the surface curvature tensor of $\delta\bm{\Sigma}$. 
This constitutes a distinctive characteristic of moving or deformable boundaries, in contrast to stationary walls. 
Substituting~\eqref{uuu6} into~\eqref{uuu5} yields the following partitioned expression:
\begin{eqnarray}\label{uuu7}
	\bm{W}_{\Sigma}^{\rm{eff}}(\bm{n}_{\Sigma})=-\bm{n}_{\Sigma}\times\left(\bm{\nabla}_{\pi}\partial_{n}\Phi+\bm{\nabla}\Phi\bm{\cdot}\bm{K}\right)-\bm{n}_{\Sigma}\times\left(\bm{\nabla}_{\pi}v_{n}+\bm{v}\bm{\cdot}\bm{K}\right).
\end{eqnarray}
The first term on the right hand side originates from the pure potential flow component, which directly influences both $\bm{\omega}_{r}$ and $\bm{\tau}_{\bm{B}}$. The second term represents the contribution from the transverse velocity component.
\section{Field description based on streamline}\label{Field description based on streamline}
Historically, vortices have often been defined intuitively based
on closed or spiraling streamlines.
For instance, \citet{Robinson1991} proposed that a vortex exists when instantaneous streamlines projected onto a plane normal to the vortex core exhibit a roughly circular or spiral pattern in a reference frame moving with the vortex core center.
However, as~\citet{Lugt1979} noted, streamlines lack Galilean and rotational invariance, meaning recirculatory streamline patterns at a certain instant in time do not necessarily reflect true vortex motions in which fluid particles rotate around a common axis. Consequently, a robust vortex identification method must satisfy Galilean invariance, keeping the vortex invariant under the choice of reference systems~\citep{LiuCQ2020}. Despite these limitations viewed from the algebraic perspective, the potential of streamline-based analysis within a field-theoretic framework remains underexplored. In particular, a systematic investigation into the interplay among streamlines, vorticity modes, and the rotational kinematics of directed material elements is still lacking.

In this section, we rigorously define the orbital rotation and spin modes of vorticity in terms of streamline geometry within a given reference frame, providing explicit vectorial expressions for each. These distinct modes are endowed with clear physical interpretations. Building on this foundation, we establish the connection between direction-dependent vorticity constituents and those in the characteristic algebraic description (e.g., Liutex) in \S\ref{Transformation between characteristic and field descriptions} and~\S\ref{sign_of_gamma}.

\subsection{Streamline geometry and Frenet-Serret intrinsic triad}\label{Streamline element and Frenet-Serret intrinsic triad}
In the field description, the counterpart of a material line element $\delta\bm{r}=\delta{r}\bm{e}$ is an infinitesimal streamline segment $\mathcal{C}$, denoted as $\delta\bar{\bm{r}}=\delta{s}\bm{t}$, where $s$ is the arc length parameter measured from a reference point on $\mathcal{C}$ and $\delta{s}$ is its infinitesimal increment. At any given instant, $\delta\bar{\bm{r}}$ coincides with $\delta\bm{r}$. 
Each point on $\mathcal{C}$ is associated with an orthonormal right-handed triad $(\bm{t},\bm{n},\bm{b})$, consisting of the unit tangent vector $\bm{t}$, the principal normal vector $\bm{n}$ (pointing toward the streamline's center of curvature), and the binormal vector $\bm{b}\equiv\bm{t}\times\bm{n}$ (figure~\ref{Field_description_streamline}). 
The local curvature of $\mathcal{C}$ is denoted by $\kappa(s)$. 
The intrinsic evolution of $(\bm{t},\bm{n},\bm{b})$ along $\mathcal{C}$  is governed by the Frenet-Serret formulas from classical differential geometry~\citep{Serrin1959,ChenWH2002}.
The osculating plane (spanned by $\bm{t}$ and $\bm{n}$) and the normal plane (spanned by $\bm{n}$ and $\bm{b}$) define the local geometry of $\mathcal{C}$.
Note that $\bm{n}$ is generally different from the unit normal vector $\bm{n}_{\Sigma}$ of an oriented surface. The velocity field along $\mathcal{C}$ is purely tangential and given by $\bm{u}=q\bm{t}$, where $q\equiv\bm{u}\bm{\cdot}\bm{t}=\lVert\bm{u}\rVert$ is the velocity magnitude. $\partial_{s}\equiv\partial/\partial s$ represents the partial derivative with respect to $s$ for a physical quantity.

\subsection{Orbital rotation mode of vorticity}
For a frozen velocity field, we observe that an orbital rotation along a streamline (segment) $\mathcal{C}$ occurs if and only if $\bm{t}\times\partial_{s}\bm{u}\neq\bm{0}$ (the non-collinearity between the tangent vector and the velocity variation in the osculating plane). Indeed, using the Frenet-Serret formula $\partial_{s}\bm{t}=\kappa\bm{n}$ and noting that $\bm{t}\times\bm{u}=\bm{0}$, we find
\begin{eqnarray}\label{C8a1}
	\bm{t}\times\partial_{s}\bm{u}=\kappa{q}\bm{b},
\end{eqnarray}
which indicates a local circular motion in the osculating plane about the binormal axis $\bm{b}$. Thus, orbital rotation arises from streamline curvature $\kappa(s)\neq{0}$ and nonzero speed $q\neq{0}$. The radius of curvature is $r_{0}\equiv\kappa^{-1}(s)$.

Since  the directional derivative of velocity along a streamline can be expressed as $\partial_{s}\bm{u}=\bm{t}\bm{\cdot}\bm{A}$,~\eqref{C8a1} can be adjusted to the form:
\begin{eqnarray}\label{C8a2}
	\bm{\psi}_{L}(\bm{t})\equiv\bm{t}\times\left(\bm{t}\bm{\cdot}\bm{A}\right)=\kappa{q}\bm{b}.
\end{eqnarray}
We emphasize that in any flow field, the binormal vector $\bm{b}$ serves as the unique physically identifiable rotation axis external to the streamline orbit. 
Substituting $\bm{e} = \bm{t}$ into~\eqref{WLeff} yields the identity $\bm{W}_{L}^{\rm{eff}}(\bm{t}) = \bm{\psi}_{L}(\bm{t})$. 
This reveals that $\bm{\psi}_{L}(\bm{t})$ in~\eqref{C8a2} can be interpreted as the effective angular velocity of rigid rotation of a material line element instantaneously aligned with $\mathcal{C}$ (i.e., when $\bm{e} = \bm{t}$). Then, the vorticity in a reference rigid body (frozen with $\bm{t}$) rotating with the angular velocity $\bm{\psi}_{L}(\bm{t})$ must be (figure~\ref{Field_description_streamline})
\begin{eqnarray}\label{C8a3}
	\bm{R}_{L}(\bm{t})=2\bm{t}\times\left(\bm{t}\bm{\cdot}\bm{A}\right)=2\kappa{q}\bm{b}.
\end{eqnarray}
We identify $\bm{R}_{L}(\bm{t})$ as the orbital rotation mode of vorticity, obtainable directly from~\eqref{pp9} with $\bm{e} = \bm{t}$. Notably, its binormal component reduces to $R_{b}\equiv\bm{R}_{L}(\bm{t})\bm{\cdot}\bm{b}=2\kappa{q}$.

\subsection{Spin mode of vorticity}
Equation~\eqref{C8a3} immediately evokes a fundamental but often overlooked vorticity representation in the Frenet-Serret frame.
This formulation appears in classical literature~\citep[\S20]{Serrin1959} when expressed in the local rectangular coordinates aligned with $(\bm{t},\bm{n},\bm{b})$:
\begin{eqnarray}\label{C8a4}
	\bm{\omega}=\xi{q}\bm{t}+\partial_{b}q\bm{n}+\left(\kappa{q}-\partial_{n}q\right)\bm{b},
\end{eqnarray}
where the scalar factor $\xi$
\begin{eqnarray*}\label{C8a5}
	\xi\equiv\bm{t}\bm{\cdot}\left(\bm{\nabla}\times\bm{t}\right)
	=\bm{b}\bm{\cdot}\partial_{n}\bm{t}-\bm{n}\bm{\cdot}\partial_{b}\bm{t}
\end{eqnarray*}
was historically termed as the abnormality (also referred to as the torsion of the curve system or torsion of neighboring vector-lines).  It was introduced as a measure of the departure of the velocity field from the property of having a normal congruence~\citep{Truesdell1954}.
The physical interpretation of this subtle parameter once perplexed prominent researchers, owing to the inherent limitation of the Frenet-Serret formalism $-$ it applies to an individual curve but not to a family of neighboring curves.
However, the ambiguity is readily resolved by recognizing that
$\xi{q}^{2}=\bm{\omega}\bm{\cdot}\bm{u}\equiv{h}$ represents the helicity density, implying $\xi=h/q^2$ (the ratio of helicity density to twice the kinetic energy density). In contrast to Serrin’s classical derivation, we can straightforwardly express the vorticity as: $\bm{\omega}=\bm{\nabla}\times\left(q\bm{t}\right)=q\bm{\nabla}\times\bm{t}+\bm{\nabla}q\times\bm{t}$. The first term expands further using the Frenet-Serret relations: $q\bm{\nabla}\times\bm{t}=\xi{q}\bm{t}+\kappa{q}\bm{b}$, and the second term captures the variation of $q$ along the principal normal and binormal directions: $\bm{\nabla}q\times\bm{t}=\bm{n}\partial_{b}q-\bm{b}\partial_{n}q$, as appeared in~\eqref{C8a4}. Note that the binormal component of vorticity is
\begin{eqnarray*}
	\omega_{b}\equiv\bm{\omega}\bm{\cdot}\bm{b}=\kappa{q}-{\partial}_{n}q,
\end{eqnarray*}
governs local rotational kinematics through its competing curvature and shear terms.

\begin{figure}[t]
	\centering
	\includegraphics[width=0.5\columnwidth,trim={0cm 0.7cm 0cm 0.0cm},clip]{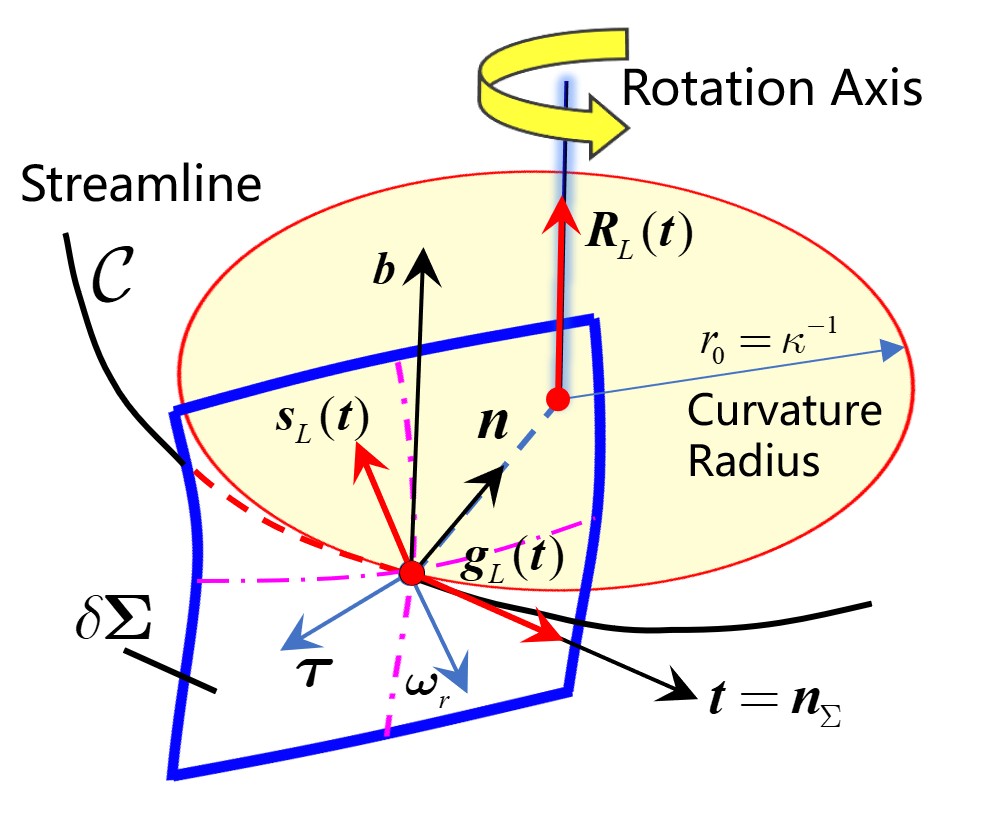}
	\caption{Schematic of streamline-based vorticity decomposition. The Frenet-Serret frame $(\bm{t},\bm{n},\bm{b})$ is attached to a regular streamline segment $\mathcal{C}$, where $\bm{t}$ is the tangent vector, $\bm{n}$ is the principal normal vector (pointing toward the curvature center), and $\bm{b}$ the binormal vector. The local curvature radius is $r_{0}=\kappa^{-1}$. The vorticity field decomposes into three physically distinct components $(\bm{R}_{L}(\bm{t}),\bm{s}_{L}(\bm{t}),\bm{g}_{L}(\bm{t}))$. A material surface element $\delta\bm{\Sigma}$, orthogonal to $\mathcal{C}$ with the unit normal vector $\bm{n}_{\Sigma}=\bm{t}$, illustrates the surface-aligned dynamics: surface shear stress $\bm{\tau}$ and relative vorticity $\bm{\omega}_{r}=-\bm{s}_{L}(\bm{n}_{\Sigma})=\bm{s}_{\Sigma}(\bm{n}_{\Sigma})$.} 
	\label{Field_description_streamline}
\end{figure}

Consider now the spin mode associated with a streamline segment $\mathcal{C}$. Owing to~\eqref{C8a3},~\eqref{C8a4} does not directly provide a vorticity decomposition into orbital and spin modes in the field description. However, a minor modification yields the desired splitting. Let $r$ denote the radial coordinate measured from the curvature center of a point on $\mathcal{C}$, with $r_{0}=\kappa^{-1}(s)$ being the local curvature radius.
Since $\partial_{n}q=-\partial_{r}q$, the binormal vorticity component becomes
\begin{eqnarray*}
	\omega_{b}=\frac{q}{r_0}+\frac{\partial q}{\partial r}=\frac{1}{r_0}\frac{\partial}{\partial r}\left(rq\right).
\end{eqnarray*}
A naive identification of $\partial_{r}q$ as the spin component in the binormal direction $(s_{b}\equiv\bm{s}_{L}(\bm{t})\bm{\cdot}\bm{b})$ leads to a paradox. Consider freezing the fluid near point $P$ into a rigid body rotating with angular velocity $W$, such that $q=Wr$. Although $\omega_{b}=2W$ correctly matches rigid-body rotation, this approach would erroneously yield $s_{b}=W$ instead of $s_{b}=0$. The resolution lies in recognizing that the principal normal $\bm{n}$ itself rotates, making $\partial_{n}q$ alone no longer characterize spin. 
The correct spin component must account for this rotation by subtracting the orbital contribution:
\begin{eqnarray*}
	s_{b}=-\partial_{n}q-\kappa{q}=r_{0}\frac{\partial}{\partial r}\left(\frac{q}{r}\right).
\end{eqnarray*} 
Furthermore, observing that $q\bm{\nabla}q=\bm{A}\bm{\cdot}\bm{u}$ (namely, $\bm{\nabla}q=\bm{A}\bm{\cdot}\bm{t}$) and~\eqref{C8a2}, we need to define, in general, the spin mode of vorticity as (figure~\ref{Field_description_streamline})
\begin{eqnarray}\label{C8a10}
	\bm{s}_{L}(\bm{t})&\equiv&\bm{\nabla}q\times\bm{t}-\bm{\psi}_{L}(\bm{t})\nonumber\\
	&=&-\bm{t}\times\left(\bm{A}\bm{\cdot}\bm{t}\right)-\bm{t}\times\left(\bm{t}\bm{\cdot}\bm{A}\right)\nonumber\\
	&=&-2\bm{t}\times\left(\bm{t}\bm{\cdot}\bm{D}\right).
\end{eqnarray}
Remarkably, $\bm{s}_{L}(\bm{t})$ in~\eqref{C8a10} exhibits identical mathematical structure to the spin vorticity term $\bm{s}_{L}(\bm{e})$
in~\eqref{pp10} when the alignment condition $\bm{e}=\bm{t}$ is imposed. Expanding~\eqref{C8a10} gives 
\begin{eqnarray*}
	\bm{s}_{L}(\bm{t})=\partial_{b}q\bm{n}+\left(-\partial_{n}q-\kappa{q}\right)\bm{b}.
\end{eqnarray*}
This correspondence reveals a fundamental consistency between the general material description and its field-specific counterpart when evaluated along streamlines.

\subsection{Streamline-based triple vorticity decomposition}\label{Streamline-based triple vorticity decomposition}
By combining~\eqref{C8a3},~\eqref{C8a4}, and~\eqref{C8a10}, we derive the~\textit{streamline-based triple decomposition of vorticity}:
\begin{eqnarray}\label{C8a12}
	\bm{\omega}=\bm{R}_{L}(\bm{t})+\bm{s}_{L}(\bm{t})+\bm{g}_{L}(\bm{t}),
\end{eqnarray}
with 
\begin{eqnarray*}
	\bm{g}_{L}(\bm{t})\equiv\left(\bm{\omega}\bm{\cdot}\bm{t}\right)\bm{t}=\xi{q}\bm{t}.
\end{eqnarray*}
The term $\bm{g}_{L}(\bm{t})$ (figure~\ref{Field_description_streamline}) retains the same form as
$\bm{g}_{L}(\bm{e})$ in~\eqref{LLL1}, though it no longer functions as a gauge term in the field description. In general, both $\bm{R}_{L}(\bm{t})$ and $\bm{s}_{L}(\bm{t})$ manifest in flows with nonzero helicity, with few exceptions.
Since the vorticity $\bm{\omega}$ at the position $\bm{x}$ can only be transported by the local background flow at $\bm{x}$--including nonlocal induction effects -- we refer to $\bm{g}_{L}(\bm{t})$ as the streaming mode of vorticity. This component combines the orbital and spin modes of vorticity composite alignment along $\bm{\omega}$.
\subsection{Relationship with Lamb formula}
We remark that~\eqref{C8a12} can also be derived directly from the Lamb formula
\begin{eqnarray}\label{C8a13}
	\bm{u}\bm{\cdot}\bm{\nabla}\bm{u}=\bm{\omega}\times\bm{u}+\bm{\nabla}\left(\frac{1}{2}q^2\right),
\end{eqnarray}
where $\bm{\omega}\times\bm{u}$ is the Lamb vector, representing the nonlinear convection contribution to the fluid acceleration.
Taking a cross product of $\bm{u}$ with both sides of~\eqref{C8a13} yields
\begin{eqnarray*}\label{C8a14}
	q^{2}\bm{\omega}=\bm{u}\left(\bm{\omega}\bm{\cdot}\bm{u}\right)+\bm{u}\times\left(\bm{u}\bm{\cdot}\bm{\nabla}\bm{u}\right)+q\bm{\nabla}q\times\bm{u}.
\end{eqnarray*}
Using the identity $\bm{\nabla}q = \bm{A}\bm{\cdot}\bm{t}$ and normalizing the equation through division by $q^2$, we obtain the final form:
\begin{eqnarray*}\label{C8a15}
	\bm{\omega}&=&(\bm{\omega}\bm{\cdot}\bm{t})\bm{t}
	+\bm{t}\times(\bm{t}\bm{\cdot}\bm{A})-\bm{t}\times(\bm{A}\bm{\cdot}\bm{t})\nonumber\\
	&=&(\bm{\omega}\bm{\cdot}\bm{t})\bm{t}
	+2\bm{t}\times(\bm{t}\bm{\cdot}\bm{A})-\bm{t}\times(\bm{A}\bm{\cdot}\bm{t})-\bm{t}\times(\bm{t}\bm{\cdot}\bm{A})\nonumber\\
	&=&(\bm{\omega}\bm{\cdot}\bm{t})\bm{t}
	+2\bm{t}\times(\bm{t}\bm{\cdot}\bm{A})
	-2\bm{t}\times(\bm{t}\bm{\cdot}\bm{D}).
\end{eqnarray*}
This result is identical to the formulation given in~\eqref{C8a12}, thereby confirming the consistency between both derivations.

\section{Field description based on streamsurface}\label{Field description based on streamsurface}
\subsection{Streamsurface-based triple vorticity decomposition}\label{Streamsurface-based triple vorticity decomposition}
At a certain time instant, a material surface element  $\delta\bm{\Sigma}=\delta\Sigma\bm{n}_{\Sigma}$ (in the Lagrangian description) corresponds to an infinitesimal surface element $\delta\overline{\bm{\Sigma}}=\delta{S}\bm{n}_{S}$ of a streamsurface $\mathcal{S}$ such that $\delta\Sigma=\delta{S}$ and $\bm{n}_{\Sigma}=\bm{n}_{S}$ at the moment.
A streamsurface is formed by the set of streamlines passing through a  reference curve in the flow field. 
Locally, $\delta\overline{\bm{\Sigma}}$ can be parameterized by two independent parameters $(x_{1},x_{2})$, with $x_{3}$ denoting the normal coordinate. The associated orthonormal basis  $(\overline{\bm{e}}_{1},\overline{\bm{e}}_{2},\overline{\bm{e}}_{3})$
defines a surface-attached reference frame, where $\overline{\bm{e}}_{3}=\overline{\bm{e}}_{1}\times\overline{\bm{e}}_{2}=\bm{n}_{S}$.
The surface curvature tensor is given by $\bm{K}=-\bm{\nabla}_{\pi}\bm{n}_{S}=b_{\alpha\beta}\overline{\bm{e}}_{\alpha}\overline{\bm{e}}_{\beta}~(\alpha,\beta=1,2)$, where $b_{\alpha\beta}$ are the tensor components. 
For conciseness, the partial derivatives with respective to the arc-length parameters $(s_{1},s_{2})$ of the coordinate curves are denoted by
\begin{eqnarray}\label{partial_i}
	\partial_{i}\equiv\frac{\partial}{\partial{s}_{i}}=\frac{1}{h_{i}}\frac{\partial}{\partial{x}_{i}},~~~i=1,2,
\end{eqnarray}
where $h_{i}=\sqrt{g_{ii}}$ are Lam\'{e} coefficients, and $g_{ii}$ are the covariant metric components of the surface.
When $(\overline{\bm{e}}_{1},\overline{\bm{e}}_{2})$ align with the two principal directions $(\widetilde{\bm{e}}_{1},\widetilde{\bm{e}}_{2})$ of $\delta\overline{\bm{\Sigma}}$ (i.e., $\widetilde{\bm{e}}_{i}=\overline{\bm{e}}_{i}$ and $\widetilde{\bm{e}}_{3}=\overline{\bm{e}}_{3}=\bm{n}_{S}$), the curvature tensor simplifies to $b_{11}=\lambda_{1}$, $b_{22}=\lambda_{2}$, and $b_{12}=b_{21}=0$, where $\lambda_{1}$ and $\lambda_{2}$ are the principal curvatures of $\mathcal{S}$. In this principal frame $(\widetilde{\bm{e}}_{1},\widetilde{\bm{e}}_{2},\widetilde{\bm{e}}_{3})$, the no-penetration condition $u_{3}=0$ on $\mathcal{S}$ leads to simplified vorticity components (see the general expressions in Appendix~\ref{AP1}): 
\begin{eqnarray*}
	\omega_{1}=-\partial_{3}u_{2}+\lambda_{2}u_{2},~\omega_{2}=\partial_{3}u_{1}-\lambda_{1}u_{1},
\end{eqnarray*}
\begin{eqnarray}
	\omega_{3}=\partial_{1}u_{2}-\partial_{2}u_{1}+\kappa_{g,1}u_{1}-\kappa_{g,2}u_{2}.
\end{eqnarray}
where $\kappa_{g,1}$ and $\kappa_{g,2}$ are the geodesic curvatures of the coordinate curves on the streamsurface, solely determined by the covariant metric components.

Analogous to the material description in~\eqref{ss9} and~\eqref{SEB}, the \textit{streamsurface-based triple vorticity decomposition} is written as
\begin{eqnarray}\label{sb1}
	\bm{\omega}=\bm{R}_{\Sigma}(\bm{n}_{S})+\bm{s}_{\Sigma}(\bm{n}_{S})+\bm{g}_{\Sigma}(\bm{n}_{S}).
\end{eqnarray}
\begin{figure}[t]
	\centering
	\includegraphics[width=0.7\columnwidth,trim={0cm 1.0cm 0cm 0.4cm},clip]{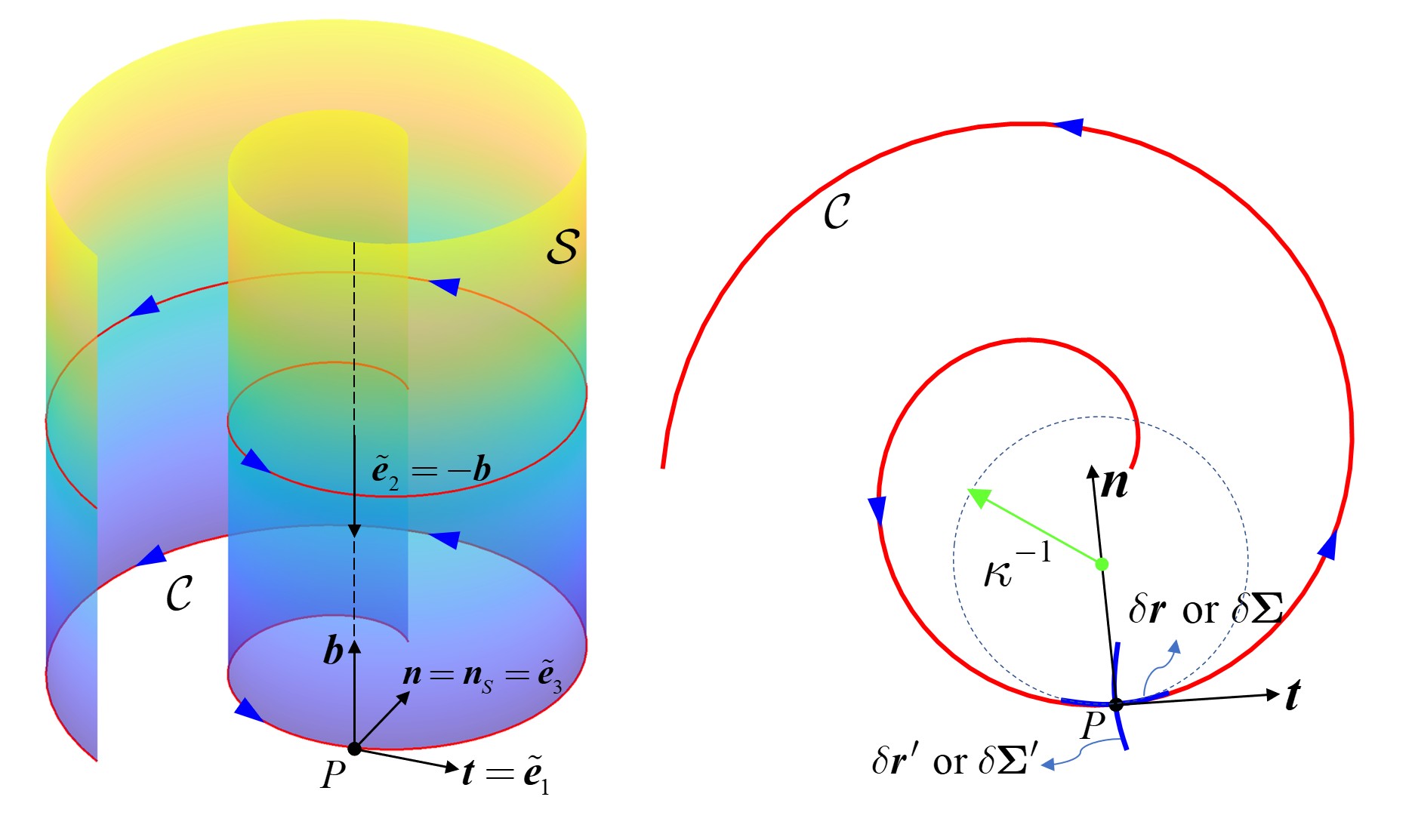}
	\caption{Sketch of a cylindrical streamsurface $\mathcal{S}$ generated by translating a regular streamline $\mathcal{C}$ (lying in the $\widetilde{\bm{e}}_{1}$~--~$\widetilde{\bm{e}}_{3}$) along its binormal direction. The Frenet-Serret intrinsic frame $(\bm{t},\bm{n},\bm{b})$ is defined at each point $P$ on $\mathcal{C}$. The streamsurface normal $\bm{n}_{S}$ coincides with the principal normal $\bm{n}$ of $\mathcal{C}$, with associated curvature radius $\kappa^{-1}$. The surface-adapted frame  $(\widetilde{\bm{e}}_{1},\widetilde{\bm{e}}_{2},\widetilde{\bm{e}}_{3})$ is oriented such that $(\widetilde{\bm{e}}_{1},\widetilde{\bm{e}}_{2})$ align with the principal directions of $\mathcal{S}$. $(\delta\bm{r},\delta\bm{r}^{\prime})$ represents a pair of orthogonal material line elements, with $\delta\bm{r}$ instantaneously coinciding with $\mathcal{C}$. They can also be viewed as an orthogonal material surface-element pair $(\delta\bm{\Sigma},\delta\bm{\Sigma}^{\prime})$ with $\delta\bm{\Sigma}$ coinciding with $\mathcal{S}$.} 
	\label{cylindrical_stream_surface}
\end{figure}
The last term in~\eqref{sb1} is equal to ${\omega_{3}}\bm{n}_{S}$ being equivalent to $2*\mathscr{A}[{\left(\bm{\nabla}_{\pi}\bm{u}\right)_{\pi}}]$, where $*$ is the Hodge-star operator (see Appendix~\ref{AP1}).
By~\eqref{ss10a} and~\eqref{uuu5}, the orbital rotation mode of vorticity in the field description is obtained as 
\begin{eqnarray}\label{C8a16}
	\bm{R}_{\Sigma}(\bm{n}_{S})&=&-2\bm{n}_{S}\times\left(\bm{A}\bm{\cdot}\bm{n}_{S}\right)\nonumber\\
	&=&-2\bm{n}_{S}\times\left(\bm{u}\bm{\cdot}\bm{K}\right)\nonumber\\
	&=&2\lambda_{2}u_{2}\widetilde{\bm{e}}_{1}-2\lambda_{1}u_{1}\widetilde{\bm{e}}_{2},
\end{eqnarray}
representing a superposition of two regular circular motions: one with curvature radius $R_{1}=\lvert\lambda_{1}\rvert^{-1}$, and the other with $R_{2}=\lvert\lambda_{2}\rvert^{-1}$.
From~\eqref{ss10b} and~\eqref{uuu2}, the spin mode of vorticity in the field description is
\begin{eqnarray}\label{D8a16}
	\bm{s}_{\Sigma}(\bm{n}_{S})&=&2\bm{n}_{S}\times\left(\bm{D}\bm{\cdot}\bm{n}_{S}\right)=\bm{\omega}_{r}(\bm{n}_{S})\nonumber\\
	&=&(\omega_{1}-2\lambda_{2}u_{2})\widetilde{\bm{e}}_{1}
	+(\omega_{2}+2\lambda_{1}u_{1})\widetilde{\bm{e}}_{2}\nonumber\\
	&=&(-\partial_{3}u_{2}-\lambda_{2}u_{2})\widetilde{\bm{e}}_{1}+(\partial_{3}u_{1}+\lambda_{1}u_{1})\widetilde{\bm{e}}_{2},
\end{eqnarray}
where each component exhibits the same form as its counterpart in planar flow. The resulting surface shear stress $\bm{\tau}=(\tau_{1},\tau_{2})$ on a streamsurface $\mathcal{S}$ is
\begin{eqnarray}
	\tau_{1}=\mu(\partial_{3}u_{1}+\lambda_{1}u_{1}),~~
	\tau_{2}=\mu(\partial_{3}u_{2}+\lambda_{2}u_{2}).
\end{eqnarray}

\subsection{Consistency between different vorticity decompositions for a cylindrical streamsurface}\label{consistency}
The vorticity decompositions based on a streamline $\mathcal{C}$ and its associated streamsurface $\mathcal{S}$ have been established in~\S\ref{Streamline-based triple vorticity decomposition} and~\S\ref{Streamsurface-based triple vorticity decomposition}, respectively. In this subsection, the consistency between these two field descriptions is examined for a cylindrical streamsurface $\mathcal{S}$ with $\mathcal{C}$ as its directrix. 
As depicted in figure~\ref{cylindrical_stream_surface}, consider a material line element $\delta\bm{r}\equiv\delta{r}\bm{e}$ and its orthogonal counterpart $\delta\bm{r}^{\prime}\equiv\delta{r}^{\prime}\bm{e}^{\prime}$, where $(\bm{e},\bm{e}^{\prime})=(\bm{t},\bm{n})$ at the current instant.
These elements can equivalently be interpreted as a pair of material surface elements $(\delta\bm{\Sigma},\delta\bm{\Sigma}^{\prime})\equiv(\delta\Sigma\bm{n}_{\Sigma},\delta\Sigma^{\prime}\bm{n}_{\Sigma}^{\prime})$ that instantaneously coincide with $\mathcal{S}$, where $\bm{n}_{\Sigma}=\bm{n}$ and $\bm{n}_{\Sigma}^{\prime}=\bm{t}$.
The curvature tensor $\bm{K}$ of $\mathcal{S}$ is given by
$\bm{K}=-\bm{\nabla}_{\pi}\bm{n}_{S}={\kappa}\bm{tt}$, where $\kappa$ is the curvature of the streamline $\mathcal{C}$ (\S~\ref{Streamline element and Frenet-Serret intrinsic triad}).
Consequently, the principal curvature of $\mathcal{S}$ reduces to $\lambda_1 = \kappa$, while $\lambda_2 = 0$.
Under this configuration, $\bm{g}_{\Sigma}(\bm{n})$ naturally vanishes.
Thus, the analysis focuses exclusively on the orbital-rotation and spin modes of vorticity in both descriptions.

Given the velocity field $\bm{u}=q\bm{t}$ (i.e., $(u_1,u_2,u_3)=(q,0,0)$ on $\mathcal{C}$), the orbital rotation mode in~\eqref{C8a16} simplifies to
\begin{eqnarray}\label{C8a16_1}
	\bm{R}_{\Sigma}(\bm{n})=-2\kappa{q}\widetilde{\bm{e}}_{2}=2\kappa{q}\bm{b}=\bm{R}_{L}(\bm{t}).
\end{eqnarray}
This result admits a clear physical interpretation: both the material line element $\delta\bm{r}$ and surface element $\delta\bm{\Sigma}$ at the point $P$ rotate with the identical angular velocities. Specifically, the following identify holds,
\begin{eqnarray}\label{New2}
	{W}_{\Sigma}(\bm{n})={W}_{L}(\bm{t})=\bm{t}\bm{\cdot}\bm{A}\bm{\cdot}\bm{n}=\kappa{q}.
\end{eqnarray}
Here, the $\bm{b}$-components of $(\bm{W}_{L}(\bm{t}),\bm{W}_{L}(\bm{n}))$ are expressed through the projections: ${W}_{L}(\bm{t})\equiv\bm{W}_{L}(\bm{t})\bm{\cdot}\bm{b}$ and ${W}_{L}(\bm{n})\equiv\bm{W}_{L}(\bm{n})\bm{\cdot}\bm{b}$. Unless otherwise specified, this convention is also suitable for other quantities.

Similarly, the spin/shear mode in~\eqref{D8a16} reduces to
\begin{eqnarray}\label{D8a16_1}
	\bm{s}_{\Sigma}(\bm{n})
	=(\partial_{n}u_{1}+\kappa{u_1})\widetilde{\bm{e}}_{2}
	=(-\partial_{n}u_{1}-\kappa{u_1})\bm{b}=\bm{s}_{L}(\bm{t}).
\end{eqnarray}
Since $q^2=u_{1}^{2}+u_{3}^{2}$ in a small neighborhood of $\mathcal{C}$ and $(u_1,u_3)=(q,0)$ on $\mathcal{C}$ itself, $\partial_{n}u_{1}$ in~\eqref{D8a16_1} can be replaced by $\partial_{n}q$ without changing the result. Furthermore, we establish the identity:
\begin{eqnarray}\label{id1}
	{W}_{L}(\bm{n})={W}_{\Sigma}(\bm{t})
	=-\bm{n}\bm{\cdot}\bm{A}\bm{\cdot}\bm{t}=-\partial_{n}q,
\end{eqnarray}
which demonstrates that the orthogonal material elements $(\delta\bm{r}^{\prime}, \delta\bm{\Sigma}^{\prime})$ rotate with identical angular velocities.

By substituting~\eqref{New2} and~\eqref{id1} into~\eqref{D8a16_1}, the spin mode can be expressed in terms of  the angular velocities characterizing the rigid rotation of the orthogonal pair $(\delta\bm{r},\delta\bm{r}^{\prime})$ or $(\delta\bm{\Sigma},\delta\bm{\Sigma}^{\prime})$:
\begin{subequations}
	\begin{eqnarray}\label{id2a}
		{s}_{\Sigma}(\bm{n})={W}_{L}(\bm{n})-{W}_{L}(\bm{t})={W}_{\Sigma}(\bm{t})-{W}_{\Sigma}(\bm{n}),
	\end{eqnarray}
	\begin{eqnarray}\label{id2b}
		{s}_{\Sigma}(\bm{n})={W}_{\Sigma}(\bm{t})-{W}_{L}(\bm{t})
		={W}_{L}(\bm{n})-{W}_{\Sigma}(\bm{n}).
	\end{eqnarray}
\end{subequations}
The spin component $s_{b}\equiv\bm{s}_{L}(\bm{t})\bm{\cdot}\bm{b}$ admits a direct geometric interpretation. Since the angular velocity represents the rate of change of rotation angle displacement, the first equality in~\eqref{id2a} implies $s_{b}=d\theta/dt$, where $\theta\equiv\langle\delta\bm{r},\delta\bm{r}^{\prime}\rangle$ is the intersection angle between $\delta\bm{r}$ and $\delta\bm{r}^{\prime}$. Furthermore, decomposing the angular velocities via
\begin{eqnarray*}
	{W}_{L}(\bm{t})=\frac{1}{2}{\omega}_{b}+{W}_{D}(\bm{t}),~~{W}_{L}(\bm{n})=\frac{1}{2}{\omega}_{b}+{W}_{D}(\bm{n}),
\end{eqnarray*}
with the identity (representing an antisymmetric duality)
\begin{eqnarray}
	{W}_{D}(\bm{t})=-{W}_{D}(\bm{n})=\bm{t}\bm{\cdot}\bm{D}\bm{\cdot}\bm{n},
\end{eqnarray}
the classical contribution $(\frac{1}{2}\omega_{b})$ cancels out identically so that the first equality in~\eqref{id2a} recovers the original definition of spin ${s}_{\Sigma}(\bm{n})=2{W}_{D}(\bm{n})$ in~\eqref{ss10b}, highlighting the special role of $\bm{D}$ in fluid. It is noted that the results in this subsection are equally valid for generic 2D flow.

\section{Physical Roles of NND Rotational Invariants: Unifying Algebraic and Field Descriptions}\label{New physical insights on NND invariants}
In this section, we establish the fundamental connections among the material/field descriptions, the characteristic algebraic description, and vortex identification methods. By unifying DVD, generalized Caswell formula, and Helmholtz decomposition, we elucidate the physical roles of rotational invariants in the algebraic description.
\subsection{Normal-Nilpotent Decomposition and Invariant Vorticity Decomposition}\label{NND_IVD}
In the region where the discriminant $\Delta$ of $\bm{A}$ is positive (i.e., $\Delta>0$), there exists a special orthonormal triad $\{\bm{e}_{1},\bm{e}_{2},\bm{e}_{3}\}$ (termed the NND triad) such that $\bm{A}$ takes a real Schur form (a block upper triangular matrix representation in the field of real numbers), leading to the normal-nilpotent decomposition (NND)~\citep{LiZhen2010,LiuCQ2018,LiZhen2024}:
\begin{subequations}\label{NND12}
	\begin{equation}\label{NND1}
		\bm{A}=
		\begin{bmatrix}
			\chi & \psi+\gamma & -\beta\\
			-\psi & \chi&\alpha\\
			0&  0& \lambda_{r}
		\end{bmatrix}
		=\bm{N}+\bm{S},
	\end{equation}
	\begin{equation}\label{NND2}
		\bm{N}=
		\begin{bmatrix}
			\chi & \psi & 0\\
			-\psi & \chi&0\\
			0&  0& \lambda_{r}
		\end{bmatrix},~~
		\bm{S}=
		\begin{bmatrix}
			0 & \gamma & -\beta\\
			0& 0&\alpha\\
			0&  0& 0
		\end{bmatrix},
	\end{equation}
\end{subequations}
where $\bm{N}$ is a normal tensor (matrix), and $\bm{S}$ is a nilpotent tensor (matrix). The real eigenvalue $\lambda_{r}$ of $\bm{A}$ corresponds to the real eigenvector $\bm{e}_{3}$, while $\bm{e}_{1}$ and $\bm{e}_{2}$ (not necessarily eigenvectors) span the invariant plane perpendicular to $\bm{e}_{3}$. 
A complex conjugate eigenvalue pair $(\lambda_{1,2}=\lambda_{\rm cr}\pm\rm{i}\lambda_{\rm ci})$ indicates a closed or spiraling trajectory pattern in this invariant plane, as claimed in the classical critical-point theory~\citep{Chong1990}.

The off-diagonal element $\psi$ describes the characteristic angular velocity of rigid rotation about the axis $\bm{e}_{3}$.
The diagonal elements of $\bm{N}$, $(\chi,\lambda_{r})$, describe the relative stretching/shrinking rates of material line elements along the axes of NND triad.
The non-trivial elements of $\bm{S}$, $(\alpha,\beta,\gamma)$, characterize spin (or shear) in the three orthogonal planes spanned by the basis vectors of NND triad (namely, the three faces of an infinitesimal cubic fluid element at the point $\bm{x}$). 
The dilatation is $\vartheta\equiv{\rm tr}(\bm{A})=\bm{\nabla}\bm{\cdot}\bm{u}=2\chi+\lambda_{r}={\rm tr}(\bm{N})$. For $\Delta\leq 0$, $\bm{A}$ retains a unique matrix representation analogous to~\eqref{NND1}, though no longer being a real Schur form, as claimed in NND-II~\citep{LiZhen2024}. 

Using~\eqref{NND1} and~\eqref{NND2}, we obtain the matrix representations of $\bm{D}$ and $\bm{\varOmega}$ in the SAD:
\begin{subequations}\label{eq2ab}
	\begin{equation}\label{strain-rate tensor}
		\bm{D}=
		\begin{bmatrix}
			\chi &\frac{1}{2}\gamma & -\frac{1}{2}\beta\\
			\frac{1}{2}\gamma & \chi&\frac{1}{2}\alpha\\
			-\frac{1}{2}\beta& \frac{1}{2}\alpha& \lambda_{r}
		\end{bmatrix},
	\end{equation}
	\begin{equation}\label{rotation-rate tensor}
		\bm{\varOmega}=
		\begin{bmatrix}
			0 &\psi+\frac{1}{2}\gamma & -\frac{1}{2}\beta\\
			-\left(\psi+\frac{1}{2}\gamma\right) & 0&\frac{1}{2}\alpha\\
			\frac{1}{2}\beta& -\frac{1}{2}\alpha& 0
		\end{bmatrix}.
	\end{equation}
\end{subequations}
The vorticity vector $\bm{\omega}$, being the dual vector of the spin tensor $\bm{\varOmega}$, admits the following double decomposition:
\begin{subequations}\label{IVD}
	\begin{eqnarray}\label{IVD1}
		\bm{\omega}=\bm{R}_{N}+\bm{s}_{N},
	\end{eqnarray}	
	\begin{eqnarray}\label{IVD2}
		\bm{R}_{N}=2\psi\bm{e}_{3},~~\bm{s}_{N}=\alpha\bm{e}_{1}+\beta\bm{e}_{2}+\gamma\bm{e}_{3}.
	\end{eqnarray}
\end{subequations}
In this article, we collectively refer to~\eqref{IVD1} and~\eqref{IVD2} as the~\textit{invariant vorticity decomposition} (IVD), where the subscript $N$ denotes the NND framework.
$\bm{R}_{N}$ represents the characteristic rigid rotation mode of vorticity, while $\bm{s}_{N}$ representing the characteristic spin/shear mode, determined by the off-diagonal elements of either the nilpotent tensor $\bm{S}$ or the strain-rate tensor $\bm{D}$. The in-plane vorticity components $(\omega_{1},\omega_{2})=(\alpha,\beta)$ stem purely from spin, whereas the invariant splitting
\begin{eqnarray*}
	\omega_{3}=2\psi+\gamma
\end{eqnarray*}
occurs exclusively for the axial component $\omega_{3}=\bm{\omega}\bm{\cdot}\bm{e}_{3}$. The rotation axis $\bm{e}_{3}$ (associated with $\lambda_{r}$)  is uniquely determined by enforcing $\omega_{3}>0$~\citep{LiuCQ2020}. However, either $\gamma>0$ or $\gamma<0$ may alternatively serve as a uniqueness condition, a point we shall further analyze in~\S\ref{sign_of_gamma}.
A positive $\gamma$ leads to the Liutex-shear decomposition while the negative $\gamma$ produces an alternative form of the IVD (\S\ref{sign_of_gamma}).

The orientation of a reference frame in three-dimensional space requires three scalars, while the second-rank tensor $\bm{A}$ possesses nine independent components. Consequently, $\bm{A}$ can be fully characterized by six intrinsic, orientation-invariant scalars. A conventional choice for these invariants comprises the three eigenvalues of $\bm{D}$ and the three vorticity components projected onto the $\bm{D}$ eigensystem, or alternatively, through properly defined tensor contractions \citep{Meneveau2011}. Working with the Schur theorem,
the NND framework introduces a new set of six rotational invariants $(\chi, \lambda_r, \alpha, \beta, \gamma, \psi)$ for generic compressible flows. In regions where $\Delta > 0$, since $(\gamma, \psi)$ are functionally dependent on $(\lambda_{\rm ci}, \omega_3)$ (see~\S\ref{sign_of_gamma}), an equivalent invariant set is $(\chi,\lambda_{r},\alpha,\beta,\lambda_{\rm ci},\omega_{3})$. This set encapsulates the complete eigenvalue information of $\bm{A}$ via $(\chi, \lambda_{\rm ci}, \lambda_r)$ (where $\chi = \lambda_{\rm cr}$ represents the real part of the complex conjugate pair), and the three vorticity components $(\alpha,\beta,\omega_{3})$ under the NND triad.

\subsection{NND rotational invariants and generalized Caswell formula}\label{NND_and_Caswell}
The generalized Caswell formula~\eqref{uuu1} for the strain-rate tensor $\bm{D}$ serves as the foundation for identifying the spin mode contribution to vorticity on $\delta\bm{\Sigma}$.
We now examine three specific material surface elements
$\delta\bm{\Sigma}^{(i)}=\delta\Sigma^{(i)}\bm{n}_{\Sigma}^{(i)}$ (where $i=1,2,3$), with the unit normals $\bm{n}_{\Sigma}^{(i)}=\bm{e}_{i}$ successively aligned with each basis vector of the NND triad. The orthogonal material line elements along $\bm{n}_{\Sigma}^{(i)}$ at the moment are denoted by $\delta\bm{r}^{(i)}=\delta{r}^{(i)}\bm{e}_{i}$. Then, we derive the NND matrix representations of both the strain-rate tensor~\eqref{uuu1} and its rotational counterpart for $\bm{\varOmega}$ in~\eqref{uuuu1}, as demonstrated in the following three cases,

\begin{subequations}
	$\circ$ Case 1: $\bm{n}_{\Sigma}^{(1)}=\bm{e}_{1}$.
	\begin{eqnarray}\label{Caswell-e1}
		\bm{D}=\begin{bmatrix}
			\chi &0 &0\\
			0 & 0&0\\
			0& 0& 0
		\end{bmatrix}+
		\begin{bmatrix}
			0 &\frac{1}{2}\gamma & -\frac{1}{2}\beta\\
			\frac{1}{2}\gamma & 0&0\\
			-\frac{1}{2}\beta&0&0
		\end{bmatrix}+
		\begin{bmatrix}
			0 &0 &0\\
			0 & \chi&\frac{1}{2}\alpha\\
			0& \frac{1}{2}\alpha& \lambda_{r}
		\end{bmatrix};
	\end{eqnarray}
	\begin{eqnarray}\label{Caswell-e1a}
		\bm{\varOmega}=\begin{bmatrix}
			0 &\frac{1}{2}\omega_{3} &-\frac{1}{2}\beta\\
			-\frac{1}{2}\omega_{3} & 0&0\\
			\frac{1}{2}\beta& 0& 0
		\end{bmatrix}+
		\begin{bmatrix}
			0 &0 &0\\
			0& 0&\frac{1}{2}\alpha\\
			0&-\frac{1}{2}\alpha&0
		\end{bmatrix}.
	\end{eqnarray}
	
	$\circ$ Case 2: $\bm{n}_{\Sigma}^{(2)}=\bm{e}_{2}$.
	\begin{eqnarray}\label{Caswell-e2}
		\bm{D}=\begin{bmatrix}
			0 &0 &0\\
			0 &\chi&0\\
			0& 0& 0
		\end{bmatrix}+
		\begin{bmatrix}
			0 &\frac{1}{2}\gamma & 0\\
			\frac{1}{2}\gamma & 0&\frac{1}{2}\alpha\\
			0&\frac{1}{2}\alpha&0
		\end{bmatrix}+
		\begin{bmatrix}
			\chi &0 &-\frac{1}{2}\beta\\
			0 &0&0\\
			-\frac{1}{2}\beta& 0& \lambda_{r}
		\end{bmatrix};
	\end{eqnarray}
	\begin{eqnarray}\label{Caswell-e2a}
		\bm{\varOmega}=\begin{bmatrix}
			0 &\frac{1}{2}\omega_{3} &0\\
			-\frac{1}{2}\omega_{3} & 0&\frac{1}{2}\alpha\\
			0& -\frac{1}{2}\alpha& 0
		\end{bmatrix}+
		\begin{bmatrix}
			0 &0 &-\frac{1}{2}\beta\\
			0& 0&0\\
			\frac{1}{2}\beta&0&0
		\end{bmatrix}.
	\end{eqnarray}
	
	$\circ$ Case 3: $\bm{n}_{\Sigma}^{(3)}=\bm{e}_{3}$.
	\begin{eqnarray}\label{Caswell-e3}
		\bm{D}=\begin{bmatrix}
			0 &0 &0\\
			0 &0&0\\
			0& 0&\lambda_{r}
		\end{bmatrix}+
		\begin{bmatrix}
			0 &0 & -\frac{1}{2}\beta\\
			0 & 0&\frac{1}{2}\alpha\\
			-\frac{1}{2}\beta&\frac{1}{2}\alpha&0
		\end{bmatrix}+
		\begin{bmatrix}
			\chi &\frac{1}{2}\gamma &0\\
			\frac{1}{2}\gamma & \chi&0\\
			0& 0& 0
		\end{bmatrix},
	\end{eqnarray}
	\begin{eqnarray}\label{Caswell-e3a}
		\bm{\varOmega}=\begin{bmatrix}
			0 &0 &-\frac{1}{2}\beta\\
			0 & 0&\frac{1}{2}\alpha\\
			\frac{1}{2}\beta& -\frac{1}{2}\alpha& 0
		\end{bmatrix}+
		\begin{bmatrix}
			0 &\frac{1}{2}\omega_{3} &0\\
			-\frac{1}{2}\omega_{3}& 0&0\\
			0&0&0
		\end{bmatrix}.
	\end{eqnarray}
\end{subequations}

On the right hand side of~\eqref{Caswell-e1},~\eqref{Caswell-e2}, and~\eqref{Caswell-e3}, the matrices (from left to right) represent the normal stretching component $\bm{n}_{\Sigma}\bm{n}_{\Sigma}\partial_{n}u_{n}$, the symmetric shear contribution $\mathscr{S}[\bm{n}_{\Sigma}(\bm{\omega}_{r}\times\bm{n}_{\Sigma})]\equiv\frac{1}{2}[\bm{n}_{\Sigma}(\bm{\omega}_{r}\times\bm{n}_{\Sigma})+(\bm{\omega}_{r}\times\bm{n}_{\Sigma})\bm{n}_{\Sigma}]$, and the in-plane strain component $\mathscr{S}[{\left(\bm{\nabla}_{\pi}\bm{u}\right)_{\pi}}]$.
Therefore, the relative vorticity $\bm{\omega}_{r}^{(i)}$ or the spin $\bm{s}_{\Sigma}(\bm{e}_{i})$, only associated with the off-diagonal components of $\bm{D}$, are derived as
\begin{subequations}\label{omegar123}
	\begin{eqnarray}\label{omegar1}
		\bm{s}_{\Sigma}(\bm{e}_{1})=\bm{\omega}_{r}^{(1)}=\beta\bm{e}_{2}+\gamma\bm{e}_{3}=-\bm{s}_{L}(\bm{e}_{1}),
	\end{eqnarray}
	\begin{eqnarray}\label{omegar2}
		\bm{s}_{\Sigma}(\bm{e}_{2})=\bm{\omega}_{r}^{(2)}=\alpha\bm{e}_{1}-\gamma\bm{e}_{3}=-\bm{s}_{L}(\bm{e}_{2}),
	\end{eqnarray}
	\begin{eqnarray}\label{omegar3}
		\bm{s}_{\Sigma}(\bm{e}_{3})=\bm{\omega}_{r}^{(3)}=-\alpha\bm{e}_{1}-\beta\bm{e}_{2}=-\bm{s}_{L}(\bm{e}_{3}).
	\end{eqnarray}
\end{subequations}
Notably, only one spin component enters the matrix form of $\mathscr{S}[{\left(\bm{\nabla}_{\pi}\bm{u}\right)_{\pi}}]$ which characterizes the surface flexibility of $\delta\bm{\Sigma}^{(i)}$.
For each $\delta\bm{\Sigma}^{(i)}$, the relative vorticity is associated with two spin components, generating the surface shear stress. While the tensor $\mathscr{S}[{\left(\bm{\nabla}_{\pi}\bm{u}\right)_{\pi}}]$ must be incorporated in the Cauchy-Poisson constitutive relation for viscous stress, it plays no dynamic role in producing tangential shear stresses. Moreover, the right-hand side matrices in~\eqref{Caswell-e1a},~\eqref{Caswell-e2a}, and~\eqref{Caswell-e3a} represent the skew-symmetric tensors: $\mathscr{A}[\bm{n}_{\Sigma}(\bm{\omega}\times\bm{n}_{\Sigma})]\equiv\frac{1}{2}\bm{n}_{\Sigma}(\bm{\omega}\times\bm{n}_{\Sigma})-\frac{1}{2}(\bm{\omega}\times\bm{n}_{\Sigma})\bm{n}_{\Sigma}$ and $\mathscr{A}[{\left(\bm{\nabla}_{\pi}\bm{u}\right)_{\pi}}]$, respectively. The former governs the tangential vorticity components relative to $\delta\bm{\Sigma}^{(i)}$, while while the second determines the remaining normal vorticity component along $\bm{n}_{\Sigma}^{(i)}$.

Furthermore, the rigid-rotation vorticity modes associated with $\delta\bm{\Sigma}^{(i)}$ and $\delta\bm{r}^{(i)}$ are derived as
\begin{subequations}\label{RLRS123}
	\begin{eqnarray}
		\bm{R}_{L}(\bm{e}_{1})=2\beta\bm{e}_{2}+2(\psi+\gamma)\bm{e}_{3},~~\bm{R}_{\Sigma}(\bm{e}_{1})=2\psi\bm{e}_{3},
	\end{eqnarray}
	\begin{eqnarray}
		\bm{R}_{L}(\bm{e}_{2})=2\alpha\bm{e}_{1}+2\psi\bm{e}_{3},~~\bm{R}_{\Sigma}(\bm{e}_{2})=2(\psi+\gamma)\bm{e}_{3},
	\end{eqnarray}
	\begin{eqnarray}
		\bm{R}_{L}(\bm{e}_{3})=\bm{0},~~\bm{R}_{\Sigma}(\bm{e}_{3})=2\alpha\bm{e}_{1}+2\beta\bm{e}_{2}.
	\end{eqnarray}
\end{subequations}
Evidently,~\eqref{omegar123} and~\eqref{RLRS123} satisfy the constraint~\eqref{uuu11}. These relations demonstrate that the rigid rotation modes derived from the analyses of material line and surface elements fundamentally differ from those in the characteristic description. The characteristic spin components can modify the rigid rotation of both material lines and surfaces. The line element aligned with $\bm{e}_{3}$ exhibits no rigid rotation, while the rigid rotation mode of $\bm{e}_{3}$-oriented surface element is exclusively determined by the two characteristic spin components $(\alpha,\beta)$.

The role of the diagonal elements, $\chi$ and $\lambda_{r}$, is interpreted as follows.
For the line elements $\delta\bm{r}^{(i)}$, we have
\begin{eqnarray}
	\left(\frac{1}{\delta{r}}\frac{D\delta{r}}{Dt}\right)^{(i)}=\bm{n}_{\Sigma}^{(i)}\bm{\cdot}\bm{D}\bm{\cdot}\bm{n}_{\Sigma}^{(i)}=(\partial_{n}u_{n})^{(i)},
\end{eqnarray} 
implying 
\begin{eqnarray}
	\left(\frac{1}{\delta{r}}\frac{D\delta{r}}{Dt}\right)^{(1)}
	=\left(\frac{1}{\delta{r}}\frac{D\delta{r}}{Dt}\right)^{(2)}=\chi,~~\left(\frac{1}{\delta{r}}\frac{D\delta{r}}{Dt}\right)^{(3)}=\lambda_{r}.
\end{eqnarray}
For the surface elements $\delta\bm{\Sigma}^{(i)}$, it holds that
\begin{eqnarray}
	\left(\frac{1}{\delta\Sigma}\frac{D\delta\Sigma}{Dt}\right)^{(i)}=(\bm{\nabla}_{\pi}\bm{\cdot}\bm{u})^{(i)}
	=\left({\rm tr}\mathscr{S}[\left(\bm{\nabla}_{\pi}\bm{u}\right)_{\pi}]\right)^{(i)},
\end{eqnarray} 
yielding
\begin{eqnarray}
	\left(\frac{1}{\delta\Sigma}\frac{D\delta\Sigma}{Dt}\right)^{(1)}
	=\left(\frac{1}{\delta\Sigma}\frac{D\delta\Sigma}{Dt}\right)^{(2)}
	=\chi+\lambda_{r},~~
	\left(\frac{1}{\delta\Sigma}\frac{D\delta\Sigma}{Dt}\right)^{(3)}=2\chi.
\end{eqnarray}
For the volume element $\delta{V}=\delta\Sigma^{(i)}\delta{r}^{(i)}$, the dilatation is
\begin{eqnarray}
	\vartheta=\left(\frac{1}{\delta r}\frac{D\delta r}{Dt}\right)^{(i)}
	+\left(\frac{1}{\delta \Sigma}\frac{D\delta \Sigma}{Dt}\right)^{(i)}
	=2\chi+\lambda_{r}.
\end{eqnarray} 

\subsection{Transformation between characteristic and field descriptions}\label{Transformation between characteristic and field descriptions}
In an infinitesimal neighborhood of a point $P$, the axial velocity component along $\bm{e}_{3}$ becomes negligible for the IVD variables $(\chi,\psi,\gamma)$, as they are independent of this direction. Consequently, the local flow can be reduced to a 2D projected velocity field in the invariant plane (i.e., $\bm{e}_{1}$-$\bm{e}_{2}$ plane). For convenience of analysis, we adopt the Frenet-Serret intrinsic triad $(\bm{t},\bm{n},\bm{b})$ associated with a regular streamline $\mathcal{C}$ in this plane, and assume $\bm{b}=\bm{e}_{3}$ aligned to the rotation axis $(\omega_{b}=\omega_{3}>0)$.

On one hand, NND-II offers a matrix representation of the VGT $\widehat{\bm{A}}$ under the basis $(\bm{e}_{1},\bm{e}_{2})$ as
\begin{eqnarray}\label{a30_1}
	\widehat{\bm{A}}
	\equiv\begin{bmatrix}
		\bm{e}_{1}\bm{\cdot}\bm{A}\bm{\cdot}\bm{e}_{1}&\bm{e}_{1}\bm{\cdot}\bm{A}\bm{\cdot}\bm{e}_{2}\\
		\bm{e}_{2}\bm{\cdot}\bm{A}\bm{\cdot}\bm{e}_{1} & \bm{e}_{2}\bm{\cdot}\bm{A}\bm{\cdot}\bm{e}_{2}
	\end{bmatrix}
	=\begin{bmatrix}
		\chi &\psi+\gamma\\
		-\psi & \chi
	\end{bmatrix},
\end{eqnarray}
which exhibits a real Schur block in the region $\Delta>0$, encoding a complex conjugate eigenvalue pair.

On the other hand, by using~\eqref{New2} and~\eqref{id1}, $\widehat{\bm{A}}$ admits a matrix representation under the intrinsic streamline basis $(\bm{t},\bm{n})$, namely,
\begin{eqnarray}\label{a30_2}
	\widehat{\bm{A}}
	\equiv\begin{bmatrix}
		\bm{t}\bm{\cdot}\bm{A}\bm{\cdot}\bm{t}&\bm{t}\bm{\cdot}\bm{A}\bm{\cdot}\bm{n}\\
		\bm{n}\bm{\cdot}\bm{A}\bm{\cdot}\bm{t} & \bm{n}\bm{\cdot}\bm{A}\bm{\cdot}\bm{n}
	\end{bmatrix}
	=\begin{bmatrix}
		\chi(\bm{t}) &W_{L}(\bm{t})\\
		-W_{L}(\bm{n}) & \chi(\bm{n})
	\end{bmatrix},
\end{eqnarray}
where the relative stretching rates along $\bm{t}$ and $\bm{n}$ are denoted by
$\chi(\bm{t})\equiv\bm{t}\bm{\cdot}\bm{D}\bm{\cdot}\bm{t}$ and $\chi(\bm{n})\equiv\bm{n}\bm{\cdot}\bm{D}\bm{\cdot}\bm{n}$, respectively.
The emergence of the irreducible real Schur form in~\eqref{a30_2} is particularly noteworthy, as it reveals the intrinsic structure of $\bm{A}$ through a streamline-based field description. Remarkably, this structure aligns with the form obtained from purely algebraic analysis in~\eqref{a30_1}.
In fact, using~\eqref{id2a}, the representation matrix of $\widehat{\bm{A}}$ under the basis $(-\bm{n},\bm{t})$ is rewritten as
\begin{eqnarray}\label{a30_2m}
	\widehat{\bm{A}}
	\equiv\begin{bmatrix}
		\bm{n}\bm{\cdot}\bm{A}\bm{\cdot}\bm{n}&-\bm{n}\bm{\cdot}\bm{A}\bm{\cdot}\bm{t}\\
		-\bm{t}\bm{\cdot}\bm{A}\bm{\cdot}\bm{n} & \bm{t}\bm{\cdot}\bm{A}\bm{\cdot}\bm{t}
	\end{bmatrix}
	=\begin{bmatrix}
		\chi(\bm{n}) &W_{L}(\bm{t})+s_{L}(\bm{t})\\
		-W_{L}(\bm{t}) & \chi(\bm{t})
	\end{bmatrix},
\end{eqnarray}
where the orbital rotation and spin modes are characterized by $W_{L}(\bm{t})=\kappa{q}$ and $s_{L}(\bm{t})=r\partial_{r}(q/r)$. In a vicinity of $\mathcal{C}$, $q$ can be replaced by  the velocity component parallel to $\mathcal{C}$ without affecting the result. It is worth pointing out that the two diagonal elements in~\eqref{a30_2} and~\eqref{a30_2m} are generally different, which however, can be made the same as the canonical NND form by an appropriate rotation operation as follows.

\begin{figure}[t]
	\centering
	\includegraphics[width=0.5\columnwidth,trim={0cm 0.2cm 0cm 0.2cm},clip]{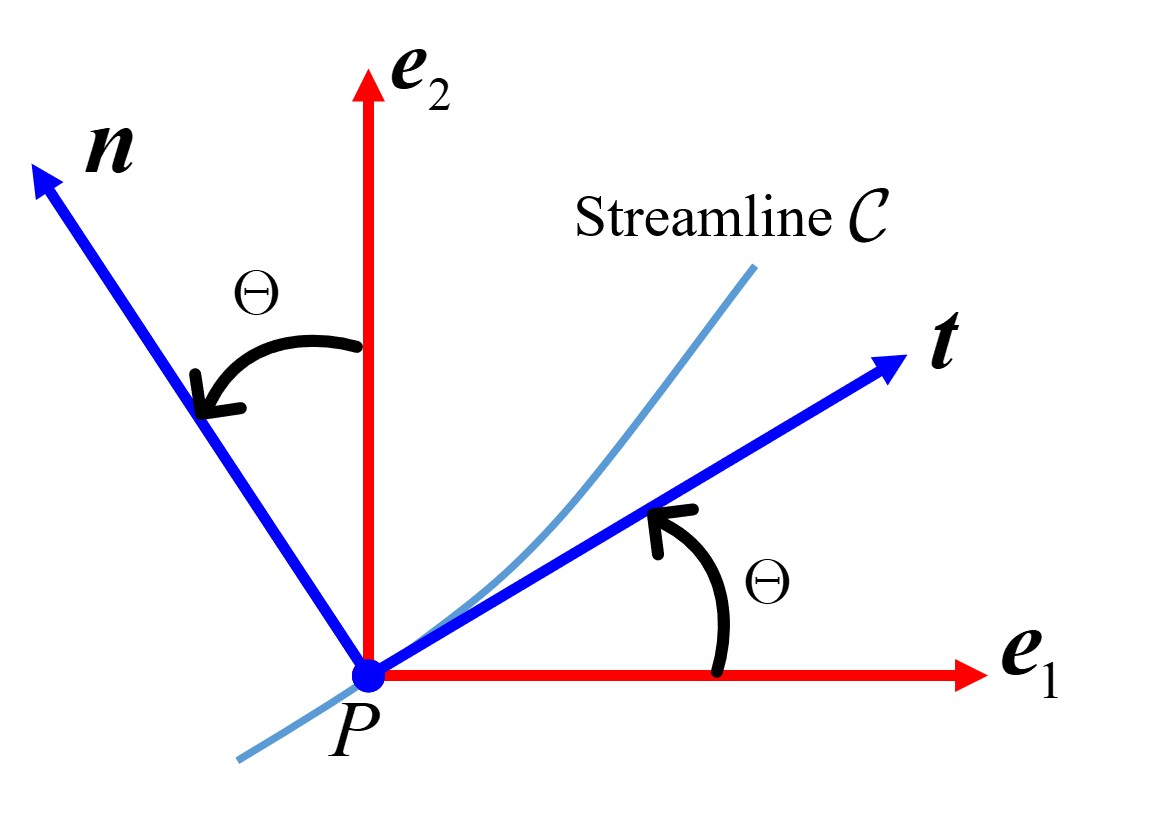}
	\caption{Two reference frames defined on the invariant plane. The planar basis $(\bm{e}_{1},\bm{e}_{2})$ originates from the NND triad, while $(\bm{t},\bm{n})$ belongs to the Frenet-Serret frame intrinsic to a streamline segment $\mathcal{C}$.} 
	\label{e1e2tn}
\end{figure}
Denoting the rotation angle between $\bm{e}_{1}$ and $\bm{t}$ as $\Theta\equiv\langle\bm{e}_{1},\bm{t}\rangle$ (figure~\ref{e1e2tn}),  $(\bm{e}_{1},\bm{e}_{2})$ and $(\bm{t},\bm{n})$ must be related through an orthogonal transformation
\begin{eqnarray}\label{a30_3}
	\begin{bmatrix}
		\bm{t} \\
		\bm{n}
	\end{bmatrix}
	=\bm{Q}(\Theta)	\begin{bmatrix}
		\bm{e}_{1} \\
		\bm{e}_{2}
	\end{bmatrix}
	=\begin{bmatrix}
		\cos\Theta &\sin\Theta\\
		-\sin\Theta & \cos\Theta
	\end{bmatrix}
	\begin{bmatrix}
		\bm{e}_{1} \\
		\bm{e}_{2}
	\end{bmatrix},
\end{eqnarray}
where $\bm{Q}(\Theta)$ is a rotation matrix.
The angle $\Theta$ is used to determine the orientation of the new reference frame, without increasing the number of independent components of $\bm{A}$.
Therefore, combining~\eqref{a30_1},~\eqref{a30_2} and~\eqref{a30_3} yields
\begin{eqnarray}\label{a30_4}
	\begin{bmatrix}
		\chi(\bm{t}) &W_{L}(\bm{t})\\
		-W_{L}(\bm{n}) & \chi(\bm{n})
	\end{bmatrix}
	=\begin{bmatrix}
		\chi+\gamma\sin\Theta\cos\Theta&\psi+\gamma\cos^{2}\Theta\\
		-(\psi+\gamma\sin^{2}\Theta) & \chi-\gamma\sin\Theta\cos\Theta
	\end{bmatrix}.
\end{eqnarray}
Equivalently, the two real Schur blocks should satisfy the following relations:
\begin{subequations}
	\begin{eqnarray}\label{a30_5}
		\chi+\frac{1}{2}\gamma\sin{2\Theta}=\chi(\bm{t}),~~
		\chi-\frac{1}{2}\gamma\sin{2\Theta}=\chi(\bm{n}),
	\end{eqnarray}
	\begin{eqnarray}\label{a30_6}
		\frac{1}{2}(\omega_{3}+\gamma\cos{2\Theta})=W_{L}(\bm{t}),~~
		\frac{1}{2}(\omega_{3}-\gamma\cos2\Theta)=W_{L}(\bm{n}).
	\end{eqnarray}
\end{subequations}

From~\eqref{a30_5}, the sum and difference of the stretching rates $(\chi(\bm{t}),\chi(\bm{n}))$ are given by
\begin{eqnarray*}
	\chi(\bm{t})+\chi(\bm{n})=2\chi,
\end{eqnarray*}
\begin{eqnarray}\label{eq81}
	\varphi(\bm{t})\equiv\chi(\bm{t})-\chi(\bm{n})=\gamma\sin2\Theta.
\end{eqnarray}
From~\eqref{id2a} and~\eqref{a30_6}, it follows that
\begin{eqnarray*}
	W_{L}(\bm{t})+W_{L}(\bm{n})=\omega_{3},
\end{eqnarray*}
\begin{eqnarray}\label{eq82}
	{s}_{L}(\bm{t})={s}_{\Sigma}(\bm{n})=-\gamma\cos2\Theta.
\end{eqnarray}
Finally, using~\eqref{eq81} and~\eqref{eq82}, the variables $(\gamma,s_{L}(\bm{t}),\varphi(\bm{t}))$ satisfy the following Pythagorean-type relation:
\begin{eqnarray}\label{eq83}
	\gamma^{2}=s_{L}^{2}(\bm{t})+\varphi^{2}(\bm{t}).
\end{eqnarray}
In addition, the rotation angle $\Theta$
is determined by
\begin{eqnarray}\label{eq84}
	\tan2\Theta=-\frac{\varphi(\bm{t})}{s_{L}(\bm{t})}.
\end{eqnarray}
We remark that~\eqref{eq83} reproduces the result reported by~\citet[(33a)]{ChenLiu2025POF} when taking $\gamma>0$. However, our derivation approach differs fundamentally from~\citet{ChenLiu2025POF} based on the frame-invariance of $\Delta$ for generic 2D flows. Moreover, the relations between the rotation angle $\Theta$ and relevant quantities were not obtained in previous approach.
\subsection{NND rotational invariants and Helmholtz decomposition}\label{NND_Helmholtz}
At a material point $P$, the NND triad $\{\bm{e}_{1},\bm{e}_{2},\bm{e}_{3}\}$ defines a local Cartesian coordinate system $\bm{x}=(x_{1},x_{2},x_{3})$ with the origin located at $P~(\bm{x}_{P}=\bm{0})$. The coordinate system can be extended throughout a neighborhood $U_{P}$ of $P$, where the NND triad at $P$ serves as the induced coordinate frame for all $\bm{x}\in U_{P}$. By acting the spatial gradient operator on both sides of~\eqref{uuu6}, we obtain
\begin{subequations}\label{HH12}
	\begin{equation}\label{HH1}
		\bm{A}\equiv\bm{\nabla}\bm{u}=\bm{\nabla\nabla}\phi+\bm{\nabla}\bm{v}.
	\end{equation}
	Working under this coordinate system, the potential Hessian $\bm{\nabla\nabla}\phi$ and the transverse velocity gradient $\bm{\nabla}\bm{v}$ take the following component forms:
	\begin{equation*}
		\bm{\nabla\nabla}\phi=\frac{\partial^{2}\phi}{\partial{x}_{i}\partial{x}_{j}}\bm{e}_{i}\bm{e}_{j}=
		\begin{bmatrix}
			\partial_{1}^{2}\phi &\partial_{1}\partial_{2}\phi  & \partial_{1}\partial_{3}\phi \\
			\partial_{2}\partial_{1}\phi & \partial_{2}^{2}\phi&\partial_{2}\partial_{3}\phi\\
			\partial_{3}\partial_{1}\phi&  \partial_{3}\partial_{2}\phi&\partial_{3}^{2}\phi
		\end{bmatrix},
	\end{equation*}
	\begin{equation}\label{HH2}
		\bm{\nabla v}=\frac{\partial v_j}{\partial x_i}\bm{e}_{i}\bm{e}_{j}=
		\begin{bmatrix}
			\partial_{1}v_{1}& \partial_{1}v_{2} & \partial_{1}v_{3}\\
			\partial_{2}v_{1}& \partial_{2}v_{2}&\partial_{2}v_{3}\\
			\partial_{3}v_{1}& \partial_{3}v_{2}&\partial_{3}v_{3}
		\end{bmatrix}.
	\end{equation}
\end{subequations}
The dilatation is determined by $\phi$: $\vartheta=\nabla^{2}\phi={\rm tr}(\bm{\nabla\nabla}\phi)$ because of $\bm{\nabla}\bm{\cdot}\bm{v}=0$. The components of vorticity are evaluated through the transverse velocity components:
\begin{eqnarray*}
	\omega_{1}=\partial_{2}v_{3}-\partial_{3}v_{2},~
	\omega_{2}=\partial_{3}v_{1}-\partial_{1}v_{3},~
	\omega_{3}=\partial_{1}v_{2}-\partial_{2}v_{1}.
\end{eqnarray*}

To ensure the consistency between the Helmholtz decomposition (\eqref{HH1} and~\eqref{HH2}) and the real Schur form (\eqref{NND1} and~\eqref{NND2}), the constraint for the normal derivative is $\partial_{3}\bm{u}_{\pi}=(\partial_{3}u_{1},\partial_{3}u_{2})=\bm{0}$, namely,
\begin{equation}\label{CS12}
	\partial_{3}\partial_{1}\phi=-\partial_{3}v_{1},~~\partial_{3}\partial_{2}\phi=-\partial_{3}v_{2},
\end{equation}
where $\bm{u}_{\pi}=(u_{1},u_{2})$ is the projected velocity field on the invariant $\bm{e}_{1}$-$\bm{e}_{2}$ plane. Actually, it can be shown that $\partial_{3}\bm{u}_{\pi}=[\bm{e}_{3}\times(\bm{e}_{3}\bm{\cdot}\bm{A})]\times\bm{e}_{3}=\bm{W}_{L}^{\rm eff}(\bm{e}_{3})\times\bm{e}_{3}$. Since the rotation axis $\bm{e}_{3}$ is the real eigenvalue vector of $\bm{A}$ that satisfies $\bm{e}_{3}\bm{\cdot}\bm{A}=\lambda_{r}\bm{e}_{3}$, the material line element aligned with $\bm{e}_{3}$ exhibits no effective rigid rotation except for the rotation about its own axis, enforcing the compatibility with the block-triangular matrix structure of $\bm{A}$.

The NND rotational invariants $(\chi,\lambda_{r},\alpha,\beta,\gamma,\psi)$ are expressed by the spatial derivatives of $(\phi,\bm{v})$ as
\begin{subequations}
	\begin{equation}\label{PPa}
		\chi=\partial_{1}^{2}\phi+\partial_{1}v_{1}=\partial_{2}^{2}\phi+\partial_{2}v_{2},
	\end{equation}
	\begin{equation}\label{PPb}
		\lambda_{r}=\partial_{3}^{2}\phi+\partial_{3}v_{3}=\partial_{3}^{2}\phi-\partial_{1}v_{1}-\partial_{2}v_{2},
	\end{equation}
	\begin{equation}\label{PPc}
		\alpha=\partial_{2}\partial_{3}\phi+\partial_{2}v_{3},~~
		\beta=-\partial_{1}\partial_{3}\phi-\partial_{1}v_{3},
	\end{equation}
	\begin{equation}\label{PPd}
		2\psi=-2\partial_{2}\partial_{1}\phi-2\partial_{2}v_{1},~~\gamma=2\partial_{1}\partial_{2}\phi+\partial_{1}v_{2}+\partial_{2}v_{1}.
	\end{equation}
\end{subequations}

By employing~\eqref{CS12} to eliminate the mixed derivatives in~\eqref{PPc}, we isolate the purely vortical contributions, yielding
\begin{eqnarray}\label{new87}
	\alpha=\partial_{2}v_{3}-\partial_{3}v_{2}=\omega_{1},~~
	\beta=\partial_{3}v_{1}-\partial_{1}v_{3}=\omega_{2},
\end{eqnarray}
which indicates that $(\alpha,\beta)$ exclusively characterize the spin mode. In contrast, $2\psi$ and $\gamma$ incorporate the contributions from both potential and vortical velocity fields. However, the potential flow terms enter with opposing signs in $2\psi$ and $\gamma$, resulting in exact cancellation when summed:
\begin{eqnarray}\label{new88}
	2\psi+\gamma=\partial_{1}v_{2}-\partial_{2}v_{1}=\omega_{3}.
\end{eqnarray}
thereby recovering the vorticity component $\omega_{3}$.

Applying the SAD to~\eqref{HH1} gives
\begin{subequations}
	\begin{eqnarray}\label{D1}
		\bm{D}=\bm{\nabla\nabla}\phi+\mathscr{S}[\bm{\nabla}\bm{v}]=\bm{\nabla\nabla}\phi+\frac{1}{2}\left(\bm{\nabla}\bm{v}+\bm{\nabla}\bm{v}^{\rm T}\right),
	\end{eqnarray}
	\begin{eqnarray}\label{Omega1}
		\bm{\varOmega}=\mathscr{A}[\bm{\nabla}\bm{v}]=\frac{1}{2}\left(\bm{\nabla}\bm{v}-\bm{\nabla}\bm{v}^{\rm T}\right).
	\end{eqnarray}	
\end{subequations}
Equation~\eqref{D1} demonstrates that the effect of the potential Hessian (namely, the gradient of the potential velocity $\bm{\nabla}\phi$) is entirely captured by $\bm{D}$: the diagonal matrix elements $(\partial_{1}^{2}\phi,\partial_{2}^{2}\phi,\partial_{3}^{2}\phi)$ contribute directly to the invariants $(\chi,\lambda_{r})$ in~\eqref{PPa} and~\eqref{PPb}, while the mixed derivative $\partial_{1}\partial_{2}\phi$ embeds in the spin invariant $\gamma$, coupling with the non-potential flow effect. Equation~\eqref{Omega1} reveals that $\bm{\varOmega}$ is solely determined by the gradients of the transverse velocity component $\bm{v}$, as revealed by~\eqref{new87} and~\eqref{new88}.

\subsection{On the sign of characteristic spin, vorticity splitting, and vortex criteria}\label{sign_of_gamma}
We notice that in NND and IVD, the sign of the spin component $\gamma$ lacks a definite sign determination. To the best of our knowledge, the fundamental ambiguity has not been explicitly recognized or analyzed in the existing literature. The characteristic equation of the reduced VGT $\widehat{\bm{A}}$ (i.e., the $2\times2$ diagonal block matrix of $\bm{A}$ in~\eqref{NND1}) is 
\begin{eqnarray}
	p(\lambda)\equiv\lambda^{2}-2\chi\lambda+\chi^{2}+\psi(\psi+\gamma)=0,
\end{eqnarray} 
where the characteristic polynomial $p(\lambda)$ is a function of the variable $\lambda$. The discriminant of $p(\lambda)$ is
\begin{eqnarray}\label{expression_delta}
	\Delta=4\psi(\psi+\gamma)=\omega_{3}^{2}-\gamma^2,
\end{eqnarray}
measuring the relative strength of $\omega_{3}$ over $\gamma$.

When the discriminant condition $\Delta>0$ is satisfied (i.e., $\lvert\omega_{3}\rvert>\lvert\gamma\rvert$), $\bm{A}$ has one real eigenvalue $\lambda_{r}$ and a complex conjugate pair $\lambda_{1,2}=\lambda_{\rm cr}\pm\rm{i}\lambda_{\rm ci}$ in which the real part $\lambda_{\rm cr}=\chi$ and the imaginary part 
$\lambda_{\rm ci}=\frac{1}{2}\sqrt{\Delta}=\frac{1}{2}\sqrt{\omega_{3}^{2}-\gamma^2}$ (being the well-known $\Delta$-criterion used to quantity the rotational strength). 
The degenerate case with $\gamma=0$ implies $\omega_{3}=2\psi$, representing purely the characteristic rigid rotation in the invariant plane.
Then, there are two possible regimes depending on the sign of $\gamma$. If (Case I) $\gamma>0$, we obtain $\omega_{3}>\gamma>0$ and thus $\psi>0$. Otherwise, if (Case II) $\gamma<0$, we get $2\psi>\omega_{3}>-\gamma>0$, implying $\psi>-\gamma>0$. For clarity, the IVD variables $(\psi,\gamma)$, respectively corresponding to Case I and II, are denoted by $(\psi^{+},\gamma^{+})$ and $(\psi^{-},\gamma^{-})$.
\begin{figure}[h!]
	\centering
	\includegraphics[width=0.9\columnwidth,trim={0cm 4.0cm 0cm 3.0cm},clip]{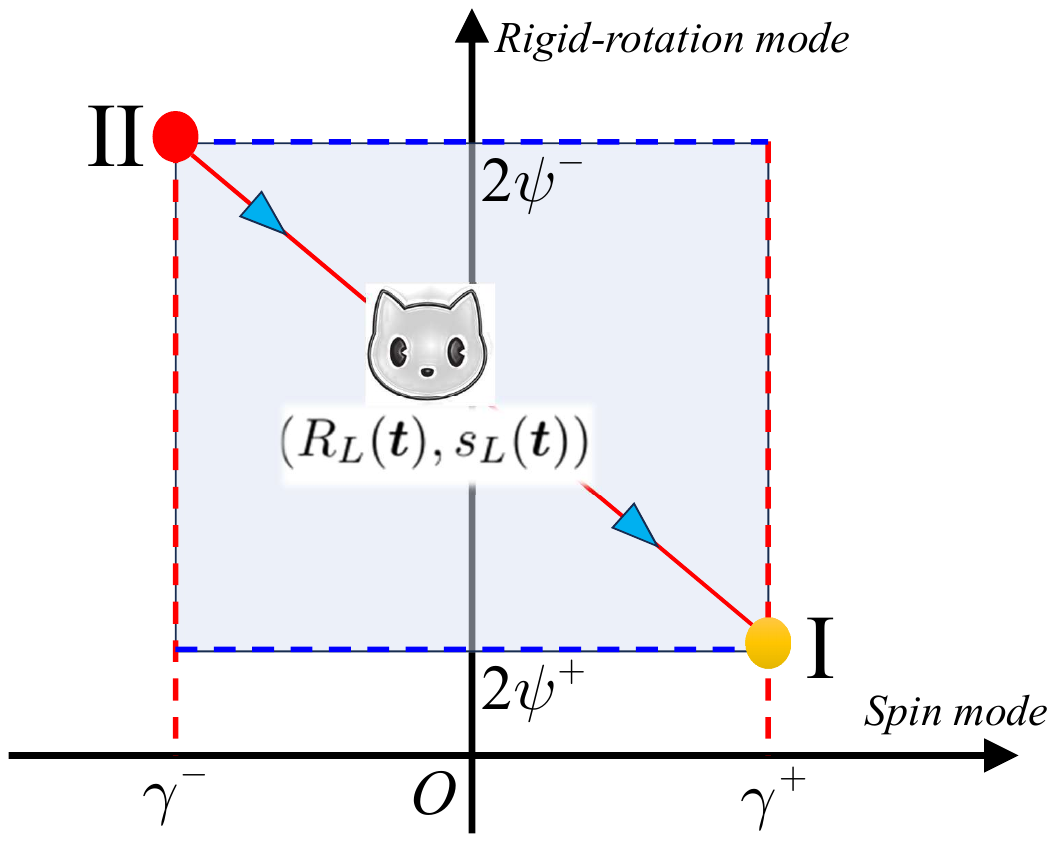}
	\caption{Illustration of the phase diagram of the distinct vorticity modes for projected flow on the invariant plane $(\Delta>0)$. The physically admissible vorticity modes $(R_{L}(\bm{t}),s_{L}(\bm{t}))$ in DVD, represented by the cat symbol, are restricted to the diagonal of the square domain $(2\psi+\gamma=\omega_{3})$ bounded by IVD vorticity constituents $(2\psi^{\pm},\gamma^{\pm})$ (indicated by the yellow and red solid circles). The IVD yields two extremal states: (Case I) $(2\psi^{+},\gamma^{+})$ and (Case II) $(2\psi^{-},\gamma^{-})$ where the former is just the Liutex-shear decomposition with minimized rigid rotation mode.} 
	\label{RS_phase}
\end{figure}

For Case I, $(\psi^{+},\gamma^{+})$ are explicitly calculated via the following expressions:
\begin{eqnarray}\label{plus_expression1}
	2\psi^{+}=\omega_{3}-\sqrt{\omega_{3}^{2}-4\lambda_{\rm ci}^{2}},~~\gamma^{+}=\sqrt{\omega_{3}^{2}-4\lambda_{\rm ci}^{2}}.
\end{eqnarray}
Note that both $\psi^{+}$ and $\gamma^{+}$ are bivariate functions of the fundamental invariants $(\omega_{3},\lambda_{\rm ci})$. The quantity $\bm{R}_{N}^{+}\equiv2\psi^{+}\bm{e}_{3}$ is precisely the Liutex vector (or Rortex) introduced by~\citet{LiuCQ2018}, which has been applied to extract the characteristic rigid rotation mode from the total vorticity to represent vortex in complex flows.
This is often the case when axial vortices are formed by wrapped shear layers, with the spin mode being transferred to the rigid-rotation mode, leading to vortex intensification.
Equation~\eqref{plus_expression1} has previously been reported in~\citet{LiuCQ2020} and~\citet{XuWenqian2019},  and also in a recent review paper by~\citet{MaoFeng2022}. The vorticity decomposition $\omega_{3}=2\psi^{+}+\gamma^{+}$ has been termed the Liutex-shear decomposition~\citep{LiuCQ2018,LiuCQ2020,Shrestha2021}. 

For Case II, one can evaluate $(\psi^{-},\gamma^{-})$ via the formulas
\begin{eqnarray}\label{plus_expression2}
	2\psi^{-}=\omega_{3}+\sqrt{\omega_{3}^{2}-4\lambda_{\rm ci}^{2}},~~\gamma^{-}=-\sqrt{\omega_{3}^{2}-4\lambda_{\rm ci}^{2}}.
\end{eqnarray}
In this case, the characteristic rigid-rotation mode is defined by $\bm{R}_{N}^{-}\equiv2\psi^{-}\bm{e}_{3}$, with the axial vorticity split as $\omega_{3}=2\psi^{-}+\gamma^{-}$. Equation~\eqref{plus_expression2} has been overlooked in the existing literature, primarily because $\gamma>0$ is usually introduced as a uniqueness condition for determining the real Schur form of $\bm{A}$. The case becomes physically relevant when the swirling strength of the axial vortex is sufficiently strong to remain coherence such that the opposing effects from spin are unable to significantly disrupt the primary vortex structure. The physical validity of this case will be demonstrated through the Burgers vortex (\S\ref{Burgers vortex}) and the vortex structures on Saturn (figure~\ref{Second_vorticity_components1}).
As a corollary, the following relations can be derived from from~\eqref{plus_expression1} and~\eqref{plus_expression2}:
\begin{eqnarray}\label{identity3}
	\psi^{+}=\psi^{-}+\gamma^{-},~~\psi^{-}=\psi^{+}+\gamma^{+},~~\gamma^{-}=-\gamma^{+}.
\end{eqnarray}

The relationship between Case I and II is elucidated as follows.
Consider a neighborhood ${U}_{P}$ of a point $P$.
The diameter of this vicinity ${\rm{diam}}{U}_{P}\equiv{\rm{sup}}\left\{\lVert\bm{x}-\bm{x}^{\prime}\rVert|\bm{x},\bm{x}^{\prime}\in{U}_{P}\right\}$ is sufficiently small to maintain the orientation of $\bm{e}_{3}$.
The projected 2D velocity field, in the invariant plane normal to $\bm{e}_{3}$, exhibits close or spiral pattern.
For Case II $(\Theta=0)$, the NND basis is notated by $(\bm{e}_{1}^{-},\bm{e}_{2}^{-})$, with the canonical block form expressed as
\begin{eqnarray}\label{a0704_1}
	\widehat{\bm{A}}
	=\begin{bmatrix}
		\chi &\psi^{-}+\gamma^{-}\\
		-\psi^{-} & \chi
	\end{bmatrix}.
\end{eqnarray}
Performing a $\frac{1}{2}\pi$ counterclockwise rotation of the right-handed basis $(\bm{e}_{1}^{-},\bm{e}_{2}^{-})$ about $\bm{e}_{3}$ yields the transformed basis $(\bm{e}_{1}^{+},\bm{e}_{2}^{+})$. Under this orthogonal transformation, we write the matrix of $\bm{A}$ under the new basis by using~\eqref{identity3}:
\begin{eqnarray}\label{a0704_2}
	\widehat{\bm{A}}
	=\begin{bmatrix}
		\chi &\psi^{-}\\
		-(\psi^{-}+\gamma^{-}) & \chi
	\end{bmatrix}
	=\begin{bmatrix}
		\chi &\psi^{+}+\gamma^{+}\\
		-\psi^{+} & \chi
	\end{bmatrix}.
\end{eqnarray}
The matrix in~\eqref{a0704_2} is identified as the canonical block representation in the real Schur form corresponding to Case I $(\Theta=\frac{\pi}{2})$. 
From~\eqref{a30_6} and~\eqref{eq82}, we derive the fundamental inequalities for the projected velocity field in the invariant plane:
\begin{eqnarray}\label{UD}
	2\psi^{+}\leq{R}_{L}(\bm{t})\leq2\psi^{-},~~\gamma^{-}\leq{s}_{L}(\bm{t})\leq\gamma^{+}.
\end{eqnarray}
Equation~\eqref{UD} reveals that the dynamic range of the DVD vorticity modes $({R}_{L}(\bm{t}),{s}_{L}(\bm{t}))$ is rigorously bounded by the IVD vorticity modes $(2\psi^{\pm},\gamma^{\pm})$. These theoretical constraints will be validated through several examples in~\S\ref{Examples}, both analytically and numerically.
As illustrated in figure~\ref{RS_phase}, physically admissible DVD vorticity modes $({R}_{L}(\bm{t}),{s}_{L}(\bm{t}))$ must reside along the principal diagonal of the square parameter space, bounded by the two critical configurations: (Case I) $(2\psi^+,\gamma^+)$ (minimum rigid rotation and maximum spin) and (Case II) $(2\psi^{-},\gamma^{-})$ (maximum rigid rotation and minimum spin).
From this perspective, the Liutex vector $\bm{R}_{N}^{+}$~\citep{LiuCQ2018} has implicitly embodied a minimization principle in selecting the characteristic rigid-rotation mode. 
Note that the NND and IVD emerge naturally from the Schur form of the VGT, revealing fundamental and coordinate-invariant kinematic constraints within an algebraic framework. For complete physical insight into vortex motions, the DVD temporal modes, rooted in intuitive streamlines, becomes indispensible in revealing the flow details.
Therefore, the coupled IVD-DVD analysis could enhance the understanding of distinct vorticity modes and vortex kinematics, as demonstrated through the following examples.

\section{Examples}\label{Examples}
\subsection{Simple shear flow and Joseph Bertrand's puzzle}\label{Simple shear flow and Bertrand's puzzle}
Consider a simple shear flow aligned with the Cartesian coordinate system $(x,y)$, where $(\bm{e}_{x},\bm{e}_{y})$ denote the streamwise and vertical unit vectors, respectively. The velocity field $\bm{u}$ is prescribed as $(u_{x},u_{y})=(ky,0)~(k>0)$. Hence, through the SAD in~\eqref{eq1a} and~\eqref{eq1b}, the VGT $\bm{A}$, the strain-rate tensor $\bm{D}$, and the rotation-rate tensor $\bm{\varOmega}$ are obtained as 
\begin{eqnarray}\label{SSF1}
	\bm{A}=k\bm{e}_{y}\bm{e}_{x},~~\bm{D}=\frac{1}{2}(k\bm{e}_{x}\bm{e}_{y}+k\bm{e}_{y}\bm{e}_{x}),~~\bm{\varOmega}=\frac{1}{2}(-k\bm{e}_{x}\bm{e}_{y}+k\bm{e}_{y}\bm{e}_{x}).
\end{eqnarray}
Then, the vorticity $\bm{\omega}$ is deduced as the dual vector of $\bm{\varOmega}$, namely, $\bm{\omega}=-k\bm{e}_{z}$. 
At the times of Cauchy and Stokes, one could only apply the classical interpretation to postulate that a volume element of fluid must perform a uniform rigid rotation with the angular velocity $\bm{W}_{V}=\frac{1}{2}\bm{\omega}=-\frac{k}{2}\bm{e}_{z}$ (see~\eqref{pp7b}). But then any material line element within a fluid volume element would obey the classical prediction, which however, fundamentally contradicts the truth that a material line element aligned with $\bm{e}_{x}$ exhibits pure translation with zero rigid rotation. This paradox naturally disappears once the specific angular velocity $\bm{W}_{D}(\bm{e})$ is recognized.
\begin{figure}[h!]
	\centering
	\subfigure[$(\bm{W}_{D}(\bm{e}),\bm{W}_{L}(\bm{e})$]{
		\begin{minipage}[t]{0.5\linewidth}
			\centering
			\includegraphics[width=1.0\columnwidth,trim={0cm 0.0cm 0.0cm 0.0cm},clip]{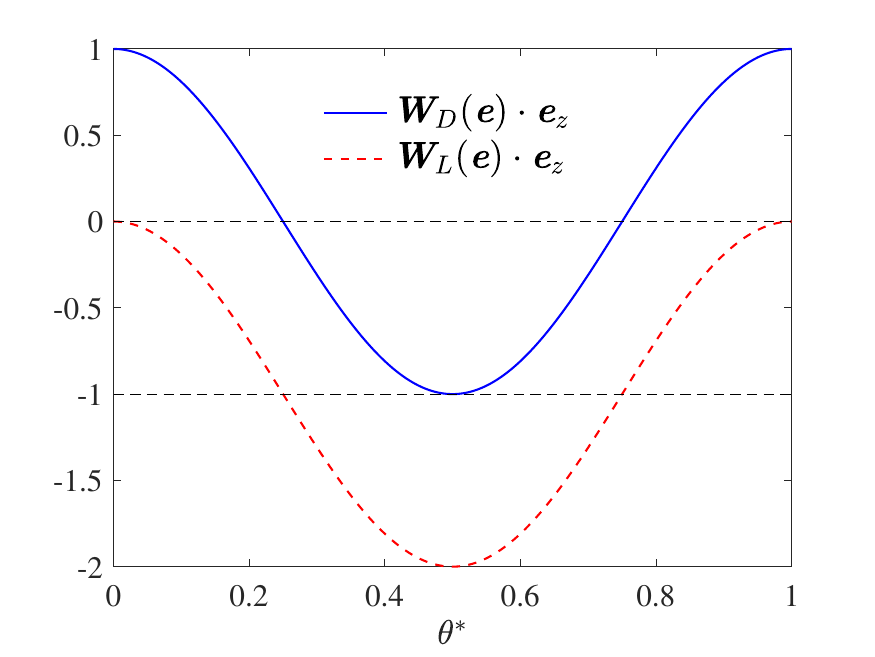}
			\label{simple_shear_flow1}
		\end{minipage}%
	}%
	\subfigure[$(\bm{R}_{L}(\bm{e}),\bm{s}_{L}(\bm{e}))$]{
		\begin{minipage}[t]{0.5\linewidth}
			\centering
			\includegraphics[width=1.0\columnwidth,trim={0cm 0.0cm 0.0cm 0.0cm},clip]{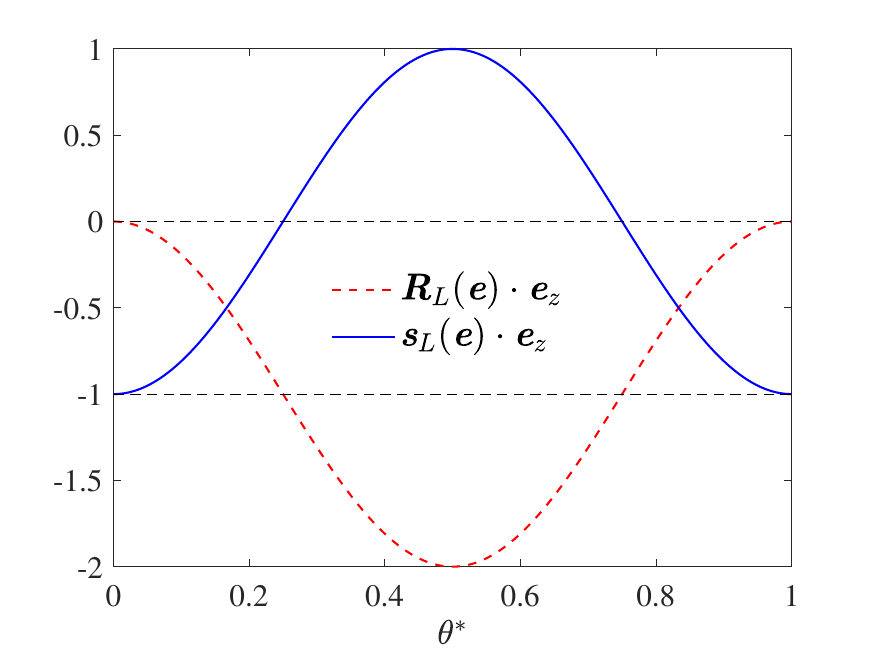}
			\label{simple_shear_flow2}
		\end{minipage}%
	}%
	\caption{The $\bm{e}_{z}$-components of (a) $(\bm{W}_{D}(\bm{e}),\bm{W}_{L}(\bm{e}))$ and (b) $(\bm{R}_{L}(\bm{e}),\bm{s}_{L}(\bm{e}))$ for a material line element $\delta\bm{r}=\delta{r}\bm{e}$, with the unit orientation vector $\bm{e}=\cos\theta\bm{e}_{x}+\sin\theta\bm{e}_{y}$.The angle $\theta$ is normalized by $\pi$ (i.e., $\theta^* \equiv \theta/\pi$), while the angular velocities and vorticity components are normalized by $\frac{k}{2}$ and $k$, respectively.} 
	\label{simple_shear_flow}
\end{figure}

For a material line element oriented at the angle $\theta$ (i.e., $\bm{e}=\cos\theta\bm{e}_{x}+\sin\theta\bm{e}_{y}$), by using~\eqref{pp3} and~\eqref{pp6}, we derive the exact expressions of its specific and total angular velocities:
\begin{eqnarray*}
	\bm{W}_{D}(\bm{e})=\bm{e}\times(\bm{e}\bm{\cdot}\bm{D})=\frac{1}{2}k\cos2\theta\bm{e}_{z},
\end{eqnarray*}
\begin{eqnarray}\label{SSF2}
	\bm{W}_{L}(\bm{e})=\bm{W}_{D}(\bm{e})+\frac{1}{2}\bm{\omega}=\frac{1}{2}k(\cos2\theta-1)\bm{e}_{z},
\end{eqnarray}
which introduces an additional rotational degree of freedom $\theta$ to the line-element rotation. Consider a material line element initially aligned with $\bm{e}=\bm{e}_{x}$ ($\theta=0$) at $t=0$. Its specific angular velocity $\bm{W}_{D}(\bm{e}_{x})=\frac{k}{2}\bm{e}_{z}$ exactly cancels out $\bm{W}_{V}=-\frac{k}{2}\bm{e}_{z}$, resulting in no net rigid rotation -- resolving the puzzle first raised by Joseph Bertrand in 1868~\citep{Truesdell1954}. 
As shown in figure~\ref{simple_shear_flow1}, when $\theta$ increases from 0 to values within $(0,\frac{1}{4}\pi)$, both angular velocity components change: $0<W_{Dz}<\frac{1}{2}k$ and $-\frac{k}{2}<W_{Lz}<0$, causing the clockwise rotation of the line element. 
At $\theta=\frac{1}{4}\pi$ (the principal stretching axis of $\bm{D}$), the specific angular velocity vanishes with $(W_{Dz},W_{Lz})=(0,-\frac{1}{2}k)$. Beyond this point $(\theta\in(\frac{1}{4}\pi,\frac{3}{4}\pi))$, the specific angular velocity component (dominated as $\cos2\theta$) becomes negative $(-\frac{1}{2}k\leq{W}_{Dz}<0)$. The rotation continues until $\theta=\frac{3}{4}\pi$ (the principal compression axis of $\bm{D}$), where $W_{Dz}$ again vanishes. 
Both components reach their minima at $\theta=\frac{1}{2}\pi$, with $(W_{Dz},W_{Lz})=(-\frac{1}{2}k,-k)$, while $W_{Lz}$ returns to zero at $\theta=\pi$. This peculiar rotation pattern provides a more complete interpretation for Joseph Bertrand's puzzle from the perspective of rotation of line element.

Using~\eqref{pp9} and~\eqref{pp10}, the DVD vorticity modes are explicitly determined as
\begin{eqnarray}\label{SSF3}
	\bm{R}_{L}(\bm{e})=k(\cos2\theta-1)\bm{e}_{z},~~
	\bm{s}_{L}(\bm{e})=-k\cos2\theta\bm{e}_{z},
\end{eqnarray}
as displayed in figure~\ref{simple_shear_flow2}.
When setting $\bm{e}=\bm{t}$ ($\theta=0$, $\bm{t}=\bm{e}_{x}$ is the unit tangent vector of a straight streamline),~\eqref{SSF3} implies $\bm{R}_{L}(\bm{t})=\bm{0}$ and $\bm{s}_{L}(\bm{t})=-k\bm{e}_{z}=\bm{\omega}$ (being consistent with the field description).
This shows that vorticity arises solely from the spin/shear mode, with no contribution from the orbital rotation mode. 
Consequently, despite a potentially high vorticity magnitude, no swirling vortex motion with curved streamlines exists, and hence vorticity is not an appropriate physical measure to represent vortex in the spin-dominant region. 

From the perspective of characteristic algebraic description $(\Delta=0)$, the condition $\omega_{3}>0$ is enforced to ensure uniqueness,  thereby determining the rotation axis as $\bm{e}_{3}=-\bm{e}_{z}$. (Case I) If $(\bm{e}_{1},\bm{e}_{2})=(\bm{e}_{y},\bm{e}_{x})$, the VGT is expressed as $\bm{A}=k\bm{e}_{1}\bm{e}_{2}$. In the Liutex-shear decomposition~\eqref{plus_expression1},
this case gives $(\psi^{+},\gamma^{+})=(0,k)$ and $\omega_{3}=\gamma^{+}=k>0$. Since $\bm{e}_{3}=-\bm{e}_{z}$, it follows that $\omega_{z}=-{\omega}_{3}=-k<0$ and $\gamma^{+}=-\bm{s}_{L}(\bm{t})\bm{\cdot}\bm{e}_{z}$.
This extremal state (Case I) satisfies the minimization principle of rigid rotation inherent in the Liutex theory, which corresponds precisely to the physically admissible solution in the field description.
Alternatively, (Case II) if $(\bm{e}_{1},\bm{e}_{2})=(\bm{e}_{x},-\bm{e}_{y})$, the VGT becomes $\bm{A}=-k\bm{e}_{2}\bm{e}_{1}$, leading to $(\psi^{-},\gamma^{-})=(2k,-k)$ in~\eqref{plus_expression2} and $\omega_{3}=2\psi^{-}+\gamma^{-}=k>0$. Due to $\bm{e}_{3}=-\bm{e}_{z}$, we obtain $\omega_{z}=-{\omega}_{3}=-k<0$ and $\gamma^{-}=\bm{s}_{L}(\bm{t})\bm{\cdot}\bm{e}_{z}$. The latter case $(\gamma<0)$ contradicts the physical scenario of rectilinear streamlines (which exhibit no rotation pattern), whereas the former $(\gamma>0)$ remains physically consistent.
In general, the sign of $\gamma$ is not inherently fixed.  To ensure physical meaningfulness, additional constraints must be imposed -- for example, enforcing $\gamma>0$ would eliminate the non-physical second case and uniquely determine the matrix representation of $\bm{A}$.

\subsection{Potential flow outside a point vortex}\label{point_vortex}
For the pure potential flow $(\Delta=-\gamma^{2}<{0}~\text{when}~\gamma\neq{0})$, the degenerate case with $\bm{\omega}=\bm{0}$ implies $\alpha=\beta=0$ and $\gamma=-2\psi=2\partial_{1}\partial_{2}\phi$. The characteristic spin $\gamma$ can coexist with the characteristic rigid rotation $\psi$, yet produces no surface shear stress on a surface element due to the absence of viscosity.
Thus, $\bm{A}$ is characterized by three rotational invariants $(\chi,\lambda_{r},\gamma)$ or $(\chi,\lambda_{r},\psi)$, where $\chi=\partial_{1}^{2}\phi=\partial_{2}^{2}\phi$ and $\lambda_{r}=\partial_{3}^{2}\phi$ (\S\ref{NND_Helmholtz}). Under the NND-II, $\bm{A}$ admits a generic matrix representation:
\begin{equation}\label{NND1_PF}
	\bm{A}=\bm{D}=\bm{\nabla\nabla}\phi=
	\begin{bmatrix}
		\chi & -\psi & 0\\
		-\psi & \chi&0\\
		0&  0& \lambda_{r}
	\end{bmatrix}
	=	\begin{bmatrix}
		\chi & \frac{1}{2}\gamma & 0\\
		\frac{1}{2}\gamma & \chi&0\\
		0&  0& \lambda_{r}
	\end{bmatrix}.
\end{equation} 
For a material line element $\delta\bm{r}=\delta{r}\bm{e}$, the following relationship generally holds:
\begin{eqnarray*}
	\bm{R}(\bm{e})=-\bm{s}(\bm{e})=2\bm{e}\times\left(\bm{e}\bm{\cdot}\bm{\nabla\nabla}\phi\right).
\end{eqnarray*}

Consider the potential flow outside a 2D point vortex with the constant circulation $\Gamma=2\pi{K} (K>0)$. Using the cylindrical coordinate system $(r,\theta,z)$, the circumferential velocity component is $u_{\theta}(r)=K/r$ for $r>0$. The VGT $\bm{A}$ under the basis $(\bm{e}_{r},\bm{e}_{\theta})$ is expressed as $\bm{A}=(\partial_{r}u_{\theta})\bm{e}_{r}\bm{e}_{\theta}-(u_{\theta}/r)\bm{e}_{\theta}\bm{e}_{r}$
from which $(\psi,\gamma)$ in~\eqref{NND1} are identified as
\begin{eqnarray*}
	\psi=\frac{u_{\theta}}{r}=\frac{K}{r^{2}}>0,
\end{eqnarray*}
\begin{eqnarray}\label{eq1031}
	\gamma=\frac{\partial{u}_{\theta}}{\partial r}-\frac{u_{\theta}}{r}=-\frac{2K}{r^{2}}<0.
\end{eqnarray}
The sum of $2\psi$ and $\gamma$ yields $\omega_{3}=\omega_{z}=2\psi+\gamma=0$. By contrast, when the Liutex method~\citep{LiuCQ2018} is applied in the case $\Delta<0$, it results in the trivial solution: $\psi=\gamma=\omega_{3}=0$, where the rigid rotation and spin effects exactly cancel. While mathematically valid, the Liutex-based analysis appears to lose information about the underlying velocity gradient structure that originally contributed to $(\psi,\gamma)$, as shown in~\eqref{eq1031}.

To fully characterize the flow physics, we examine the line-element-based vorticity decomposition (i.e.,~\eqref{pp8} and~\eqref{eq14}). Along a closed streamline $\mathcal{C}$, its holds that $(\bm{t},\bm{n},\bm{b})=(\bm{e}_{\theta},-\bm{e}_{r},\bm{e}_{z})$. For a material line element aligned with $\bm{t}$, $(\bm{R}_{L}(\bm{t}),\bm{s}_{L}(\bm{t}))$ are just the DVD modes (\eqref{C8a3} and~\eqref{C8a10}), evaluated as
\begin{eqnarray*}
	\bm{R}_{L}(\bm{t})=2\frac{u_{\theta}}{r}\bm{e}_{z}=2\psi\bm{e}_{z},
\end{eqnarray*}
\begin{eqnarray}
	\bm{s}_{L}(\bm{t})=\left(\frac{\partial u_{\theta}}{\partial r}-\frac{u_{\theta}}{r}\right)\bm{e}_{z}=\gamma\bm{e}_{z}.
\end{eqnarray}
In this case, the DVD modes are consistent with the IVD modes (in \eqref{IVD}), i.e., $\bm{R}_{L}(\bm{t})=\bm{R}_{N}$ and $\bm{s}_{L}(\bm{t})=\bm{s}_{N}$. $\bm{R}_{L}(\bm{t})$ characterizes a classical circular motion with the speed $u_{\theta}$ and the curvature radius $r$.
For the complementary material line element aligned with $\bm{n}$, the following relations hold:
\begin{eqnarray*}
	\bm{R}_{L}(\bm{n})=2\frac{\partial u_{\theta}}{\partial r}\bm{e}_{z}=2(\psi+\gamma)\bm{e}_{z},
\end{eqnarray*}
\begin{eqnarray}\label{eq105}
	\bm{s}_{L}(\bm{n})=-\left(\frac{\partial u_{\theta}}{\partial r}-\frac{u_{\theta}}{r}\right)\bm{e}_{z}=-\gamma\bm{e}_{z}.
\end{eqnarray}
It is seen that the characteristic spin $\gamma$ affects both the rigid rotation and spin modes in the DVD. Due to $\gamma=-2\psi$, the first equality in~\eqref{eq105} is simplified as $-2\psi\bm{e}_{z}$.
\subsection{Burgers vortex}\label{Burgers vortex}
\begin{figure}[t]
	\centering
	\subfigure[]{
		\begin{minipage}[t]{0.5\linewidth}
			\centering
			\includegraphics[width=1.0\columnwidth,trim={0cm 0.0cm 0.0cm 0.0cm},clip]{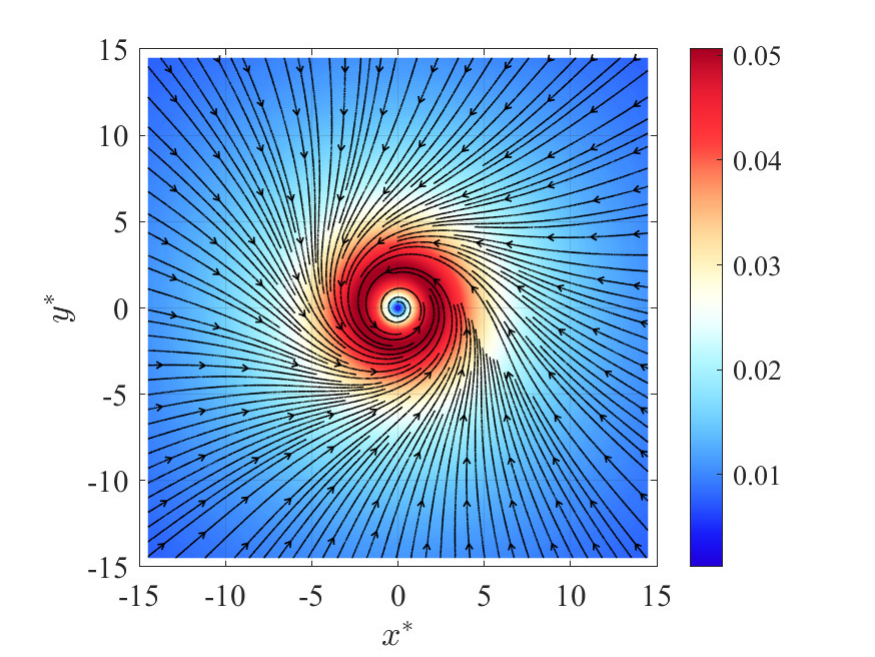}
			\label{CT_Burgers_utheta}
		\end{minipage}%
	}%
	\subfigure[]{
		\begin{minipage}[t]{0.5\linewidth}
			\centering
			\includegraphics[width=1.0\columnwidth,trim={0cm 0.0cm 0.0cm 0.0cm},clip]{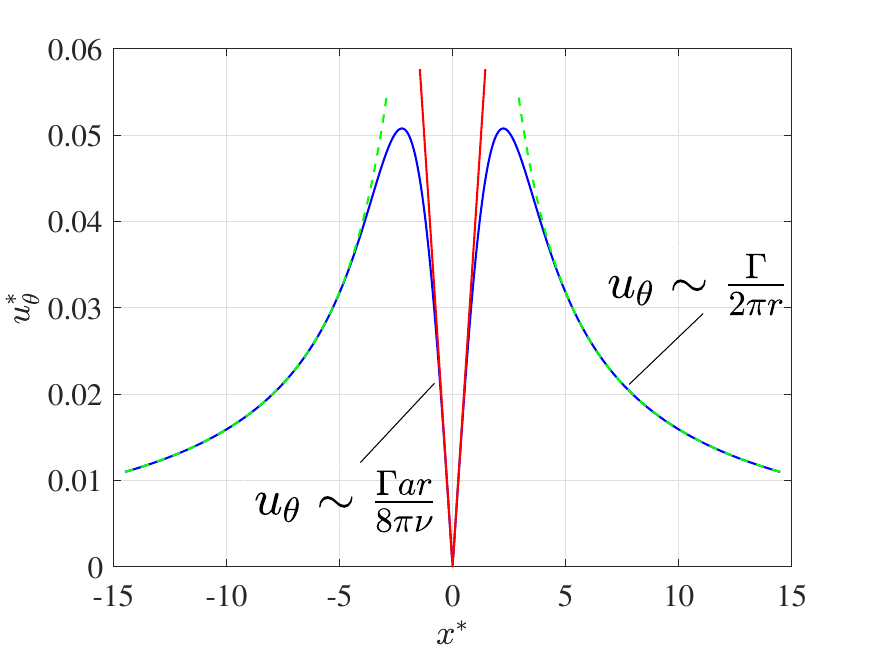}
			\label{Burgers_utheta_r_line}
		\end{minipage}%
	}%
	
	\subfigure[]{
		\begin{minipage}[t]{0.5\linewidth}
			\centering
			\includegraphics[width=1.0\columnwidth,trim={0cm 0.0cm 0.0cm 0.0cm},clip]{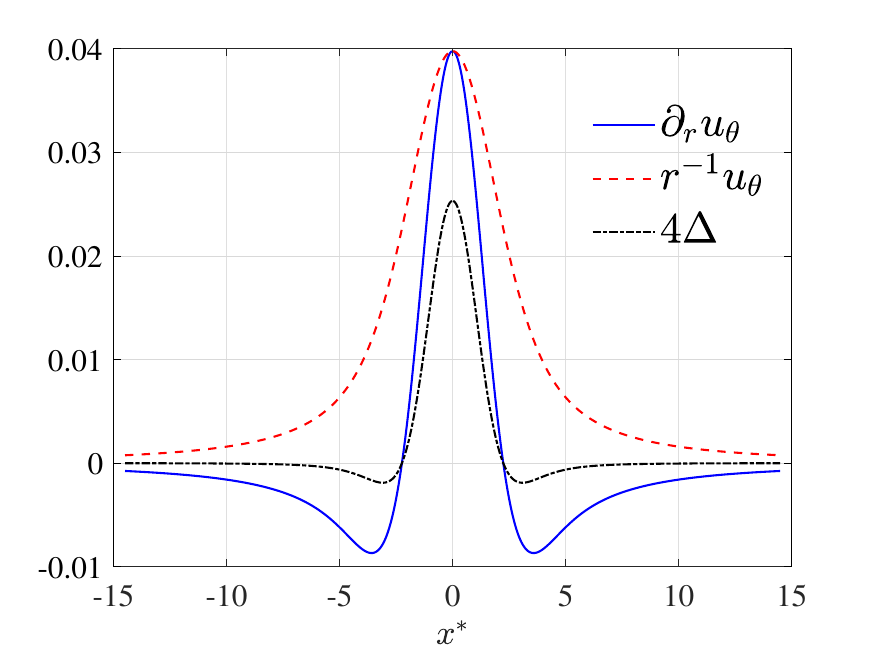}
			\label{CT_compare_drdu_udr_delta}
		\end{minipage}%
	}%
	\caption{(a) Contour map and (b) radial profile of the normalized circumferential velocity component $u_{\theta}(r)$ for the Burgers vortex. (c) Comparison of normalized quantities: the radial gradient of the circumferential velocity $\partial_{r}u_{\theta}$, the angular velocity $u_{\theta}/r$, and the discriminant $\Delta = 4(u_{\theta}/r)(\partial_{r}u_{\theta})$. For visual clarity, the normalized discriminant $\Delta^{*} \equiv \Delta/\omega_{\rm ref}^{2}$ has been scaled by a factor of 4.} 
	\label{HHH1}
\end{figure}

\begin{figure}[t]
	\centering
	\subfigure[$2\psi^{-}$]{
		\begin{minipage}[t]{0.5\linewidth}
			\centering
			\includegraphics[width=1.0\columnwidth,trim={0cm 0.0cm 0.0cm 0.0cm},clip]{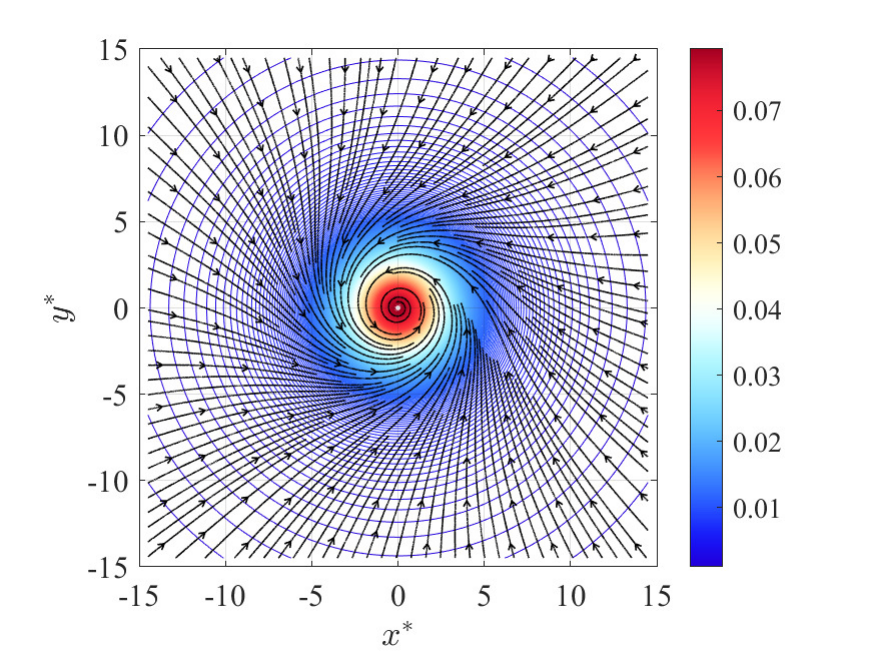}
			\label{CT_Burgers_Psi_square}
		\end{minipage}%
	}%
	\subfigure[$\gamma^{-}$]{
		\begin{minipage}[t]{0.5\linewidth}
			\centering
			\includegraphics[width=1.0\columnwidth,trim={0cm 0.0cm 0.0cm 0.0cm},clip]{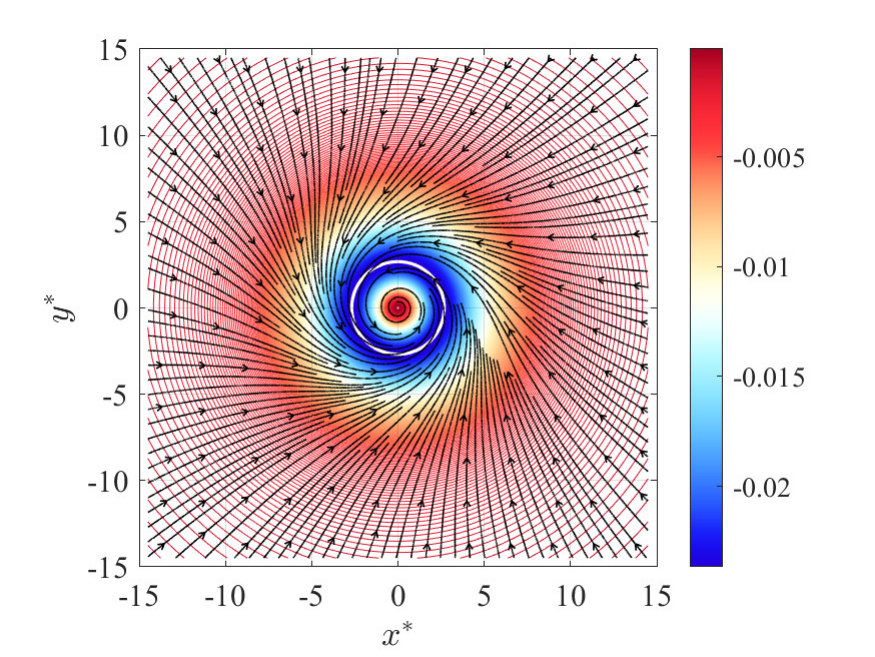}
			\label{CT_Burgers_gamma}
		\end{minipage}%
	}%
	
	\subfigure[$\omega_{3}$]{
		\begin{minipage}[t]{0.5\linewidth}
			\centering
			\includegraphics[width=1.0\columnwidth,trim={0cm 0.0cm 0.0cm 0.0cm},clip]{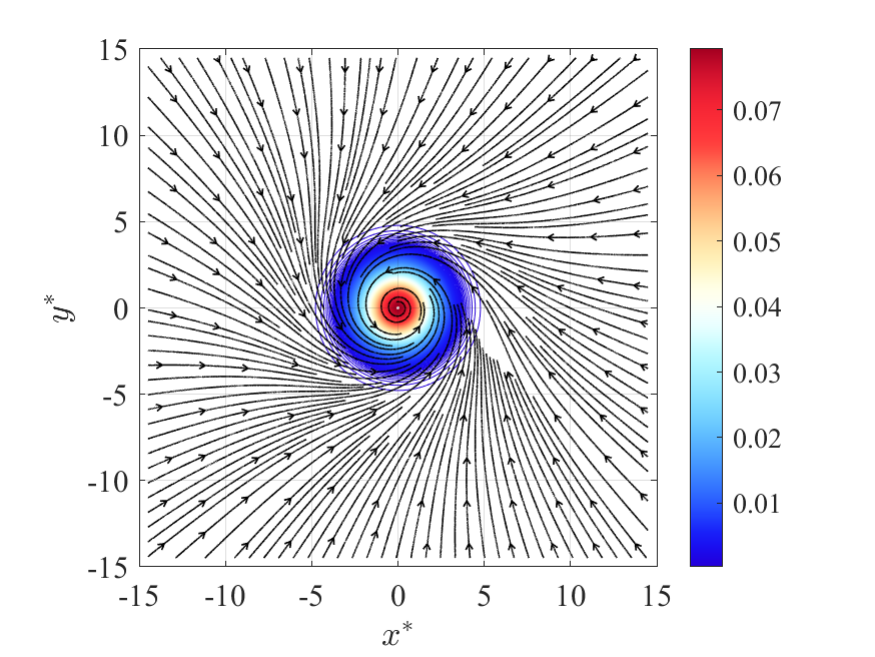}
			\label{CT_Burgers_omegaz}
		\end{minipage}%
	}%
	\subfigure[$(2\psi^{-},\gamma^{-},\omega_{3})$]{
		\begin{minipage}[t]{0.5\linewidth}
			\centering
			\includegraphics[width=1.0\columnwidth,trim={0cm 0.0cm 0.0cm 0.0cm},clip]{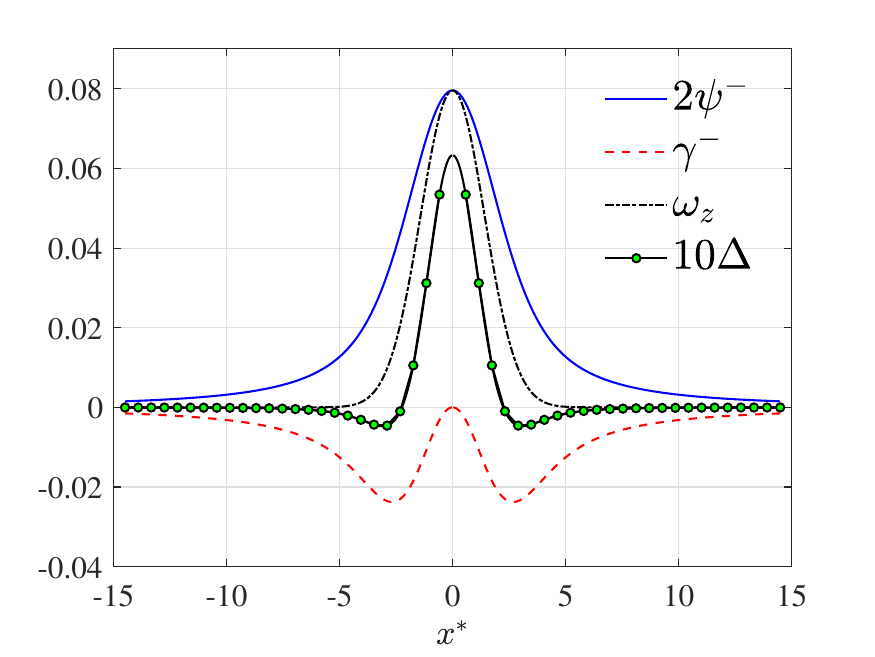}
			\label{CT_Burgers_compare_IVD}
		\end{minipage}%
	}%
	\caption{Contour maps of the IVD vorticity modes: (a) rigid rotation $2\psi^{-}$, (b) spin $\gamma^{-}$, and (c) axial vorticity $\omega_{3}$ for the Burgers vortex. (d) Comparison of the radial profiles of $(2\psi^{-},\gamma^{-},\omega_{3})$. For visual clarity, the normalized discriminant $\Delta^{*} \equiv \Delta/\omega_{\rm ref}^{2}$ has been scaled by a factor of 10 in panel (d).} 
	\label{HHH2}
\end{figure}

\begin{figure}[t]
	\centering
	\subfigure[$\bm{R}_{L}(\bm{t})\bm{\cdot}\bm{e}_{z}$]{
		\begin{minipage}[t]{0.5\linewidth}
			\centering
			\includegraphics[width=1.0\columnwidth,trim={0cm 0.0cm 0.0cm 0.0cm},clip]{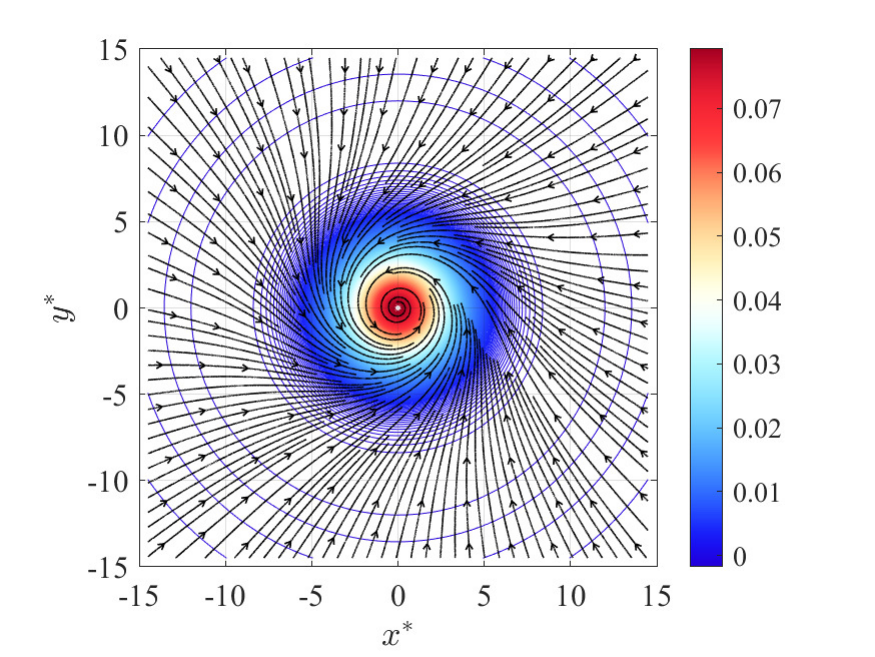}
			\label{CT_Burgers_Rtz_z0}
		\end{minipage}%
	}%
	\subfigure[$\bm{s}_{L}(\bm{t})\bm{\cdot}\bm{e}_{z}$]{
		\begin{minipage}[t]{0.5\linewidth}
			\centering
			\includegraphics[width=1.0\columnwidth,trim={0cm 0.0cm 0.0cm 0.0cm},clip]{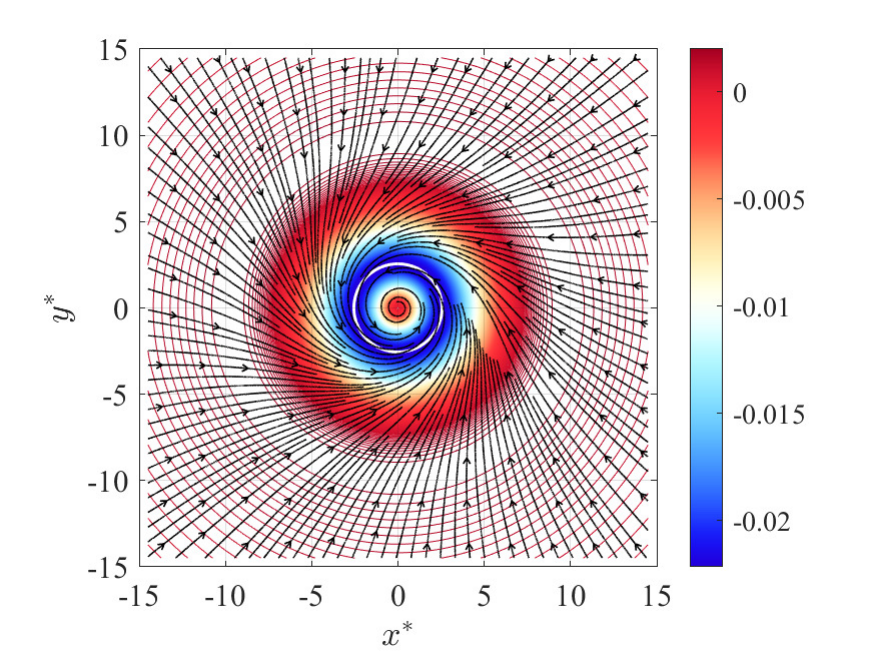}
			\label{CT_Burgers_stz_z0}
		\end{minipage}%
	}%
	
	\subfigure[$(\bm{R}_{L}(\bm{t})\bm{\cdot}\bm{e}_{z},\bm{s}_{L}(\bm{t})\bm{\cdot}\bm{e}_{z},\omega_{z})$]{
		\begin{minipage}[t]{0.5\linewidth}
			\centering
			\includegraphics[width=1.0\columnwidth,trim={0cm 0.0cm 0.0cm 0.0cm},clip]{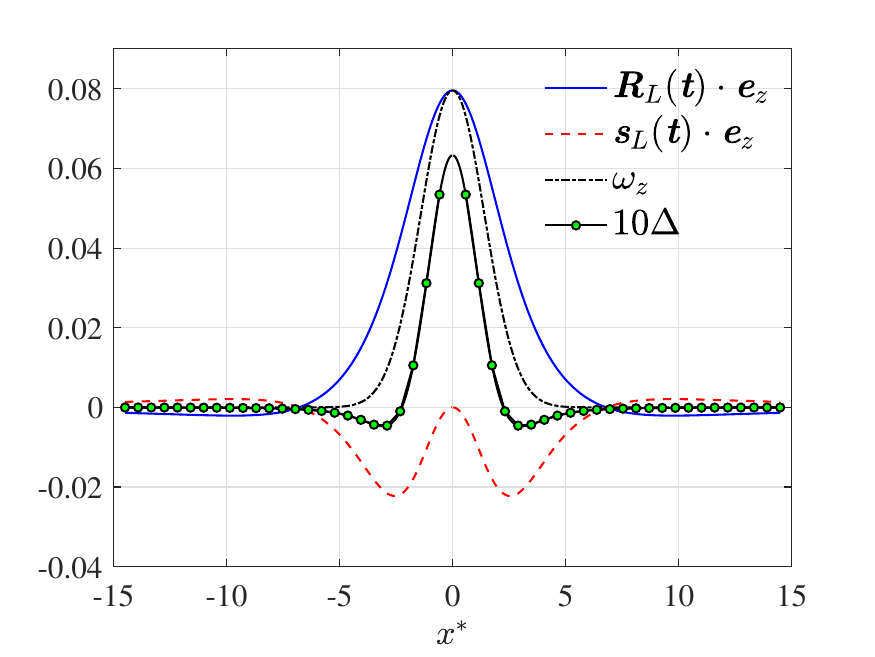}
			\label{CT_Burgers_compare_DVD}
		\end{minipage}%
	}%
	\caption{Contour maps of the DVD vorticity modes at $z=0$: (a) orbital rotation $\bm{R}_{L}(\bm{t})\bm{\cdot}\bm{e}_{z}$ and (b) spin $\bm{s}_{L}(\bm{t})\bm{\cdot}\bm{e}_{z}$. (c) Comparison of the radial profiles of $(\bm{R}_{L}(\bm{t})\bm{\cdot}\bm{e}_{z},\bm{s}_{L}(\bm{t})\bm{\cdot}\bm{e}_{z},\omega_{z})$. For visual clarity, the normalized discriminant $\Delta^{*} \equiv \Delta/\omega_{\rm ref}^{2}$ has been scaled by a factor of 10 in panel (c).} 
	\label{HHH3}
\end{figure}
\begin{figure}[t]
	\centering
	\subfigure[$(2\psi^{-},\bm{R}_{L}(\bm{t})\bm{\cdot}\bm{e}_{z})$]{
		\begin{minipage}[t]{0.5\linewidth}
			\centering
			\includegraphics[width=1.0\columnwidth,trim={0cm 0.0cm 0.0cm 0.0cm},clip]{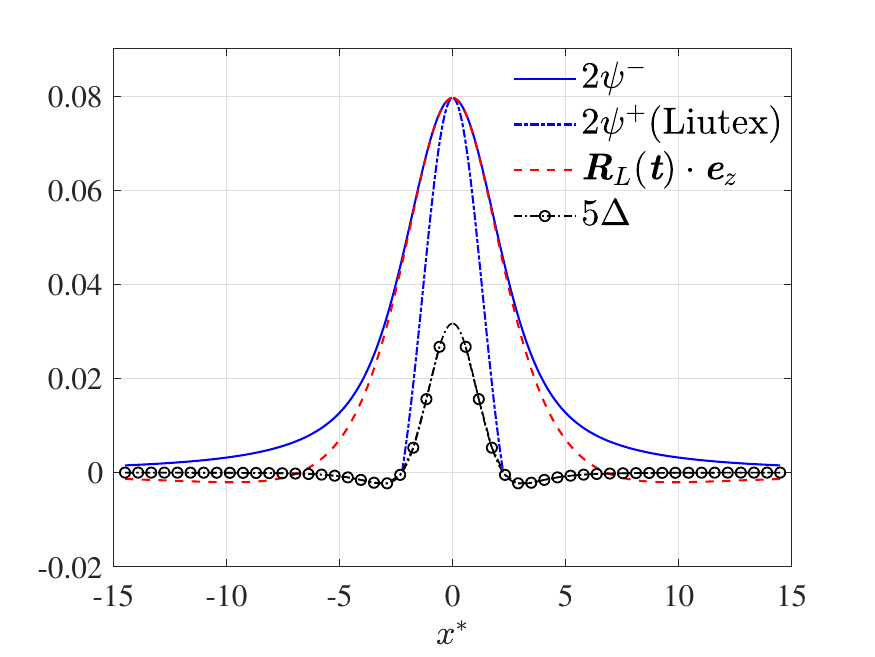}
			\label{CT_compare_2Psi_Rtz}
		\end{minipage}%
	}%
	\subfigure[$(\gamma^{-},\bm{s}_{L}(\bm{t})\bm{\cdot}\bm{e}_{z})$]{
		\begin{minipage}[t]{0.5\linewidth}
			\centering
			\includegraphics[width=1.0\columnwidth,trim={0cm 0.0cm 0.0cm 0.0cm},clip]{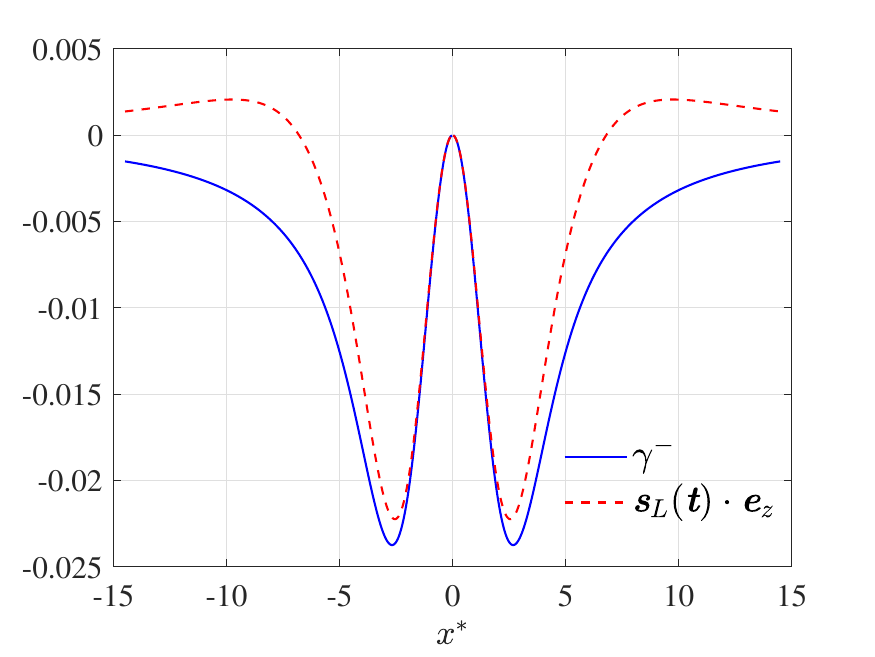}
			\label{CT_compare_gamma_stz}
		\end{minipage}%
	}%
	\caption{Comparison of (a) $(2\psi^{\pm},\bm{R}_{L}(\bm{t})\bm{\cdot}\bm{e}_{z})$ with $R_{N}^{\pm}=2\psi^{\pm}$, and (b) $(\gamma^{-},\bm{s}_{L}(\bm{t})\bm{\cdot}\bm{e}_{z})$ for the Burgers vortex. For visual clarity, the normalized discriminant $\Delta^{*} \equiv \Delta/\omega_{\rm ref}^{2}$ has been scaled by a factor of 5.} 
	\label{HHH4}
\end{figure}
The celebrated Burgers vortex is a steady, axisymmetric exact solution of the Navier-Stokes (NS) equations, widely used to model fine-scale turbulence structures owing to its finite and viscosity-independent total dissipation rate~\citep{Burgers1948,WuJZ2015book}. 
It consists of an inward radial flow that concentrates and rotates around the symmetry axis, while the fluid moves outward in opposite directions along the axial dimension.
In the following analysis, we adopt the cylindrical coordinate system $(r,\theta,z)$ with the unit basis vectors $\{\bm{e}_{r},\bm{e}_{\theta},\bm{e}_{z}\}$. The velocity field of the Burgers vortex $\bm{u}=u_{r}\bm{e}_{r}+u_{\theta}\bm{e}_{\theta}+u_{z}\bm{e}_{z}$, can then be expressed as follows:
\begin{subequations}\label{Burgers_123}
	\begin{eqnarray}\label{Burgers_r}
		u_{r}=-\frac{1}{2}ar,
	\end{eqnarray}
	\begin{eqnarray}\label{Burgers_theta}
		u_{\theta}=\frac{\Gamma}{2\pi r}\left[1-\exp\left(-\frac{r^2a}{4\nu}\right)\right],
	\end{eqnarray}
	\begin{eqnarray}\label{Burgers_z}
		u_{z}=az.
	\end{eqnarray}
\end{subequations}
Here, $\Gamma$ denotes the total circulation that quantities the vortex strength; the constant ${a} (>0)$ represents the axial stretching rate, $\nu$ is the kinematic viscosity, and ${r}$ is the radial coordinate measured from the $z$-axis (i.e., the symmetry axis of the Burgers vortex).

Using~\eqref{Burgers_123}, the tensors $\bm{A}$ and $\bm{D}$ in the coordinate system $(r,\theta,z)$ are given by
\begin{eqnarray}\label{Burgers_VGT}
	\bm{A}=
	\begin{bmatrix}
		-\frac{1}{2}a &\partial_{r}u_{\theta}& 0\\
		-\frac{1}{r}u_{\theta} &-\frac{1}{2}a&0\\
		0&0& a
	\end{bmatrix},~~
	\bm{D}=
	\begin{bmatrix}
		-\frac{1}{2}a &\frac{1}{2}\gamma^{-}& 0\\
		\frac{1}{2}\gamma^{-} &-\frac{1}{2}a&0\\
		0&0& a
	\end{bmatrix}.
\end{eqnarray}
Interestingly, the VGT $\bm{A}$ in~\eqref{Burgers_VGT} exhibits exactly the canonical form of the NND matrix in~\eqref{NND1}.
Consequently, the orthonormal triad $\{\bm{e}_{1},\bm{e}_{2},\bm{e}_{3}\}$ naturally aligns with the standard cylindrical basis vectors $\{\bm{e}_{r},\bm{e}_{\theta},\bm{e}_{z}\}$. Through direct comparison, we obtain the following parameter correspondence: $\chi=-\frac{1}{2}a$, $\lambda_{r}=a$, and $\alpha=\beta=0$. The IVD variables $(\psi^{-},\gamma^{-},\omega_{3})$ are subsequently identified as
\begin{subequations}\label{97abc}
	\begin{eqnarray}\label{new000}
		\psi^{-}=\frac{u_{\theta}}{r},
	\end{eqnarray}
	\begin{eqnarray}\label{new00}
		\gamma^{-}=\frac{\partial{u}_{\theta}}{\partial r}-\frac{u_{\theta}}{r}
		=r\frac{\partial}{\partial r}\left(\frac{u_{\theta}}{r}\right),
	\end{eqnarray}
	\begin{eqnarray}\label{new0}
		\omega_{3}=2\psi^{-}+\gamma^{-}=\frac{\partial{u}_{\theta}}{\partial r}+\frac{u_{\theta}}{r}
		=\frac{1}{r}\frac{\partial}{\partial r}\left(ru_{\theta}\right).
	\end{eqnarray}
\end{subequations}
From~\eqref{new000}, the characteristic rigid-rotation mode is given by
\begin{eqnarray}\label{RN-}
	\bm{R}_{N}^{-}\equiv2\psi^{-}\bm{e}_{3}=2\frac{{u}_{\theta}}{r}\bm{e}_{z}.
\end{eqnarray}
Locally, $\psi^{-}$ describes a virtual circular motion characterized by the circumferential velocity $u_{\theta}$ and the curvature radius $r$, contributing $2\psi^{-}$ to the vorticity. While this simplified representation does not capture all the flow details, it highlights a key physical feature of a vortex: the surrounding streamlines spiral around the local rotation axis in a consistent direction.
From~\eqref{expression_delta}, the discriminant $\Delta$ of the reduced VGT $\widehat{\bm{A}}$ is 
\begin{eqnarray}\label{XX1}
	\Delta=4\psi^{-}(\psi^{-}+\gamma^{-})=4\frac{u_{\theta}}{r}\frac{\partial{u}_{\theta}}{\partial r}.
\end{eqnarray}
Using~\eqref{Burgers_theta}, the expressions in~\eqref{97abc} are explicitly evaluated as
\begin{eqnarray*}
	\psi^{-}=\frac{\Gamma}{2\pi r^2}\left[1-\exp\left(-\frac{r^2a}{4\nu}\right)\right],
\end{eqnarray*}
\begin{eqnarray*}
	\gamma^{-}=\frac{\Gamma a}{4\pi\nu}\exp\left(-\frac{r^2a}{4\nu}\right)-\frac{\Gamma}{\pi r^2}\left[1-\exp\left(-\frac{r^2a}{4\nu}\right)\right],
\end{eqnarray*}
\begin{eqnarray*}
	\omega_{3}=\frac{\Gamma a}{4\pi\nu}\exp\left(-\frac{r^2a}{4\nu}\right).
\end{eqnarray*}
It should be noted that the Burgers vortex represents a canonical example of Case II $(\gamma<0)$, as established in~\S\ref{sign_of_gamma} and confirmed by the negative spin in figure~\ref{CT_Burgers_compare_IVD}. Consequently, we consistently employ the notation $(\psi^{-},\gamma^{-})$ in the analysis to emphasize this characteristic feature. In the region $\Delta>0$, since $\Delta=4\lambda_{\rm ci}^{2}$, substituting~\eqref{new0} and~\eqref{XX1} into~\eqref{plus_expression2} yields~$(\psi^{-},\gamma^{-})$ in~\eqref{new000} and~\eqref{new00}, and $\bm{R}_{N}^{-}$ in~\eqref{RN-}. In addition, the Helmholtz decomposition of the Burgers vortex solution is given in Appendix~\ref{AP3}.

The Burgers vortex emerges from a dynamic equilibrium among vorticity convection, vortex stretching, and viscous diffusion. 
The reference characteristic scales for length, velocity and vorticity are therefore introduced as $L_{\rm{ref}}\equiv({\nu}/{a})^{1/2}$, $U_{\rm{ref}}\equiv{\Gamma}/{L_{\rm{ref}}}$, and $\omega_{\rm{ref}}\equiv{\Gamma}/{L_{\rm{ref}}^2}$.
Normalized quantities are denoted by a superscript asterisk.
Using the parameters ($a=0.084$, $\Gamma=1.45$, $\nu=0.01$) consistent with those in~\citet{Shrestha2021}, we analyze the flow structure.
Figure~\ref{CT_Burgers_utheta} presents the distribution of the normalized circumferential velocity component $u_{\theta}^{*}$ on the $(x,y)$ plane (i.e., the invariant plane), with the radial profile (along $y=0$) shown in figure~\ref{Burgers_utheta_r_line}.
The observed spiral streamline pattern indicates a three-dimensional vortex structure with axial stretching, where the mass conservation law requires the outward axial flow to balance the inward radial motion.
We observe that $u_{\theta}$ vanishes linearly with $r$ as $r\rightarrow{0}$, following the asymptotic relation $u_{\theta}\sim{(\Gamma ar)}/{(8\pi\nu)}$ (or $u_{\theta}^{*}\sim{r^*}/{8\pi}$ in normalized form), which is consistent with the behavior of the rigid-body rotation. 
The velocity profile attains its global maximum at the critical radius $r_{1}=2.242{L}_{\rm{ref}}$ (or $r_{1}^{*}=2.242$), determined by the extremum condition $\partial_{r}u_{\theta}(r_{1})=0$.
Beyond this peak, the velocity decays asymptotically as
$u_{\theta}\sim{\Gamma}/{(2\pi r)}$ (or $u_{\theta}^{*}\sim{1}/{(2\pi{r}^*)}$), demonstrating the typical characteristic of a potential vortex flow in the far field.
The critical radius $r_{1}$ naturally defines the vortex core radius $({r}_{c}\equiv{r}_{1})$, marking the transition between distinct flow behaviors.
Figure~\ref{CT_compare_drdu_udr_delta} presents the radial distributions of the key kinematic quantities $\partial_{r}u_{\theta}$ and ${u_{\theta}}/{r}$, along with the discriminant $\Delta$ of $\widehat{\bm{A}}$. These quantities govern the fundamental properties of the IVD variables in~\eqref{97abc}. As $r\rightarrow{0}$, both $\partial_{r}u_{\theta}$ and ${u_{\theta}}/{r}$ converge to ${\Gamma{a}}/{(8\pi\nu)}$, demonstrating rigid-body rotation characteristics in the vortex core. The quantity ${u_{\theta}}/{r}$ maintains positive-definite behavior throughout the domain, decaying to zero as $r\rightarrow{+\infty}$. Both $\partial_{r}u_{\theta}$ and $\Delta$ exhibit sign reversal at the critical radius $r_c$: positive for  $r < r_c$ (vortex core region) and negative for $r > r_c$ (outer region). These results demonstrate that the rigid-rotation characteristics are physically meaningful within the vortex core ($r < r_c$), where $\Delta$ remains positive.

Figure~\ref{HHH2} presents the contour plots of the IVD variables $(2\psi^{-},\gamma^{-},\omega_{3})$, with their radial profiles shown in figure~\ref{CT_Burgers_compare_IVD}. 
In figure~\ref{CT_Burgers_Psi_square}, the characteristic rigid rotation mode $(R_{N}^{-}=2\psi^{-})$ is determined by the angular velocity $\psi^{-}={u_{\theta}}/{r}$, which remains strictly positive while vanishing at infinity.
In contrast, the characteristic spin mode $(s_{N}^{-}=\gamma^{-})$ in figure~\ref{CT_Burgers_gamma} exhibits exclusively negative values with two features: a null point at the vortex axis $r_{\gamma}^{(1)}=0$, and a pronounced minimum at $r_{\gamma}^{(2)}$ positioned within the characteristic range $r_{c}=r_{1}<r_{\gamma}^{(2)}<r_{2}$, where
$r_{2}$ is the inflection point of $u_{\theta}$ that satisfies $\partial_{r}^{2}u_{\theta}(r_2)=0$.
This minimum corresponds to the twin negative peaks visible in figure~\ref{CT_Burgers_compare_IVD}, revealing a distinctive dual-cored spin structure of the vortex.
The existence of the characteristic radius $r_{\gamma}^{(2)}$ can be qualitatively analyzed as follows.
At the core boundary $(r=r_{1})$, while $\partial_{r}u_{\theta}(r_{1})=0$, figure~\ref{CT_compare_drdu_udr_delta} reveals the negative slopes for both $\partial_{r}u_{\theta}$ and ${u_{\theta}}/{r}$, that is, $\partial_{r}^{2}u_{\theta}(r_1)<\partial_{r}({u_{\theta}}/{r})(r_{1})$, leading to $\partial_{r}\gamma(r_{1})<0$. 
At the inflection point $r=r_{2}$, the slope reversal manifests as $\partial_{r}\gamma(r_{2})=-\partial_{r}({u_{\theta}}/{r})(r_{2})>0$. By the intermediate value theorem, there must exist a critical point $r_{\gamma}^{(2)}\in(r_{1},r_{2})$ of $\partial_{r}\gamma$ where $\gamma$ attains a local extreme (minimum).
The superposition of $2\psi^{-}$ and $\gamma^{-}$ produces a significantly more localized vorticity distribution $\omega_z (=\omega_3)$ than $2\psi^{-}$, as evidenced in both figures~\ref{CT_Burgers_omegaz} and~\ref{CT_Burgers_compare_IVD}.

Substituting~\eqref{Burgers_VGT} into~\eqref{C8a3},~\eqref{C8a10}, and~\eqref{C8a12} yields the DVD vorticity modes:
\begin{eqnarray*}\label{99a}
	q^2\bm{R}_{L}(\bm{t})&=&a^2rz\left(\omega_{z}+2\psi^{-}\right)\bm{e}_{r}+arz\left(\frac{3}{2}a^2-2(\psi^-)^2\right)\bm{e}_{\theta}\nonumber\\
	& &+r^2\left(\frac{1}{2}a^2(\psi^{-}+\gamma^{-})+2(\psi^{-})^3\right)\bm{e}_{z},
\end{eqnarray*}
\begin{eqnarray*}\label{99b}
	q^2\bm{s}_{L}(\bm{t})&=&-a^{2}rz\left(\frac{1}{2}\omega_{z}+2\psi^{-}\right)\bm{e}_{r}
	-arz\left(\frac{3}{2}a^2+\psi^{-}\gamma^{-}\right)\bm{e}_{\theta}\nonumber\\
	& &+r^{2}\left((\psi^{-})^2-\frac{1}{4}a^2\right)\gamma^{-}\bm{e}_{z},
\end{eqnarray*}
\begin{eqnarray*}\label{99c}
	q^2\bm{g}_{L}(\bm{t})=az\omega_{z}\left(-\frac{1}{2}ar\bm{e}_{r}+r\psi^{-}\bm{e}_{\theta}+az\bm{e}_{z}\right).
\end{eqnarray*}
Note that for any fixed axial position $z$, the Galilean invariance allows the reduction of analysis to $z=0$ through a co-moving frame transformation $(\bm{u}\rightarrow\bm{u}-u_{z}\bm{e}_{z})$, preserving all essential vortex kinematics while simplifying the velocity field to the invariant plane without loss of generality. At $z=0$, noting the two facts that $q^{2}=u_{\theta}^{2}+u_{r}^{2}$ and $\bm{g}_{L}(\bm{t})=\bm{0}$,  the above expressions are simplified as
\begin{eqnarray}\label{zequal0_Rs}
	{R}_{L}(\bm{t})=2\psi^{-}+2\frac{u_{r}^{2}}{q^{2}}\gamma^{-},~~
	{s}_{L}(\bm{t})=\frac{u_{\theta}^{2}-u_{r}^{2}}{q^2}\gamma^{-}.
\end{eqnarray}
Let $\Theta$ denote the rotation angle between $\bm{e}_{1}=\bm{e}_{r}$ (the radial unit basis vector) and $\bm{t}$ (the unit tangent vector of a streamline). The velocity components in the invariant plane admit the polar representation: $(u_{r},u_{\theta})=(q\cos\Theta,q\sin\Theta)$. This parametrization allows us to reformulate~\eqref{zequal0_Rs} equivalently as:
\begin{eqnarray*}
	{R}_{L}(\bm{t})=2\psi^{-}+2\gamma^{-}\cos^2{\Theta}=\omega_{3}+\gamma^{-}\cos2\Theta,
\end{eqnarray*}
\begin{eqnarray}
	{s}_{L}(\bm{t})=(\sin^{2}\Theta-\cos^{2}\Theta)\gamma^{-}=-\gamma^{-}\cos2\Theta,
\end{eqnarray}
which are consistent with the general theory in~\eqref{a30_6} and~\eqref{eq82}. 

Figure~\ref{HHH3} presents the axial components of the DVD variables $({R}_{L}(\bm{t}),{s}_{L}(\bm{t}),{\omega}_{z})$, with the radial distributions shown in figure~\ref{CT_Burgers_compare_DVD}.
Figures~\ref{CT_Burgers_Rtz_z0} and~\ref{CT_Burgers_stz_z0} reveal that the DVD vorticity modes exhibit qualitatively similar patterns to their IVD counterparts (figures~\ref{CT_Burgers_Psi_square} and~\ref{CT_Burgers_gamma}): ${R}_L(\bm{t})$ maintains the monopole structure of $2\psi^{-}$, and ${s}_L(\bm{t})$ preserves the dipole character of $\gamma^{-}$ (figures~\ref{CT_Burgers_compare_IVD} and~\ref{CT_Burgers_compare_DVD}).
Figure~\ref{CT_compare_2Psi_Rtz} presents a comparative analysis of the radial distributions for $(2\psi^{-},{R}_{L}(\bm{t}))$.
The results demonstrate excellent agreement within the vortex core $(r<{r_c})$, with noticeable discrepancy in the far field $(r^{*}>5)$, particularly in the tail regions. 
Figure~\ref{CT_compare_gamma_stz} examines the spin components $(\gamma^{-},s_L(\bm{t}))$, which also show strongest agreement in the vortex core region. The IVD spin $\gamma^{-}$ maintains negative values throughout the domain, while the DVD component $s_L(\bm{t})$ exhibits sign reversal, that is, negative values in the central core region and positive values in the far-field tail region. 
Note that the streaming vorticity $\bm{g}_{L}(\bm{t})$ admits a fundamental kinematic decomposition
\begin{eqnarray*}
	\bm{g}_{L}(\bm{t})=\bm{g}_{L}^{R}(\bm{t})+\bm{g}_{L}^{s}(\bm{t}),
\end{eqnarray*}
\begin{eqnarray*}
	\bm{g}_{L}^{R}(\bm{t})\equiv\bm{R}_{N}-\bm{R}_{L}(\bm{t}),~~\bm{g}_{L}^{s}(\bm{t})\equiv\bm{s}_{N}-\bm{s}_{L}(\bm{t}),
\end{eqnarray*}
where $(\bm{g}_{L}^{R}(\bm{t}),\bm{g}_{L}^{s}(\bm{t}))$ provide rigorous measures of the differences between IVD and DVD modes, characterizing the frame-dependence of vorticity decomposition. For the Burgers vortex in the invariant plane ($z=0$), these components are explicitly given by:
\begin{eqnarray*}	
	{g}_{L}^{R}(\bm{t})=-2\gamma^{-}\cos^{2}\Theta=-\gamma^{-}(1+\cos2\Theta),
\end{eqnarray*}
\begin{eqnarray}\label{eq103}
	{g}_{L}^{s}(\bm{t})=2\gamma^{-}\cos^{2}\Theta=\gamma^{-}(1+\cos2\Theta).
\end{eqnarray}
Obviously, the sum of ${g}_{L}^{R}(\bm{t})$ and ${g}_{L}^{s}(\bm{t})$ is equal to zero. Equation~\eqref{eq103} reveals that the observed deviations in figures~\ref{CT_compare_2Psi_Rtz} and~\ref{CT_compare_gamma_stz}, when normalized by the spin $\gamma^{-}$, are solely determined by the directional cosine $\cos2\Theta$.
Both IVD and DVD effectively characterize the vorticity modes of the Burgers vortex. The IVD provides a robust representation of fundamental vortex characteristics from an algebraic perspective. In contrast, the DVD offers enhanced flow details through streamline geometry analysis, redistributing the vorticity constituents while maintaining the total circulation.
This comparative study highlights the complementary nature of these two decomposition approaches for physical insights in vortex motions.

In addition, we have shown in~\S\ref{sign_of_gamma} that the physically admissible DVD vorticity modes $({R}_{L}(\bm{t}),{s}_{L}(\bm{t}))$ are bounded by the two IVD modes $(R_{N}^{\pm},s_{N}^{\pm})=(2\psi^{\pm},\gamma^{\pm})$ for the projected 2D velocity field in the invariant plane.
Using~\eqref{plus_expression1},~\eqref{new0}, and~\eqref{XX1}, the quantities $(\psi^{+},\gamma^{+})$ are evaluated as
\begin{eqnarray*}
	\psi^{+}=\frac{\partial{u}_{\theta}}{\partial r},
\end{eqnarray*}
\begin{eqnarray}\label{XX3}
	\gamma^{+}=\frac{u_{\theta}}{r}-\frac{\partial{u}_{\theta}}{\partial r}.
\end{eqnarray}
From~\eqref{XX3}, the Liutex vector is derived as
\begin{eqnarray}\label{RN+}
	\bm{R}_{N}^{+}\equiv2\psi^{+}\bm{e}_{3}=2\frac{\partial{u}_{\theta}}{\partial r}\bm{e}_{z}.
\end{eqnarray}
The discriminant $\Delta$ remains invariant for $(\psi^{+},\gamma^{+})$, namely, $\Delta=4\psi^{+}(\psi^{+}+\gamma^{+})$.
In contrast to $\bm{R}_{N}^{-}$ in~\eqref{RN-}, which is governed by the angular velocity ${u_{\theta}}/{r}$, $\bm{R}_{N}^{+}$ is determined by the radial velocity gradient $\partial_{r}u_{\theta}$.
This distinction renders $\bm{R}_{N}^{+}$ devoid of a direct kinematic interpretation via orbital motion, as its mathematical form does not intrinsically link to closed or spiral streamline topologies. As demonstrated in figure~\ref{CT_compare_2Psi_Rtz}, $R_{N}^{+}=2\psi^{+}$, exhibits a significantly larger deviation from the DVD spin mode $R_{L}(\bm{t})$ compared to $R_{N}^{-}$.
Nevertheless, both rigid-rotation modes asymptotically converge to the same limit ${\Gamma a}/{(4\pi\nu)}$ as $r\rightarrow{0}$, demonstrating their equivalence in characterizing rigid-body rotation at the vortex core. This positive limiting value further indicates the increase of $u_{\theta}$ along the radial direction.
To the best of our knowledge, while analytical solutions for the Burgers vortex have been known since the pioneering work of~\citet{Burgers1948}, this particular vorticity decomposition analysis -- highlighting the contrasting roles of $\bm{R}_{N}^{+}$ and $\bm{R}_{N}^{-}$ -- represents a novel contribution to vortex dynamics literature.

\subsection{Instability of a vortex sheet to small disturbance}\label{Vortex-sheet instability sec}
The study of dynamical instability in flow systems has become a critical and coherent branch of fluid mechanics.
It is well documented that at the interface between two approximately uniform streams with different velocities, undulations of increasing amplitude tend to develop, typically leading to turbulent flow~\citep{Batchelor1967}.
Consider an idealized vortex sheet $\Sigma$ with infinitesimal thickness $(\delta\rightarrow{0})$, representing a material surface that forms the common boundary between two streams of the same fluid.
The velocity fields $(\bm{u}_{1},\bm{u}_{2})$ on the upper and lower sides of $\Sigma$
are derived from the corresponding velocity potentials $(\Phi_{1},\Phi_{2})$, which satisfy the Laplace equations
$\nabla^{2}\Phi_{i}=0~(i=1,2)$.
These velocity fields are determined by the gradient of the potentials via
$\bm{u}_{i}=\bm{\nabla}{\Phi}_{i}$, with the velocity jump across the sheet defined as
$\left[\!\left[ \bm{u} \right]\!\right]\equiv\bm{u}_{1}-\bm{u}_{2}$. The local strength density of the vortex sheet is then quantified by integrating the vorticity across the thin transition layer at a given point:
\begin{eqnarray}\label{Gamma}
	\bm{\Gamma}\equiv\lim_{\delta\rightarrow{0}}\int_{-\delta/2}^{\delta/2}\bm{\omega}d\zeta
	=\bm{n}_{\Sigma}\times\left[\!\left[ \bm{u} \right]\!\right],~\zeta\in\left[-\frac{\delta}{2},\frac{\delta}{2}\right],
\end{eqnarray}
where the unit normal vector $\bm{n}_{\Sigma}$ is oriented from Region 2 to Region 1. Equation~\eqref{Gamma} shows that $\bm{\Gamma}$  is entirely determined by the velocity jump discontinuity across $\Sigma$.
\begin{figure}[t]
	\centering
	\includegraphics[width=0.7\columnwidth,trim={0cm 0.0cm 0cm 0.0cm},clip]{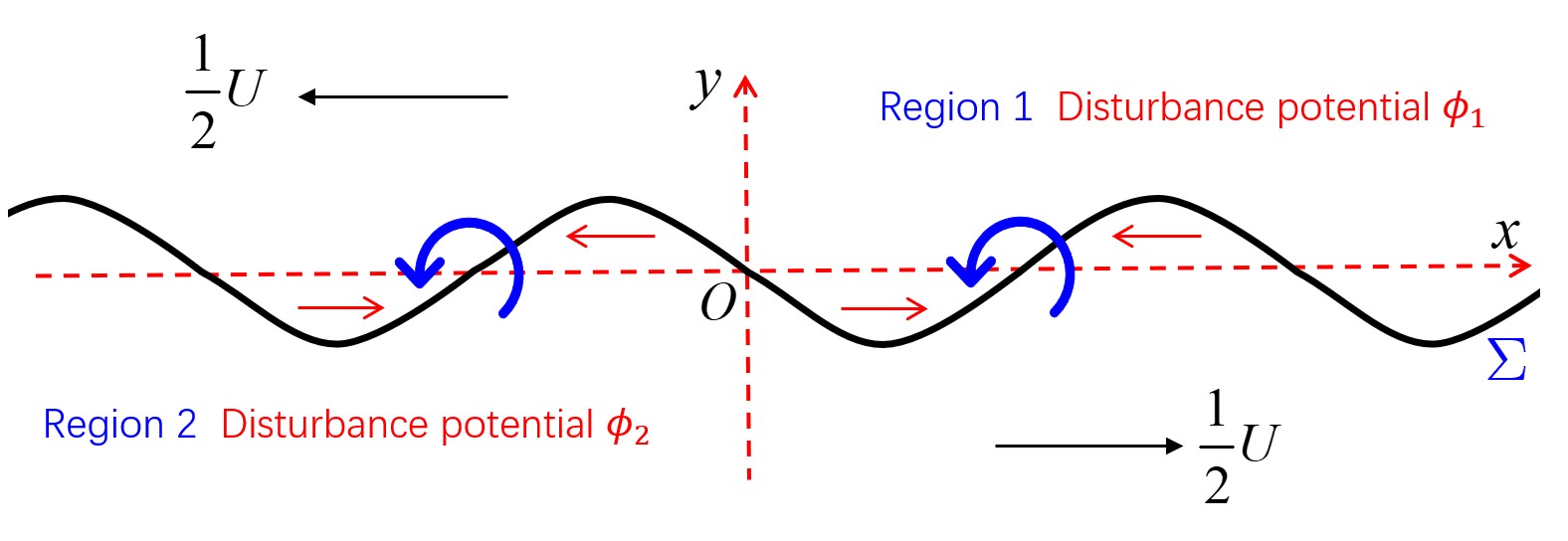}
	\caption{Schematic of instability in a plane vortex sheet under small-amplitude sinusoidal perturbation. The blue arrows denote the self-induced fluid rotation direction at the discrete points satisfying $(\eta=0,\partial_{x}\eta>0)$, and the red arrows represent the process of vorticity accumulation at these points.} 
	\label{Vortex_sheet_instability}
\end{figure}

Within the framework of the linearized normal-mode theory, the classical Kelvin-Helmholtz instability predicts that an initially straight vortex sheet $\Sigma$ with the uniform strength density $\bm{\Gamma}_{0}$ is inherently unstable to infinitesimal perturbations.
Consider a planar vortex sheet initially positioned at $y=0$, separating two irrotational uniform streams $\bm{u}_{1}=-\frac{1}{2}U\bm{e}_{x}$ for the upper region $y>0$ and $\bm{u}_{2}=\frac{1}{2}U\bm{e}_{x}$ for the lower region $y<0$. This configuration yields an initial vortex sheet strength $\bm{\Gamma}_{0}\equiv\lim\limits_{\delta\rightarrow{0}}\int_{-\delta/2}^{\delta/2}\bm{\omega}_{0}d\zeta=U\bm{e}_{z}$ (figure~\ref{Vortex_sheet_instability}).
A generic small disturbance superposed on the base flow can be expressed as a superposition of distinct Fourier modes. 
For a sinusoidal perturbation mode with the wavelength $\lambda$, the wavenumber is given by $k=2\pi/\lambda~(k>0)$.
The disturbance displacement $\eta(x,t)=\mathcal{O}(\epsilon)$ of the deformed vortex sheet must maintain a characteristic amplitude scaling, with the small parameter $\epsilon$ satisfying $\epsilon\ll\lambda$. Under the assumption that the vortex sheet thickness $\delta$ remains constant (neglecting any growth during the relevant time interval), we require $\delta$ to be much smaller than the disturbance amplitude scale $\epsilon$ throughout the evolution $(\delta\ll\epsilon)$ (For convenience, setting $\lambda=\mathcal{O}(1)$ yields $\delta\ll\epsilon\ll{1}$).
The total velocity potentials are decomposed into their base and perturbation components as $\Phi_{1}=-\frac{1}{2}Ux+\phi_{1}$ and  $\Phi_{2}=\frac{1}{2}Ux+\phi_{2}$, where $(\phi_{1},\phi_{2})$ represent the disturbance velocity potentials. $\bm{\Gamma}$ admits the decomposition
\begin{eqnarray}
	\bm{\Gamma}=\bm{\Gamma}_{0}+\bm{\Gamma}^{\prime},
\end{eqnarray}
where $\bm{\Gamma}^{\prime}=\Gamma^{\prime}_{z}\bm{e}_{z}=\bm{n}_{\Sigma}\times\left[\!\left[ \bm{\nabla}\phi \right]\!\right]$ denotes the perturbation-induced local strength density. It is noted that $\bm{\Gamma}_{0}=\mathcal{O}(1)$ and $\bm{\Gamma}^{\prime}=\mathcal{O}(\epsilon)$.

\citet[pp. 511-517]{Batchelor1967} derived the solutions for the perturbed vortex sheet system: $\eta$, $\phi_{1,2}$ and  $\Gamma^{\prime}_{z}\approx-\partial_{x}(\phi_{1}-\phi_{2})_{y=0}$ (only their real parts are physically meaningful):
\begin{subequations}
	\begin{eqnarray}
		\eta(x,t)=C\exp(\sigma t)\exp(ikx),
	\end{eqnarray}
	\begin{eqnarray}
		\Gamma_{z}^{\prime}={2i}\sigma\eta={2}\sigma{C}\exp(\sigma t)\exp\left[i\left(kx+\frac{\pi}{2}\right)\right].
	\end{eqnarray}
\end{subequations}
Here, $C>0$ represents the initial disturbance magnitude, and $\sigma=d\log\exp(\sigma{t})/dt=kU/2>0$ denotes the temporal growth rate, characterizing the exponential amplification of perturbations.
The solution reveals that $\Gamma_{z}^{\prime}$ leads the phase of $\eta$ by $\pi/2$, indicating a quarter-period phase difference between vorticity modulation and sheet displacement.
As vividly sketched in~\citet[figure 7.1.3]{Batchelor1967} and figure~\ref{Vortex_sheet_instability}, the phase disparity drives the instability through two coupled processes occurring simultaneously:
vorticity accumulation near the nodal points satisfying $(\eta=0,\partial_{x}\eta>0)$, and induced rotation of adjacent sheet segments. These mutually reinforcing mechanisms constitute the essential dynamics of the Kelvin-Helmholtz instability, explaining both the exponential growth of disturbances and the subsequent formation of discrete vortices.
Building upon the vorticity decomposition~\eqref{C8a12} in the field description, this work provides a novel reinterpretation of the vortex sheet dynamics by distinctly analyzing the physical roles of the orbital rotation and spin vorticity modes. 

\textcolor{red}{$\circ$ Orbital rotation mode $\bm{R}_{L}(\bm{t})$}

In regions with positive displacement ($\eta>0$, corresponding to $\cos(kx)>0$), since the streamline curvature $\kappa=\lvert\partial_{xx}\eta\rvert+\mathcal{O}(\epsilon^3)=\mathcal{O}(\epsilon)$, the velocity magnitude $q=U/2+\mathcal{O}(\epsilon)$, and the binormal vector $\bm{b}=\bm{e}_{z}$ (with negligible $\mathcal{O}(\delta)$ terms replaced by $\mathcal{O}(\epsilon)$ as $\delta\ll\epsilon$), we have
\begin{eqnarray}
	\bm{R}_{L}(\bm{t})&=&2\kappa{q}\bm{b}\nonumber\\
	&=&\lvert\partial_{xx}\eta\rvert{U}\bm{e}_{z}+\mathcal{O}(\epsilon^2)\nonumber\\
	&=&k^{2}C\exp(\sigma t)\cos(kx)U\bm{e}_{z}+\mathcal{O}(\epsilon^2),
\end{eqnarray}
whose $z$-component maintains phase coherence with the displacement $\eta$. 
The orbital rotation attains its maximum amplitude at  positions where $\cos(kx)=1$.
Conversely, in regions with negative displacement ($\eta<0$, corresponding to $\cos(kx)<0$), we obtain
\begin{eqnarray}
	\bm{R}_{L}(\bm{t})&=&-k^{2}C\exp(\sigma t)\cos(kx)U\bm{e}_{z}+\mathcal{O}(\epsilon^2)\nonumber\\
	&=&k^{2}C\exp(\sigma t)\cos(kx\pm\pi)U\bm{e}_{z}+\mathcal{O}(\epsilon^2),
\end{eqnarray}
whose $z$-component exhibits a $\pi$-phase difference relative to $\eta$. The orbital rotation attains its minimum magnitude at the valley points where $\cos(kx)=-1$.
Note that the displacement $\eta$ alternates sign with $x$,
the projection $\bm{R}_{L}(\bm{t})\bm{\cdot}\bm{e}_{z}$ remains non-negative for all $x$ for both upper and lower regions. This implies that  ${R}_{L}(\bm{t})$ maintains phase synchronization with the absolute displacement  $\lvert\eta\rvert=C\exp(\sigma t)\lvert\cos(kx)\rvert$, consistent with physical intuition. The magnitude of orbital rotation $\bm{R}_{L}(\bm{t})$ scales as $\mathcal{O}(\epsilon)$, and its integral across the vortex sheet yields
\begin{eqnarray}\label{FF2}
	\int_{-\delta/2}^{\delta/2}\bm{R}_{L}(\bm{t})d\zeta=\mathcal{O}(\epsilon\delta)\ll\mathcal{O}(\epsilon^2),~\delta\rightarrow{0}.
\end{eqnarray}

\textcolor{red}{$\circ$ Spin mode $\bm{s}_{L}(\bm{t})$}

We approach the problem fundamentally by considering $\bm{s}_{L}(\bm{t})$ in~\eqref{C8a10}:
\begin{eqnarray}\label{Spin_X1}
	\bm{s}_{L}(\bm{t})&=&-2\bm{t}\times\left(\bm{t}\bm{\cdot}\bm{D}\right)\nonumber\\
	&=&-\bm{t}\times\left(\bm{t}\bm{\cdot}\bm{A}\right)-\bm{t}\times\left(\bm{A}\bm{\cdot}\bm{t}\right).
\end{eqnarray}
On the right hand side of~\eqref{Spin_X1}, the first term is similar to the contribution from the orbital rotation, which scales with $\mathcal{O}(\epsilon)$.
For the second term, it can be estimated inside the vortex sheet as
\begin{eqnarray}
	-\bm{t}\times\left(\bm{A}\bm{\cdot}\bm{t}\right)&=&	-\bm{t}\times\left(\bm{\nabla}_{\pi}\bm{u}\bm{\cdot}\bm{t}+\bm{n}_{\Sigma}\partial_{\zeta}\bm{u}\bm{\cdot}\bm{t}\right)\nonumber\\
	&\approx&-\bm{t}\times\left(\bm{n}_{\Sigma}\partial_{\zeta}\bm{u}\bm{\cdot}\bm{t}\right).
\end{eqnarray}

We first analyze the perturbed flow in the region where $\eta>0$; the case for $\eta<0$ follows through analogous arguments. Given the first-order approximations $\bm{t}=-\bm{e}_{x}+\mathcal{O}(\epsilon)$, $\bm{n}_{\Sigma}=\bm{e}_{y}+\mathcal{O}(\epsilon)$, and $\bm{u}=-\frac{1}{2}U\bm{e}_{x}+\mathcal{O}(\epsilon)$, we derive
\begin{eqnarray}\label{mm0}
	-\bm{t}\times\left(\bm{n}_{\Sigma}\partial_{\zeta}\bm{u}\bm{\cdot}\bm{t}\right)&=&-[-\bm{e}_{x}+\mathcal{O}(\epsilon)]\times\left[\bm{n}_{\Sigma}\partial_{\zeta}\bm{u}\bm{\cdot}(-\bm{e}_{x}+\mathcal{O}(\epsilon))\right]\nonumber\\
	&=&	-\bm{e}_{x}\times\left(\bm{e}_{y}\partial_{\zeta}{u}_{x}\right)+\mathcal{O}(\epsilon)\nonumber\\
	&=&-\partial_{\zeta}u_{x}\bm{e}_{z}+\mathcal{O}(\epsilon),
\end{eqnarray}
which scales with $\mathcal{O}(\delta^{-1})\gg\mathcal{O}(1)\gg\mathcal{O}(\epsilon)$ due to the asymptotic ordering $\delta \ll \epsilon \ll 1$. Consequently, it dominates the spin vorticity $\bm{s}_{L}(\bm{t})$ in the vortex sheet dynamics.

Although the variation details of $u_{x}$ inside the sheet are not available, its values at the boundaries $\zeta=\pm\frac{1}{2}\delta$ of the transition layer are already known. Using $\bm{n}_{\Sigma}=\bm{e}_{y}+\mathcal{O}(\epsilon)$ and denoting
\begin{eqnarray*}
	\bm{r}^{\pm}=\bm{r}\pm\frac{\delta}{2}\bm{n}_{\Sigma},
\end{eqnarray*}
$\bm{\Gamma}^{\prime}$ can be estimated as 
\begin{eqnarray}
	\bm{\Gamma}^{\prime}=\bm{n}_{\Sigma}\times\left[\!\left[ \bm{\nabla}\phi \right]\!\right]=\left[(\partial_{x}\phi_{2})(\bm{r}^{+})-(\partial_{x}\phi_{1})(\bm{r}^{-})\right]\bm{e}_{z}+\mathcal{O}(\epsilon^2),
\end{eqnarray}
where the sheet centerline position is parameterized as $\bm{r}(x,t)={x}\bm{e}_{x}+\eta(x,t)\bm{e}_{y}$. By integrating equation~\eqref{mm0} across the vortex sheet thickness, we obtain
\begin{eqnarray}\label{FF1}
	\int_{-\delta/2}^{\delta/2}	-\bm{t}\times\left(\bm{A}\bm{\cdot}\bm{t}\right)d\zeta
	&=&-\left[\!\left[u_{x} \right]\!\right]\bm{e}_{z}+\mathcal{O}(\epsilon\delta)\nonumber\\
	&=&\left[\frac{U}{2}+\partial_{x}\phi_{2}(\bm{r}^{+})\right]\bm{e}_{z}-\left[-\frac{U}{2}+\partial_{x}\phi_{1}(\bm{r}^{-})\right]\bm{e}_{z}+\mathcal{O}(\epsilon^2)\nonumber\\
	&=&U\bm{e}_{z}+\left[(\partial_{x}\phi_{2})(\bm{r}^{+})-(\partial_{x}\phi_{1})(\bm{r}^{-})\right]\bm{e}_{z}+\mathcal{O}(\epsilon^2)\nonumber\\
	&=&\bm{\Gamma}_{0}+\bm{\Gamma}^{\prime}+\mathcal{O}(\epsilon^2)\nonumber\\
	&=&\bm{\Gamma}+\mathcal{O}(\epsilon^2).
\end{eqnarray}

From~\eqref{FF2} and~\eqref{FF1}, we find
\begin{eqnarray}\label{integral_sL}
	\int_{-\delta/2}^{\delta/2}\bm{s}_{L}(\bm{t})d\zeta=\bm{\Gamma}+\mathcal{O}(\epsilon^2),~~\delta\rightarrow{0}.
\end{eqnarray}
For the unperturbed initial state, there is no orbital rotation, which means that the initial vorticity $\bm{\omega}_{0}$ arises solely from the spin (i.e., $\bm{\omega}_{0}=\bm{s}_{L}(\bm{t})(t=0)$), determined by the velocity distribution within the sheet.
By defining the disturbance spin as $\bm{s}_{L}^{\prime}(\bm{t})\equiv\bm{s}_{L}(\bm{t})-\bm{\omega}_{0}$,we obtain an equivalent representation of~\eqref{integral_sL}:
\begin{eqnarray}\label{integral_sLp}	\int_{-\delta/2}^{\delta/2}\bm{s}_{L}^{\prime}(\bm{t})d\zeta=\bm{\Gamma}^{\prime}+\mathcal{O}(\epsilon^2),~~\delta\rightarrow{0}.
\end{eqnarray}
Equation~\eqref{integral_sL} demonstrates that $\bm{\Gamma}$ can be approximated as the integrated spin $\bm{s}_{L}(\bm{t})$ over the vortex sheet under small perturbations. Furthermore, $\bm{s}_{L}^{\prime}(\bm{t})$ is the dominant contributor to $\bm{\Gamma}^{\prime}$, as evidence by~\eqref{integral_sLp}.
Consequently, the $\frac{1}{2}\pi$ phase difference between $\bm{\Gamma}^{\prime}$ and $\eta$ is primarily governed by the disturbance spin mode, with minor modulation from the orbital rotation mode. This analysis offers a novel vorticity-dynamical interpretation of vortex sheet instability, complementing the classical framework established by~\citet{Batchelor1967}.

\subsection{Moffatt-Kida-Ohkitani asymptotic vortex solution}\label{Moffatt-Kida-Ohkitani vortex solution}
\subsubsection{Background: Stretched vortices in triaxial strain field}
\citet{Moffatt1994} (MKO94) proposed that stretched vortices constitute the sinews of turbulence. To substantiate this claim, they developed a theory for a high-Reynolds-number asymptotic vortex solution in a generic uniform non-axisymmetric irrotational 
strain field.
Consider a background straining flow of the form
\begin{eqnarray*}
	\bm{U}=(\bar{\alpha} x,\bar{\beta} y,\bar{\gamma}z)=\bm{\nabla}\Phi,
\end{eqnarray*}
\begin{eqnarray}\label{ss89}
	\bar{\alpha}+\bar{\beta}+\bar{\gamma}=0,~~~\bar{\alpha}<0<\bar{\gamma},~~\bar{\beta}\geq\bar{\alpha},
\end{eqnarray}
where $\bar{\gamma}$ is the positive strain rate along the principal stretching axis $\bm{e}_{z}$, and the velocity potential is $\Phi=\frac{1}{2}(\bar{\alpha}x^2+\bar{\beta}y^2+\bar{\gamma}z^2)$. Note that $(\bar{\alpha},\bar{\beta},\bar{\gamma})$ in~\eqref{ss89} are distinct from the spin components $(\alpha,\beta,\gamma)$ in~\eqref{IVD2}. Different strain configurations can be characterized by the single dimensionless parameter:
\begin{eqnarray*}\label{ss90}
	\lambda=\frac{\bar{\alpha}-\bar{\beta}}{\bar{\alpha}+\bar{\beta}}=\frac{\bar{\beta}-\bar{\alpha}}{\bar{\gamma}}\geq{0}
\end{eqnarray*}
such that
\begin{eqnarray}\label{ss91}
	\bar{\alpha}=-\frac{1}{2}\bar{\gamma}(1+\lambda),~~\bar{\beta}=-\frac{1}{2}\bar{\gamma}(1-\lambda).
\end{eqnarray}
When $\alpha\leq\beta < 0$ ($0 \leq \lambda < 1$), the system exhibits a single positive principal strain rate ($\bar{\gamma}$), promoting the formation of vortex tubes aligned with the $z$-axis. This configuration is known as \textit{axial strain}. The special case where $\alpha = \beta < 0$ ($\lambda = 0$) corresponds to \textit{axisymmetric axial strain}. $\lambda=1$ and $\lambda>0$ correspond to the plane strain and biaxial strain, respectively.

\begin{figure}[h!]
	\centering
	\subfigure[$\omega(r,\theta)$]{
		\begin{minipage}[t]{0.5\linewidth}
			\centering
			\includegraphics[width=1.0\columnwidth,trim={0cm 0.0cm 0.0cm 0.0cm},clip]{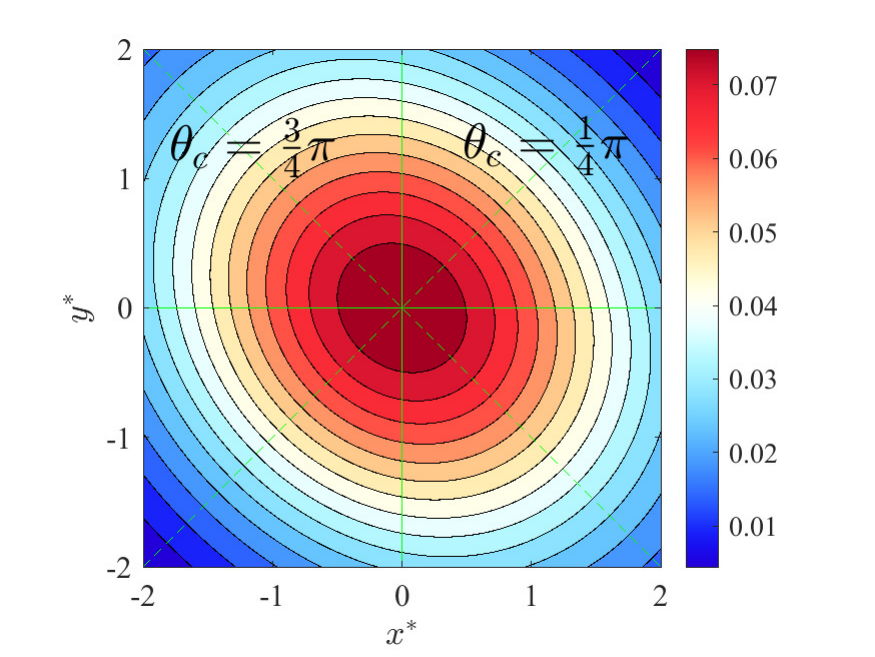}
			\label{omega_r_theta}
		\end{minipage}%
	}%
	\subfigure[$\sin(2\phi)\sin\theta$]{
		\begin{minipage}[t]{0.5\linewidth}
			\centering
			\includegraphics[width=1.0\columnwidth,trim={0cm 0.0cm 0.0cm 0.0cm},clip]{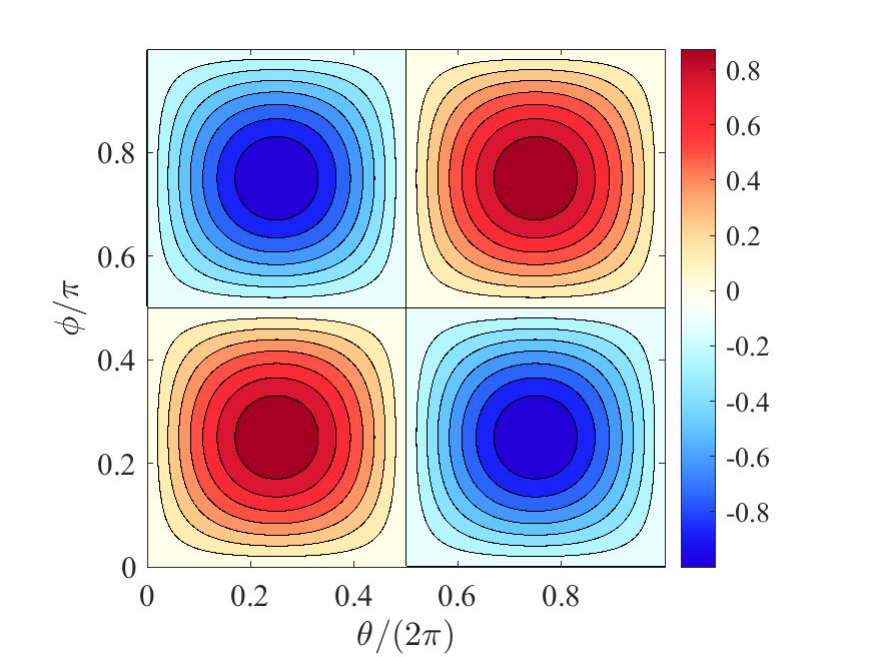}
			\label{Mode1_sin2phi_sintheta}
		\end{minipage}%
	}%
	
	\subfigure[$\sin(2\phi)\cos\theta$]{
		\begin{minipage}[t]{0.5\linewidth}
			\centering
			\includegraphics[width=1.0\columnwidth,trim={0cm 0.0cm 0.0cm 0.0cm},clip]{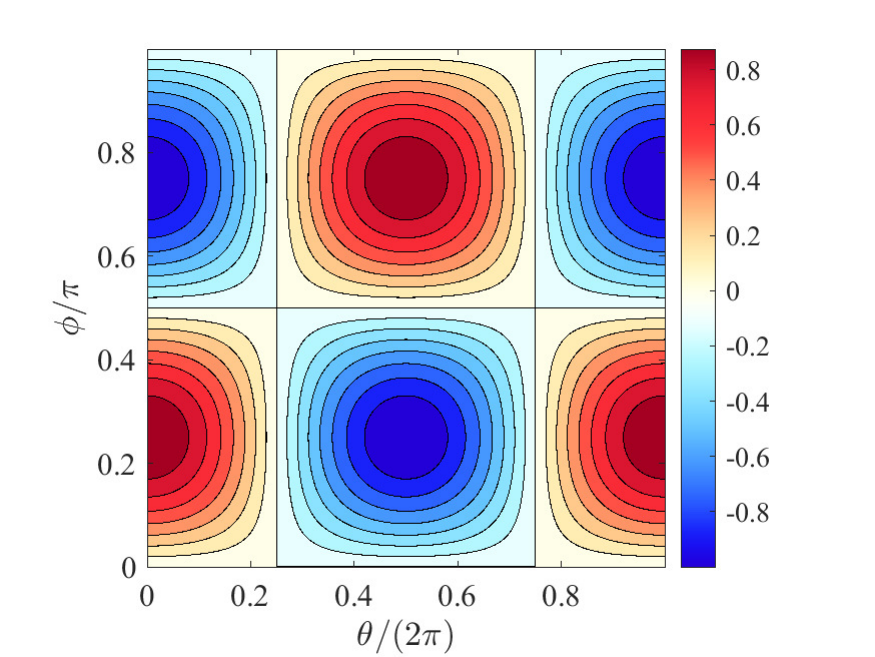}
			\label{Mode1_sin2phi_costheta}
		\end{minipage}%
	}%
	\subfigure[$(1-\cos(2\phi))\sin(2\theta)$]{
		\begin{minipage}[t]{0.5\linewidth}
			\centering
			\includegraphics[width=1.0\columnwidth,trim={0cm 0.0cm 0.0cm 0.0cm},clip]{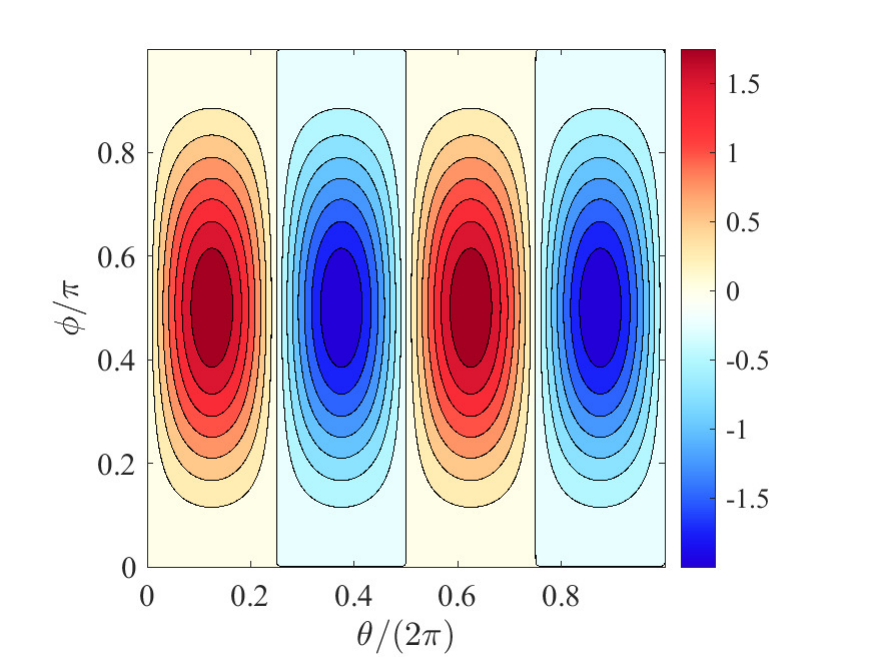}
			\label{Mode3}
		\end{minipage}%
	}%
	\caption{(a) Contour plot of the normalized vorticity field $\omega(r,\theta)$ for the MKO solution~\eqref{ss93} with $\lambda=0.725$, where the green dashed lines mark the critical polar angles $\theta_c = \pi/4, 3\pi/4$. Typical modes of the specific angular velocity $\bm{W}_{D}(\bm{e})$ in $(\theta,\phi)$ space (see~\eqref{eq135}): (b) $\sin(2\phi)\sin\theta$, (c) $\sin(2\phi)\cos\theta$, and (d) $(1-\cos(2\phi))\sin(2\theta)$.} 
	\label{flag12}
\end{figure}

For an incompressible fluid, the stream function $f(x,y)$ is introduced to determine the velocity
through $u_{x}=\partial_{y}f$ and $u_{y}=-\partial_{x}f$, with the vorticity given by $\omega=-\nabla^{2}f$.
Assuming the total circulation $\Gamma=\iint\omega(x,y)dxdy$ associated with the vortex is finite, and the Reynolds number
$R_{\Gamma}\equiv\Gamma/\nu\gg{1}$ ($\nu$ is the kinematic viscosity), we have $\epsilon\equiv{1}/R_{\Gamma}\ll{1}$. 
The characteristic radial length scale is $\delta=(\nu/\bar{\gamma})^{1/2}$, and the maximum velocity 
in the vortex region is of the order $\mathcal{O}(\Gamma/\delta)$. After normalizing $(x,y)$,$(\bar{\alpha},\bar{\beta})$, and $f$ using $\delta$,$\bar{\gamma}$, and $\Gamma$, respectively, the dimensionless steady vorticity equation becomes
\begin{eqnarray*}\label{ss92}
	\frac{\partial(f,\omega)}{\partial(x,y)}=\epsilon\left[\left(\bar{\alpha}{x}\frac{\partial}{\partial x}+\bar{\beta}{y}\frac{\partial}{\partial y}\right)f-\omega-\nabla^{2}\omega\right],
\end{eqnarray*}
which permits a series expansion solution $f=f_{0}+\epsilon{f}_{1}+\epsilon^{2}{f}_{2}+\cdots$.
Using the plane polar coordinates $(r,\theta)$ in the $(x,y)$ plane, the vorticity to the first order in $\epsilon$ is given by (figure~\ref{omega_r_theta})
\begin{eqnarray}\label{ss93}
	\omega(r,\theta)=\omega_{0}(r)+\epsilon\lambda\Omega(r)\sin(2\theta),~\omega_{0}(r)=\frac{1}{4\pi}\exp\left(-\frac{1}{4}r^2\right),
\end{eqnarray}
where $\Omega(r)=\left(\frac{1}{4}r^2-f(r)\right)\eta(r)$ is a well-behaved function. For small $r$, $f(r)$ and $\eta(r)$ can be expanded as
\begin{eqnarray*}
	f(r)=Cr^2+\frac{1}{12}\left(\frac{1}{4}-C\right)\left(r^4-\frac{5}{64}r^6+\cdots\right),~~C=-0.381475,
\end{eqnarray*}
\begin{eqnarray*}
	\eta(r)=-1+\frac{1}{8}r^{2}-\frac{1}{192}r^4+\cdots.
\end{eqnarray*}
At the leading order, $\omega_{0}(r)$ corresponds to the Burgers vortex solution (in~\S\ref{Burgers vortex}) normalized by $\Gamma/\delta^2=\bar{\gamma}/\epsilon$, which describes an axisymmetric flow under axial strain with $\bar{\alpha}=\bar{\beta}=-\bar{\gamma}/2=-a/2~(\lambda=0)$.

\subsubsection{Rotation of material line element in a background strain field}
We now analyze the rotation of a material line element ${\delta}\bm{r}=\delta{r}\bm{e}$, retaining only the leading-order term from the MKO94 vortex solution, where the background strain field has been modified by the non-axisymmetry of strain $(\lambda>0)$ from the higher orders. Let $\phi$ denote the angle between $\bm{e}$ and the axial direction $\bm{e}_{z}$, and $\theta$ the angle between the projection of $\bm{e}$ onto the $(x,y)$ plane and $\bm{e}_{x}$, such that $\bm{e}=\sin\phi(\cos\theta\bm{e}_{x}+\sin\theta\bm{e}_{y})+\cos\phi\bm{e}_{z}=\sin\phi\bm{e}_{r}+\cos\phi\bm{e}_{z}$.
From~\eqref{ss89} and~\eqref{ss91}, the VGT $\bm{A}_{\Phi}$ and strain-rate tensor $\bm{D}_{\Phi}$ caused by the background potential flow are given by
\begin{eqnarray}\label{ss94}
	\bm{A}_{\Phi}=\bm{D}_{\Phi}=\frac{1}{2}\bar{\gamma}\left[-(1+\lambda)\bm{e}_{x}\bm{e}_{x}-(1-\lambda)\bm{e}_{y}\bm{e}_{y}+2\bm{e}_{z}\bm{e}_{z}\right].
\end{eqnarray}
It follows that the specific angular velocity induced by $\bm{D}_{\Phi}$ is
\begin{eqnarray}\label{ss95}
	\bm{W}_{D\Phi}(\bm{e})&=&\bm{e}\times(\bm{e}\bm{\cdot}\bm{D}_{\Phi})\nonumber\\
	&=&\frac{1}{4}\bar{\gamma}\sin(2\phi)\left[(3-\lambda)\sin\theta\bm{e}_{x}-(3+\lambda)\cos\theta\bm{e}_{y}\right]\nonumber\\
	& &+\frac{1}{4}\bar{\gamma}\lambda(1-\cos(2\phi))\sin(2\theta)\bm{e}_{z}.
\end{eqnarray}

Noting the identity $\bm{e}_{r}\bm{e}_{r}+\bm{e}_{\theta}\bm{e}_{\theta}=\bm{e}_{x}\bm{e}_{x}+\bm{e}_{y}\bm{e}_{y}=\bm{I}-\bm{e}_{z}\bm{e}_{z}$, and subtracting the diagonal matrix elements in~\eqref{Burgers_VGT} induced by the background straining flow, the disturbance VGT $\bm{A}_{\omega_{0}}$ and strain-rate tensor $\bm{D}_{\omega_{0}}$ arising from the leading-order solution (i.e., the Burgers vortex), are given as
\begin{eqnarray*}
	\bm{A}_{\omega_{0}}
	=\begin{bmatrix}
		0 &\psi^{-}+\gamma^{-}& 0\\
		-\psi^{-} &0&0\\
		0&0& 0
	\end{bmatrix}=(\psi^{-}+\gamma^{-})\bm{e}_{r}\bm{e}_{\theta}-\psi^{-}\bm{e}_{\theta}\bm{e}_{r},
\end{eqnarray*}
\begin{eqnarray}\label{ss96}
	\bm{D}_{\omega_{0}}
	=\begin{bmatrix}
		0 &\frac{1}{2}\gamma^{-}& 0\\
		\frac{1}{2}\gamma^{-} &0&0\\
		0&0& 0
	\end{bmatrix}=\frac{1}{2}\gamma^{-}(\bm{e}_{r}\bm{e}_{\theta}+\bm{e}_{\theta}\bm{e}_{r}).
\end{eqnarray}
Using~\eqref{ss96}, the dot product of $\bm{e}$ and $\bm{D}_{\omega_{0}}$ is evaluated as
\begin{eqnarray*}
	\bm{e}\bm{\cdot}\bm{D}_{\omega_{0}}=\frac{1}{2}\gamma^{-}\sin\phi\bm{e}_{\theta}.
\end{eqnarray*}
Consequently, the specific angular velocity induced by $\bm{D}_{\omega_{0}}$ is
\begin{eqnarray}\label{ss97}
	\bm{W}_{D\omega_{0}}(\bm{e})&=&\bm{e}\times(\bm{e}\bm{\cdot}\bm{D}_{\omega_{0}})\nonumber\\
	&=&-\frac{1}{4}\gamma^{-}\sin(2\phi)\bm{e}_{r}+\frac{1}{4}\gamma^{-}(1-\cos(2\phi))\bm{e}_{z}\nonumber\\
	&=&-\frac{1}{4}\gamma^{-}\sin(2\phi)\cos\theta\bm{e}_{x}
	-\frac{1}{4}\gamma^{-}\sin(2\phi)\sin\theta\bm{e}_{y}\nonumber\\
	& &+\frac{1}{4}\gamma^{-}(1-\cos(2\phi))\bm{e}_{z}.
\end{eqnarray}

The superposition of~\eqref{ss95} and~\eqref{ss97} yields the specific angular velocity at the leading order:
\begin{eqnarray*}
	\bm{W}_{D}(\bm{e})=\bm{e}\times(\bm{e}\bm{\cdot}\bm{D})\simeq\bm{W}_{D\perp}(\bm{e})+\bm{W}_{Dz}(\bm{e}),
\end{eqnarray*}
\begin{eqnarray*}
	\bm{W}_{D\perp}(\bm{e})&=&\frac{1}{4}\bar{\gamma}\sin(2\phi)\left[(3-\lambda)\sin\theta-\gamma^{*}\cos\theta\right]\bm{e}_{x}\nonumber\\
	& &-\frac{1}{4}\bar{\gamma}\sin(2\phi)\left[(3+\lambda)\cos\theta+\gamma^{*}\sin\theta\right]\bm{e}_{y},
\end{eqnarray*}
\begin{eqnarray}\label{eq135}
	\bm{W}_{Dz}(\bm{e})=\frac{1}{4}\bar{\gamma}(1-\cos(2\phi))(\gamma^{*}+\lambda\sin(2\theta))\bm{e}_{z},
\end{eqnarray}
where $\gamma^{*}\equiv\gamma^{-}/\bar{\gamma}$.
Therefore, the total angular velocity $\bm{W}_{L}(\bm{e})$ is obtained as 
\begin{eqnarray*}
	\bm{W}_{L}(\bm{e})=\bm{W}_{D\perp}(\bm{e})+\bm{W}_{Lz}(\bm{e}),~W_{Lz}=W_{Dz}+\frac{1}{2}\omega_{z}, 
\end{eqnarray*}
\begin{eqnarray}\label{eq136}
	W_{Lz}\simeq\frac{1}{2}\omega_{0}(r)+\frac{1}{4}\gamma^{-}(1-\cos(2\phi))+\frac{1}{4}\bar{\gamma}\lambda(1-\cos(2\phi))\sin(2\theta).
\end{eqnarray}

Both $\bm{W}_{D}(\bm{e})$ and $\bm{W}_{L}(\bm{e})$ demonstrate nontrivial dependence on $(\phi,\theta,\lambda)$. Typical modes of $\bm{W}_{D}(\bm{e})$ in the $(\theta,\phi)$ space exhibit periodic roll patterns, as displayed in figures~\ref{Mode1_sin2phi_sintheta}--\ref{Mode3}.
We analyze two special orientations of the unit vector $\bm{e}$.
When $\bm{e}=\bm{e}_{z}$ (i.e., $\phi=0$, axial alignment), $\bm{W}_{D}(\bm{e})=\bm{0}$ indicates the disappearance of specific rotation of $\delta\bm{r}$.
When $\bm{e}\perp\bm{e}_{z}$ (i.e., $\phi=\pi/2$, planar alignment), $\delta\bm{r}$ is confined to the $(x,y)$ plane so that
$\bm{W}_{D\perp}(\bm{e})=\bm{0}$, and $\bm{W}_{D}(\bm{e})=\bm{W}_{Dz}(\bm{e})=\frac{1}{2}(\gamma^{-}+\bar{\gamma}\lambda\sin(2\theta))\bm{e}_{z}$, revealing explicit dependence on $(\theta,\lambda)$.
For a given $\lambda$, $W_{Dz}$ attains the maximum at $\theta=\pi/4$ (i.e., $y=x$) and the minimum at $\theta=3\pi/4$ (i.e., $y=-x$). Interestingly, the two special angles just coincide with the principal axes of the iso-vorticity lines ($\omega=\text{constant}$), as evidenced by~\eqref{ss93} and figure~\ref{omega_r_theta}. In addition, the rate of viscous dissipation $2\nu{D}_{ij}D_{ij}$, despite evolving in a rather complex manner, always exhibits two maxima near $\theta=3\pi/4$ (i.e., $y=-x$) in contour maps across varying $R_{\Gamma}$ and symmetrically displaced from the vortex center~\citep{KO1992,Moffatt1994}.

\subsubsection{Disturbed vortex-filament rotation}
An axial vortex in fluid dynamics cannot be simply approximated as a flexible rubber bar with limited degrees of freedom. Instead, this vortical structure comprises a bundle of innumerable thin vortex filaments, analogous to a brush used for traditional Chinese painting or calligraphy, made up of a large amount of wool fibers bundled in parallel alignment. Under constrained conditions, each vortex filament exhibits independent degrees of freedom and may tilt under external disturbance. Once a rectilinear vortex filament deviates from the $\bm{e}_{z}$-direction due to some perturbation, it undergoes complex rotational motions governed by $\bm{W}_{L}(\bm{e})$, particularly the transverse component $\bm{W}_{D\perp}(\bm{e})$.

Assume initially a rectilinear vortex filament $\delta\bm{r}=\delta{z}\bm{e}_{z}$ positioned the original point, where the feedback of its circulation $\delta\Gamma$ on the flow is negligible.
When a small perturbation tilt $\delta\bm{r}$ within the meridian plane $(\theta=0,\phi=\mathcal{O}(\epsilon)\ll{1})$, it will be driven by $W_{Lz}$ (dominated by the term $\frac{1}{2}\omega_{0}(r)$) to rotate around $\bm{e}_{z}$. Over a very short time interval, this rotation generates a small but finite $\theta=\mathcal{O}(\epsilon)$. Consequently, using the fact that $\gamma^{-}(0)=\lim\limits_{r\rightarrow{0}}\gamma^{-}(r)=0$ (see figure~\ref{CT_Burgers_compare_IVD}), we simplify~\eqref{eq135} to obtain
\begin{eqnarray}
	\bm{W}_{D\perp}(\bm{e})=-\frac{1}{2}\bar{\gamma}\phi(3+\lambda)\bm{e}_{y}+\mathcal{O}(\epsilon^2),~~\bm{W}_{Dz}(\bm{e})=\mathcal{O}(\epsilon^3).
\end{eqnarray}
Notably, $\bm{W}_{Dz}(\bm{e})=\mathcal{O}(\epsilon^3)$ is two orders of magnitude smaller than $\bm{W}_{D\perp}(\bm{e})=\mathcal{O}(\epsilon)$ which depends on $(\phi,\lambda)$. Hence, the latter dominates $\bm{W}_{D}(\bm{e})$. From~\eqref{eq136}, $W_{Lz}$ reduces to
\begin{eqnarray}
	W_{Lz}=\frac{1}{2}{\omega}_{0}(r)+\mathcal{O}(\epsilon^2).
\end{eqnarray}
Therefore, over an ensuing short time interval, $\bm{W}_{D\perp}(\bm{e})$ derives $\delta\bm{r}$ to rotate about $\bm{e}_{y}$, while $W_{Lz}$ simultaneously governs its rotation about $\bm{e}_{z}$.
If a small perturbation tilts $\delta\bm{r}$ within the meridian plane $(\theta=\pi/4,\phi=\mathcal{O}(\epsilon)\ll{1})$, then~\eqref{eq135} gives
\begin{eqnarray*}
	\bm{W}_{D\perp}(\bm{e})=\frac{\sqrt{2}}{4}\bar{\gamma}\phi\left[(3-\lambda)\bm{e}_{x}+(3+\lambda)\bm{e}_{y}\right]+\mathcal{O}(\epsilon^2),
\end{eqnarray*}
\begin{eqnarray}
	\bm{W}_{Dz}(\bm{e})=\frac{1}{2}\bar{\gamma}\phi^2\lambda\bm{e}_{z}+\mathcal{O}(\epsilon^{3}).
\end{eqnarray}
\begin{figure}[t]
	\centering
	\includegraphics[width=0.6\columnwidth,trim={0cm 0.0cm 0cm 0.0cm},clip]{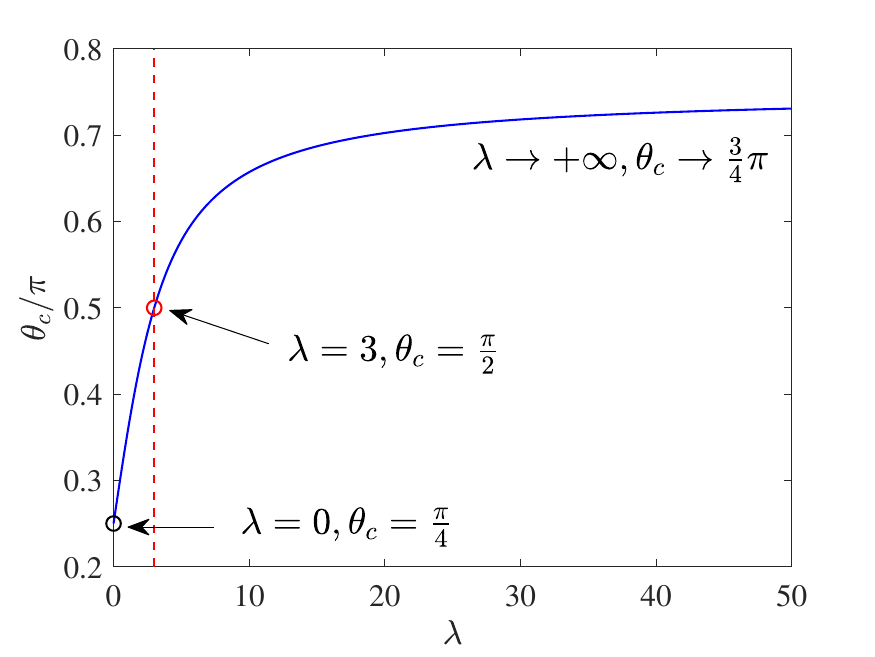}
	\caption{The polar angle $\theta_{c}$ of the composite axis in the $(x,y)$ plane as a strictly increasing function of the parameter $\lambda$.} 
	\label{thetac_lambda}
\end{figure}
Typically, $\bm{W}_{Dz}(\bm{e})=\mathcal{O}(\epsilon^2)$ is one order of magnitude smaller than $\bm{W}_{D\perp}(\bm{e})=\mathcal{O}(\epsilon)$.
Beyond driving rotation about $\bm{e}_{z}$, $\bm{W}_{D\perp}(\bm{e})$ induces a complicated dancing motion (nutation) about a composite axis lying in the $(x,y)$ plane: the $\bm{e}_{y}$-component is always positive whereas the $\bm{e}_{x}$-component remains positive for $\lambda\in[0,3)$ but becomes negative for $\lambda\in(3,+\infty)$. As shown in figure~\ref{thetac_lambda}, the polar angle $\theta_{c}$ between this composite axis (restricted to the upper half-plane) and $\bm{e}_{x}$ exhibits a unique dependence on $\lambda$. This parameter characterizes the departure from axisymmetry in the
strain field, with $\theta_{c}$ determined by
\begin{eqnarray*}
	\tan\theta_{c}=\frac{3+\lambda}{3-\lambda},~\theta_{c}\in\left[\frac{1}{4}\pi,\frac{3}{4}\pi\right),~\lambda\in[0,+\infty).
\end{eqnarray*}
$\lambda=0$ and $3$ correspond to the rotation about $\theta_{c}=\pi/4$ and $\bm{e}_{y} (\theta_{c}=\pi/2)$, respectively. For a high value of $\lambda$, the rotation axis approaches $\theta_{c}=3\pi/4$ asymptotically.

The peculiar dancing patterns of vortex filaments bring additional degrees of freedom to the theoretical analysis. A fundamental issue arises: could these supplementary dynamical modes give rise to novel vortical-flow instability mechanisms -- potentially even chaotic behavior -- in both axial vortices and shear layers? This consideration is particularly significant as conventional stability analyses rely exclusively on volume-element formulations.
It is worth noticing that if to these $\bm{e}$-dependent rotation modes, one adds the coupled effects of filament stretching and self-induced velocity field governed by the Biot-Savart law, a complete theoretical framework for vortex filament evolution could be established. 
For example, the numerical study by~\citet{Siggia1985} on the self-induced evolution of an initially elliptical thin vortex ring clearly revealed these complex $\bm{e}$-dependent rotation patterns by solving the Euler equations, even without external strain fields or additional vortices. These findings raise several important questions worthy of further investigation: whether such intricate filament motions could initiate strongly nonlinear processes like vortex breakdown, and how these  rotation patterns evolve during 
the development of breakdown. These promising directions extend beyond the present study and merit dedicated exploration. 

\subsection{Hexagon and North Polar Vortex on Saturn}\label{Hexagon and North Polar Vortex on Saturn}
\subsubsection{Extraction of velocity field from cloud images}
\begin{figure}[t]
	\centering
	\subfigure[Downsampled image 1]{
		\begin{minipage}[t]{0.5\linewidth}
			\centering
			\includegraphics[width=1.0\columnwidth,trim={0cm 0.0cm 0.0cm 0.0cm},clip]{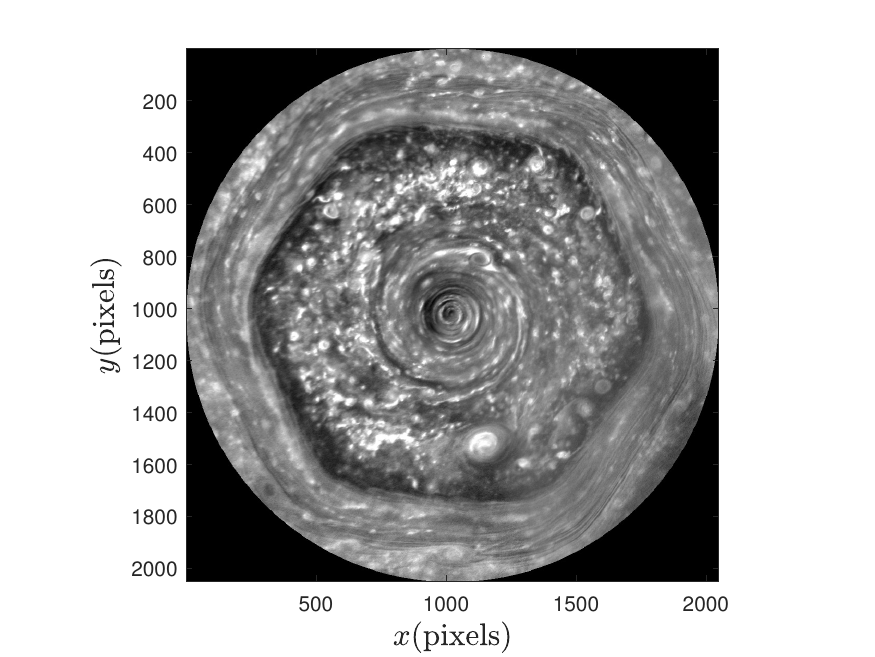}
			\label{SNPVS1}
		\end{minipage}%
	}%
	\subfigure[Downsampled image 2]{
		\begin{minipage}[t]{0.5\linewidth}
			\centering
			\includegraphics[width=1.0\columnwidth,trim={0cm 0.0cm 0.0cm 0.0cm},clip]{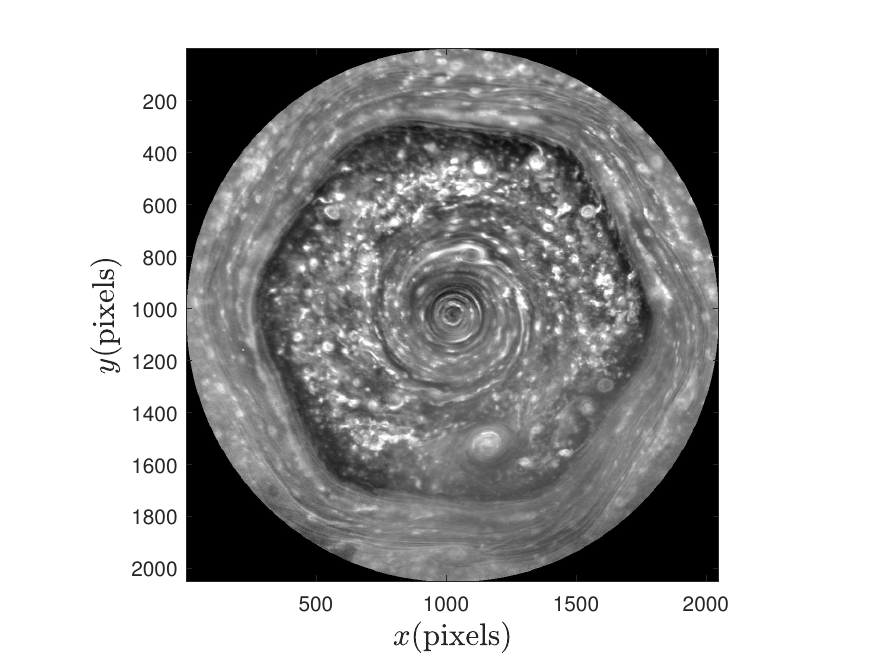}
			\label{SNPVS2}
		\end{minipage}%
	}%
	
	\subfigure[Extracted velocity vectors]{
		\begin{minipage}[t]{0.5\linewidth}
			\centering
			\includegraphics[width=1.0\columnwidth,trim={0cm 0.0cm 0.0cm 0.0cm},clip]{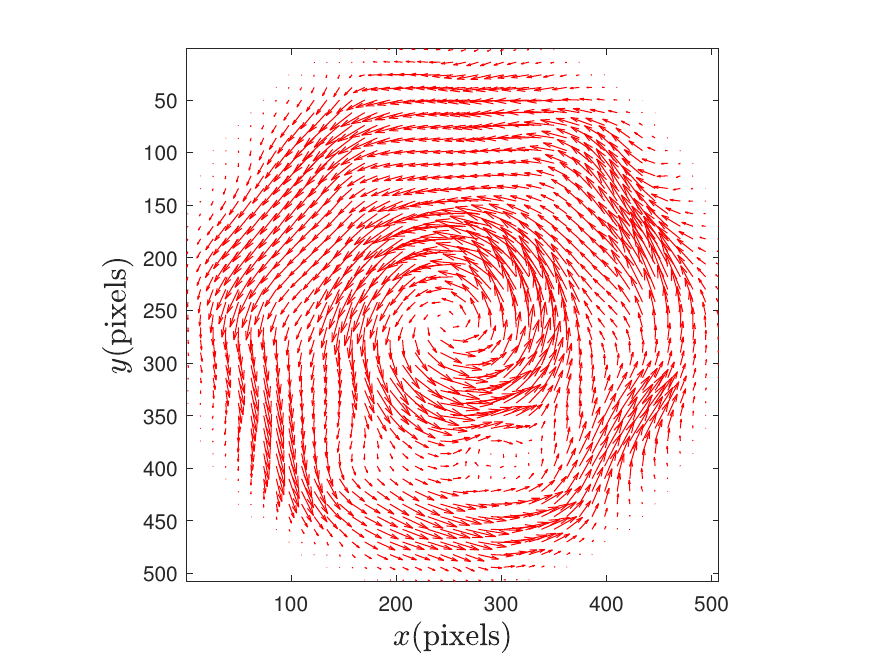}
			\label{SNPVS_vectors}
		\end{minipage}%
	}%
	\caption{Two consecutive cloud images of Saturn's Hexagon and North Polar Vortex (NPV) captured by Cassini on December 10, 2012: (a, b) raw images, and (c) extracted mean velocity vectors within the hexagon boundary using OFM.} 
	\label{Hexagon_polar_structure_images}
\end{figure}

The atmospheric flow structures near Saturn's north pole (NP) have attracted considerable attention in the scientific community and public. In particular, the north polar vortex (NPV) and the hexagon 
circulating jet (Hexagon) around the NPV were observed.
The NPV is a hurricane-like vortex that has the radius of about 54,000-60,000 km and a maximum wind speed of about 150 m/s, while the Hexagon is a jet stream of gas exhibiting a hexagon shape. 
The Cassini probe that orbited Saturn from 2004-2017 gave the best views of the Hexagon and NPV, heightening the mysterious phenomenon.  
The color-filter movie of the hexagon was assembled from eight images captured by Cassini over 10 hours on December 10, 2012, which shows the hexagon in color filters and a complete view from the north pole down to about 70 degrees north latitude.~\footnote{Please refer to the webpage \url{https://science.nasa.gov/mission/cassini/science/saturn/hexagon-in-motion} for more information.}
Figures~\ref{SNPVS1} and~\ref{SNPVS2} show two consecutive black-and-white images (2048 pixels $\times$ 2048 pixels) of the movie near Saturn's NP, in which a wide variety of cloud structures within the Hexagon are observed.
The NPV is centered at the north pole, surrounded by numerous small vortices. The Hexagon and NPV rotate counterclockwise, while some of the smaller vortices traveling along with the jet stream exhibit clockwise rotation. The biggest distinct one of these vortices, which is seen near the lower right corner of the Hexagon in the color image, spans about 3,500 km.

We will employ the optical flow method (OFM) to extract high-resolution velocity fields from the eight black-and-white images of the NPV and Hexagon obtained during the NASA flight missions. 
Originally developed for visual image measurement and pattern  recognition in computer vision, the OFM solves the Horn-Schunck brightness constraint equation in the image plane~\citep{Horn1981}. 
The general relationship between the optical flow and fluid flow was first established by~\citet{LiuShen2008}, who derived a physics-based optical flow equation, providing a rigorous foundation for applying the OFM to flow visualization images.
As a differential-equation-based method, the OFM is particularly well-suited for extracting small displacement vectors from high-resolution images with continuous patterns (yielding one vector per pixel). Detailed variational formulations and numerical algorithms for optical flow computations can be found in~\citet{LiuShen2008} and~\citet{WangBo2015}.
As a special form of global flow diagnostics, the physics-based OFM has been successfully applied to track fascinating coherent structures in planetary atmospheric flows based on the available images from space flight missions~\citep{Liu2012POF,LiuTS2019JGR,LiuTS2022}.
\begin{figure}[t]
	\centering
	\subfigure[]{
		\begin{minipage}[t]{0.5\linewidth}
			\centering
			\includegraphics[width=1.0\columnwidth,trim={0cm 0.0cm 0.0cm 0.0cm},clip]{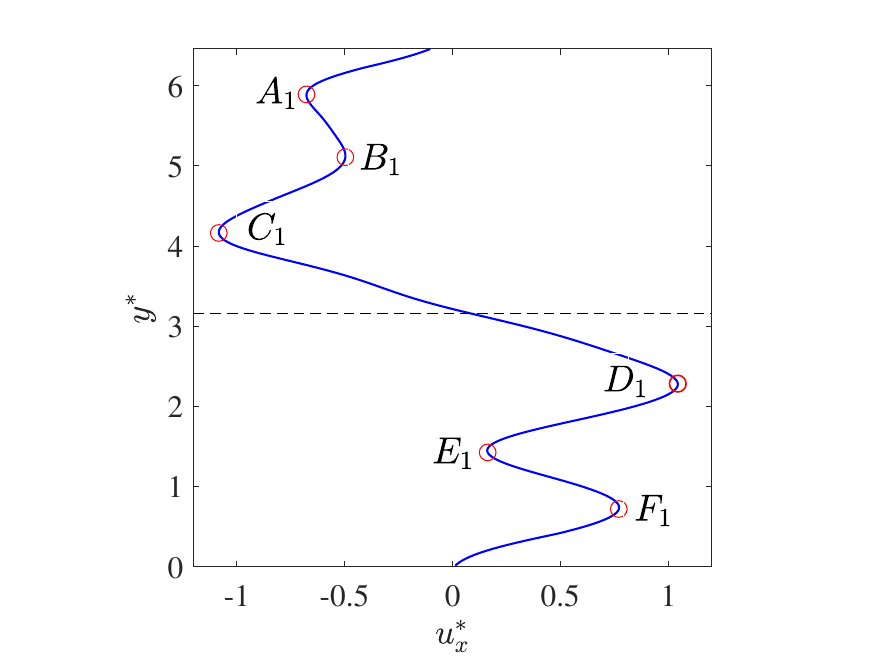}
			\label{ux_y}
		\end{minipage}%
	}%
	\subfigure[]{
		\begin{minipage}[t]{0.5\linewidth}
			\centering
			\includegraphics[width=1.0\columnwidth,trim={0cm 0.0cm 0.0cm 0.0cm},clip]{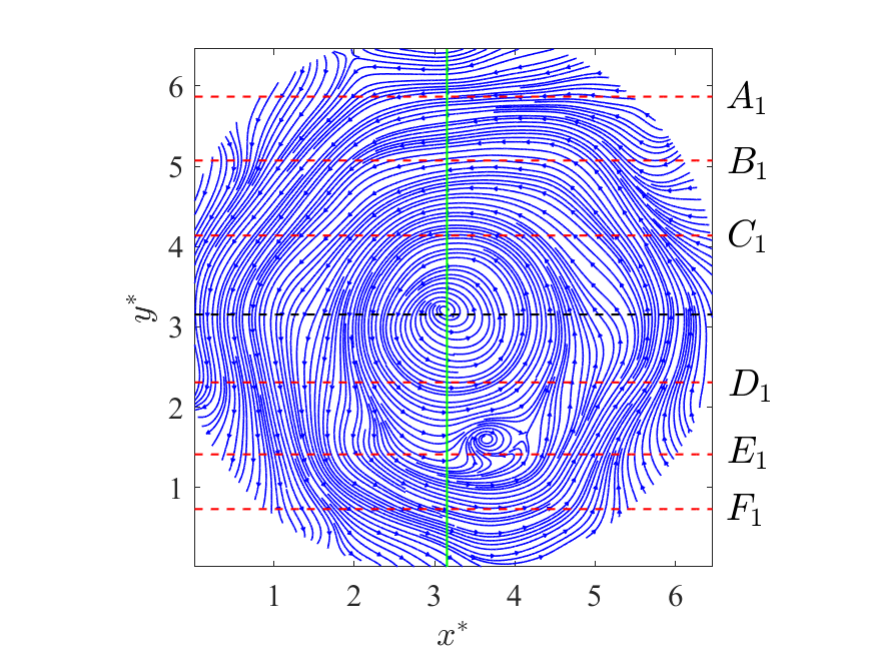}
			\label{Streamlines_ux_y}
		\end{minipage}%
	}%
	
	\subfigure[]{
		\begin{minipage}[t]{0.5\linewidth}
			\centering
			\includegraphics[width=1.0\columnwidth,trim={0cm 0.0cm 0.0cm 0.0cm},clip]{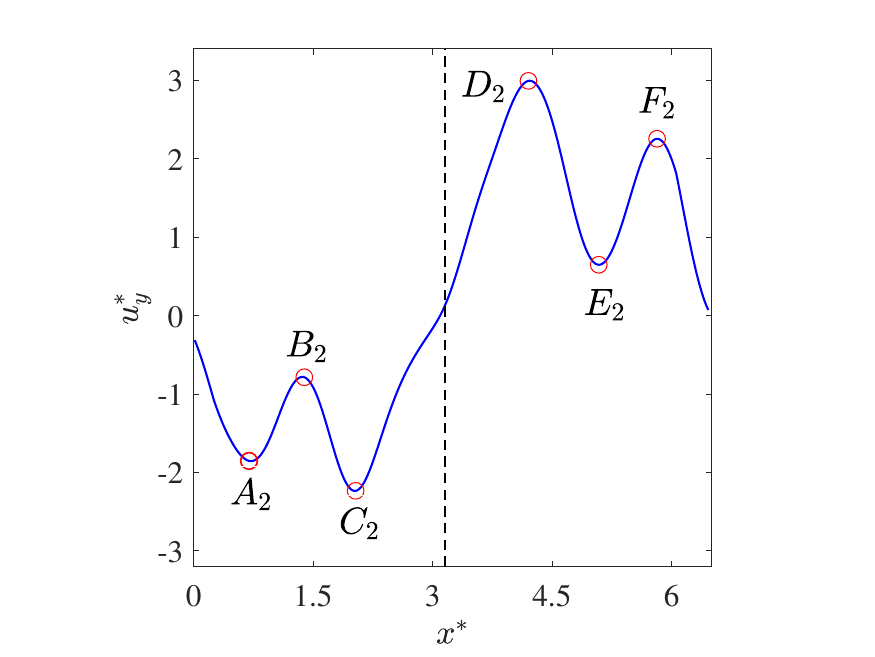}
			\label{uy_x}
		\end{minipage}%
	}%
	\subfigure[]{
		\begin{minipage}[t]{0.5\linewidth}
			\centering
			\includegraphics[width=1.0\columnwidth,trim={0cm 0.0cm 0.0cm 0.0cm},clip]{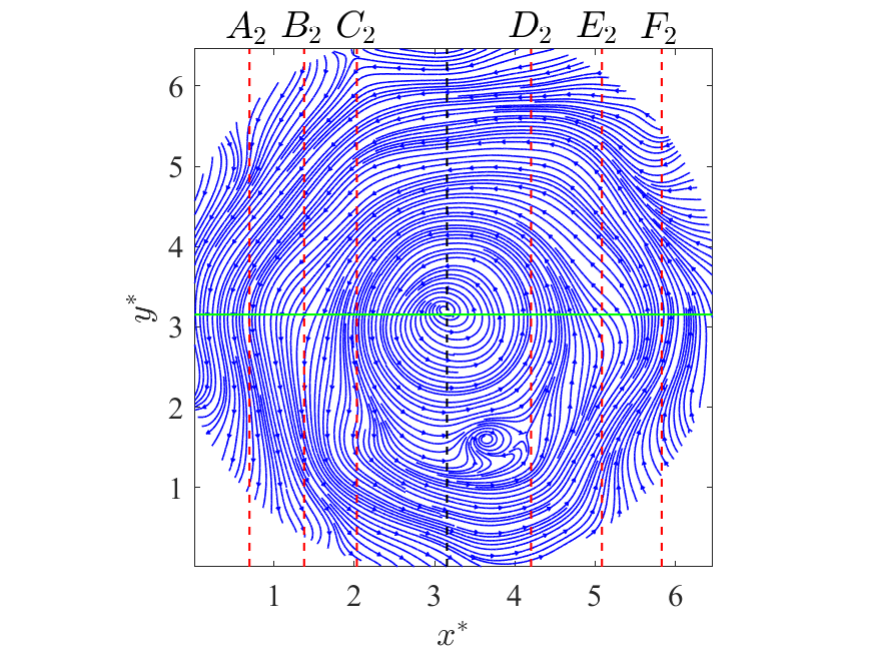}
			\label{Streamlines_uy_x}
		\end{minipage}%
	}%
	\caption{Spatial distributions of normalized velocity components in the Hexagon and NPV: (a,b): $u_{x}^{*}$ along $x^{*}=3.155$ (green vertical line); (c,d): $u_{y}^{*}$ along $y^{*}=3.155$ (green horizontal line). Peak and valley points of the velocity profiles are marked by red circles in (a,b), with their corresponding positions indicated by red dashed lines in (c,d).} 
	\label{Velocity_Profiles}
\end{figure}

The working principle of the OFM is briefly described as follows. When projected onto the image plane $(x,y)$, the governing optical flow equation is expressed as~\citep{LiuShen2008}
\begin{eqnarray}
	\frac{\partial g}{\partial t}+\bm{\nabla}\bm{\cdot}(g\bm{u})=S_{g},
\end{eqnarray}
where $g$ is the normalized image intensity associated with a measured physical quantity, $\bm{u}=(u_{x},u_{y})$ represents the two-component optical flow velocity field, and $\nabla=(\partial_{x},\partial_{y})$ is the spatial gradient operator. In practical applications, the source term $S_{g}$ is typically neglected as a first-order approximation. For a given $g$ and $S_{g}$, determining the optical flow velocity field from images can be treated as an inverse problem.  Incorporating the first-order Tikhonov regularization as a constraint, the approximate solution of the velocity field $\bm{u}$ can be obtained by minimizing the following functional:
\begin{eqnarray}
	\mathcal{L}(\bm{u})=\int_{\mathcal{D}}\left[\frac{\partial g}{\partial t}+\bm{\nabla}\bm{\cdot}(g\bm{u})-S_{g}\right]dxdy+\alpha\int_{\mathcal{D}}\bm{\nabla}\bm{u}\bm{:}\bm{\nabla}\bm{u}dxdy,
\end{eqnarray}
where $\mathcal{D}$ is the integral domain, and $\alpha$ is a constant Lagrange multiplier.
Applying the first-order variational principle to $\mathcal{L}(\bm{u})$ yields the corresponding Euler-Lagrange equation:
\begin{eqnarray}
g\left[\frac{\partial g}{\partial t}+\bm{\nabla}\bm{\cdot}(g\bm{u})-S_{g}\right]+\alpha\nabla^{2}\bm{u}=\bm{0}.
\end{eqnarray}
The velocity field solution is then obtained numerically using finite difference methods, subject to appropriate Neumann boundary conditions enforced on the image plane~\citep{WangBo2015}.
As shown in figure~\ref{SNPVS_vectors}, this approach successfully generates seven high-resolution snapshots of the velocity field ($512\times512$ vectors). The mean velocity field computed within the hexagon boundary clearly validates the effectiveness of the OFM for flow field analysis. In the following analysis, the reference radius $R_{\rm ref}$ and velocity $U_{\rm ref}$ are defined as the mean radius of the rigid-rotation vortex core and the mean maximum circumferential velocity, respectively. Consequently, the reference vorticity is given by $U_{\rm ref}/R_{\rm ref}$.

\subsubsection{Results and analysis}
\begin{figure}[t]
	\centering
	\subfigure[$\bm{R}_{L}(\bm{t})\bm{\cdot}\bm{e}_{z}$]{
		\begin{minipage}[t]{0.5\linewidth}
			\centering
			\includegraphics[width=1.0\columnwidth,trim={0cm 0.0cm 0.0cm 0.0cm},clip]{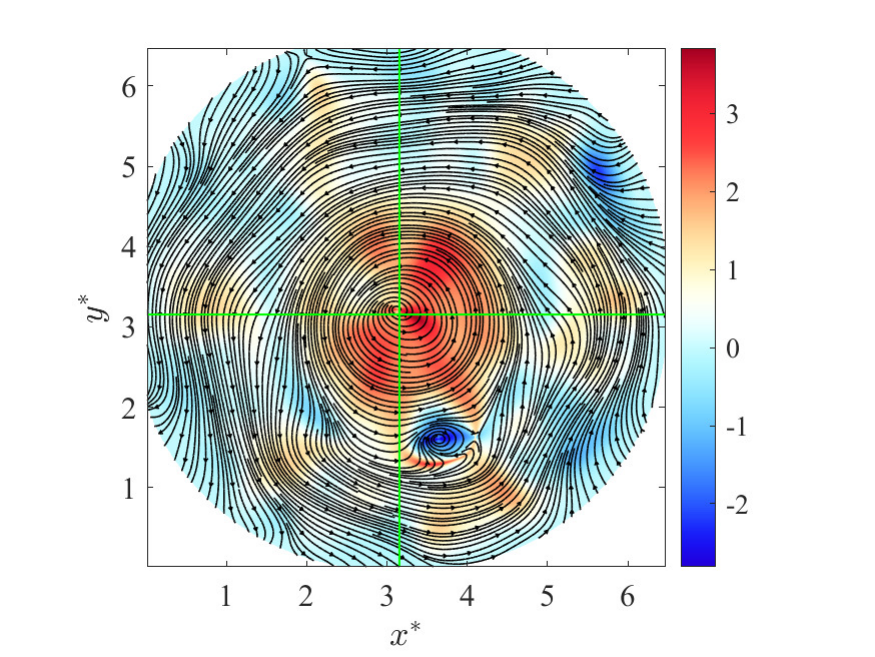}
			\label{Global_Rtz_contour_streamlines}
		\end{minipage}%
	}%
	\subfigure[$\bm{s}_{L}(\bm{t})\bm{\cdot}\bm{e}_{z}$]{
		\begin{minipage}[t]{0.5\linewidth}
			\centering
			\includegraphics[width=1.0\columnwidth,trim={0cm 0.0cm 0.0cm 0.0cm},clip]{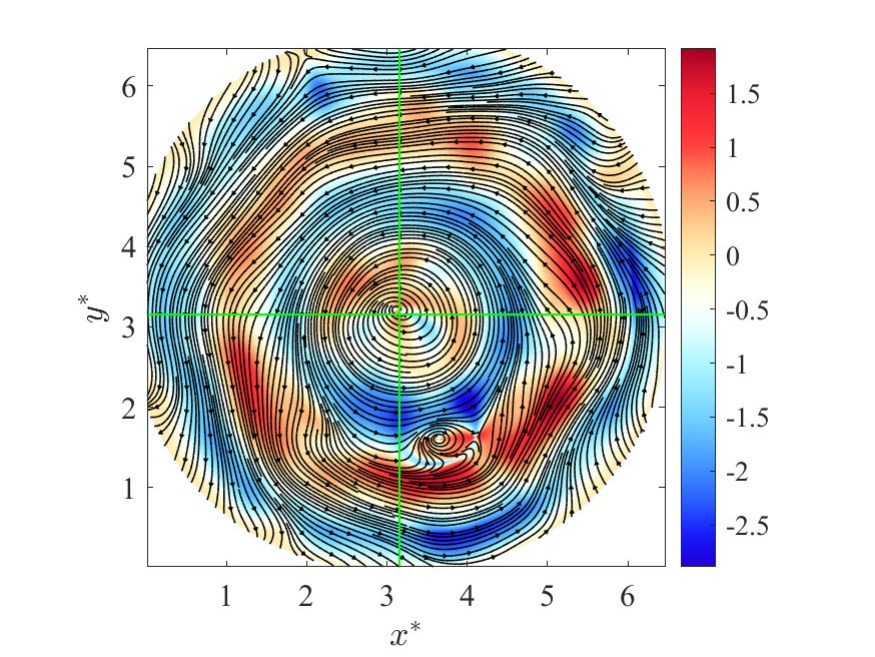}
			\label{Global_Stz_contour_streamlines}
		\end{minipage}%
	}%
	
	\subfigure[$\omega_{z}$]{
		\begin{minipage}[t]{0.5\linewidth}
			\centering
			\includegraphics[width=1.0\columnwidth,trim={0cm 0.0cm 0.0cm 0.0cm},clip]{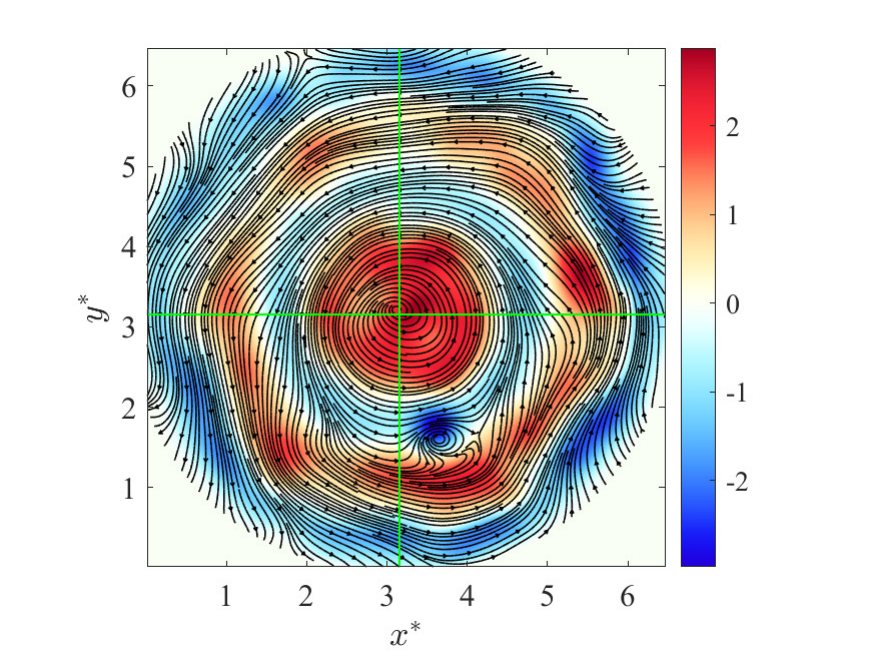}
			\label{Global_oz_contour_streamlines}
		\end{minipage}%
	}%
	\subfigure[$-Q$]{
		\begin{minipage}[t]{0.5\linewidth}
			\centering
			\includegraphics[width=1.0\columnwidth,trim={0cm 0.0cm 0.0cm 0.0cm},clip]{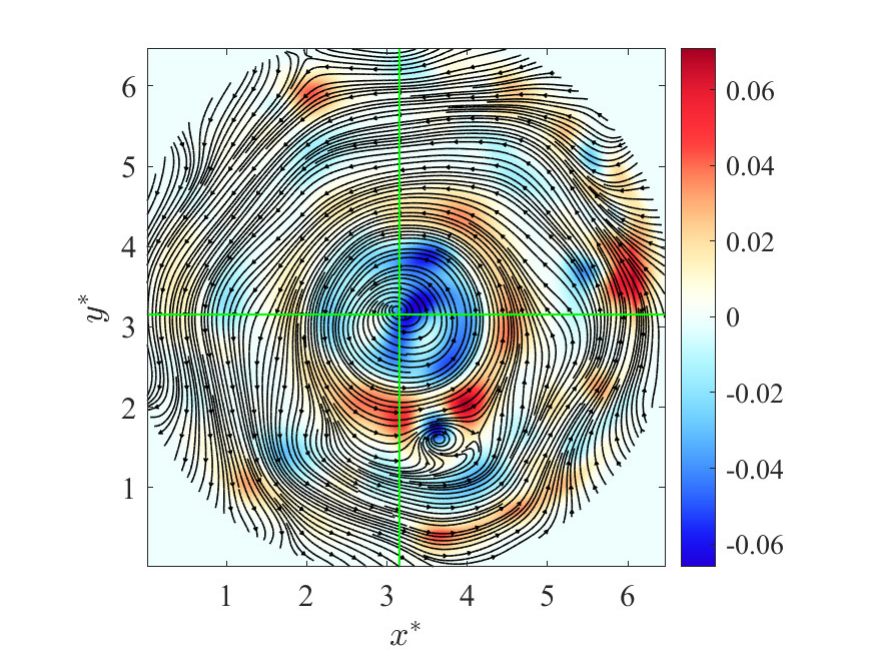}
			\label{Minus_Qc_global}
		\end{minipage}%
	}%
	\caption{Global contour maps of vortex characteristics in the Hexagon and NPV: (a) the orbital rotation $\bm{R}_{L}(\bm{t})\bm{\cdot}\bm{e}_{z}$, (b) the spin $\bm{s}_{L}(\bm{t})\bm{\cdot}\bm{e}_{z}$, (c) the vorticity $\omega_{z}$, and (d) the second principal invariant of the VGT $Q$ (with a minus sign). The green lines indicate the cross-sections $x^{*}=3.155$ and $y^{*}=3.155$, respectively.} 
	\label{Global_vorticity_components}
\end{figure}
\begin{figure}[t]
	\centering
	\subfigure[]{
		\begin{minipage}[t]{0.5\linewidth}
			\centering
			\includegraphics[width=1.0\columnwidth,trim={0cm 0.0cm 0.0cm 0.0cm},clip]{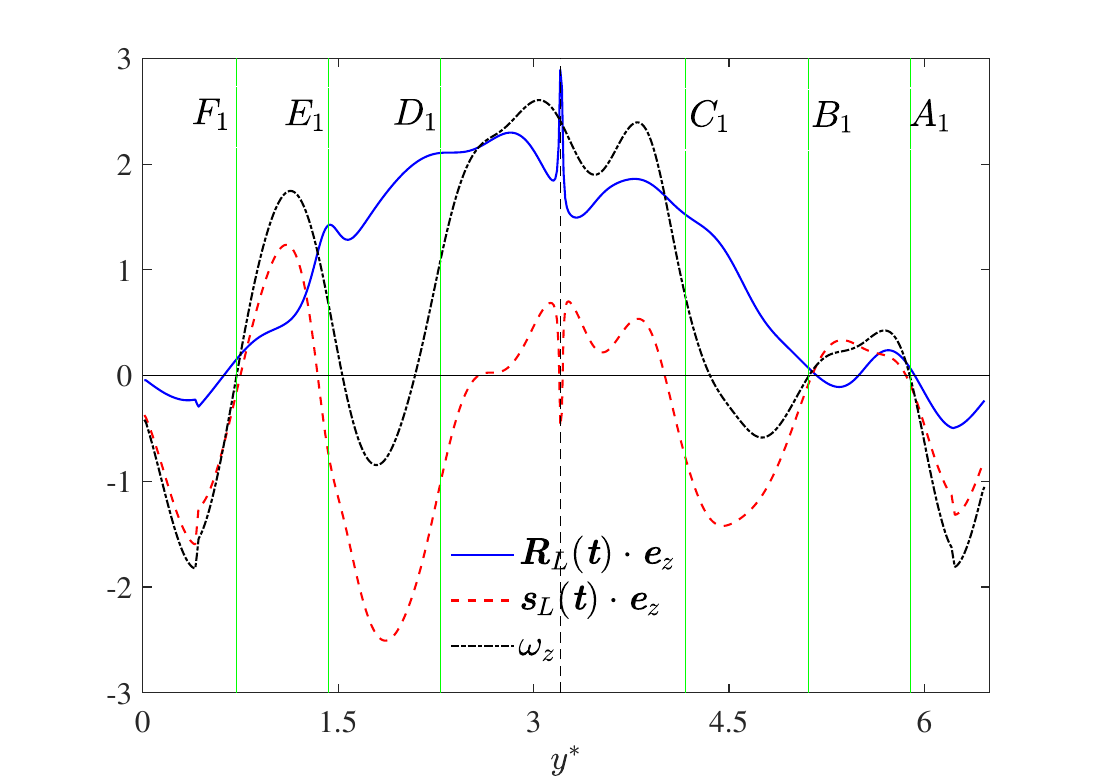}
			\label{Global_RSO2}
		\end{minipage}%
	}%
	\subfigure[]{
		\begin{minipage}[t]{0.5\linewidth}
			\centering
			\includegraphics[width=1.0\columnwidth,trim={0cm 0.0cm 0.0cm 0.0cm},clip]{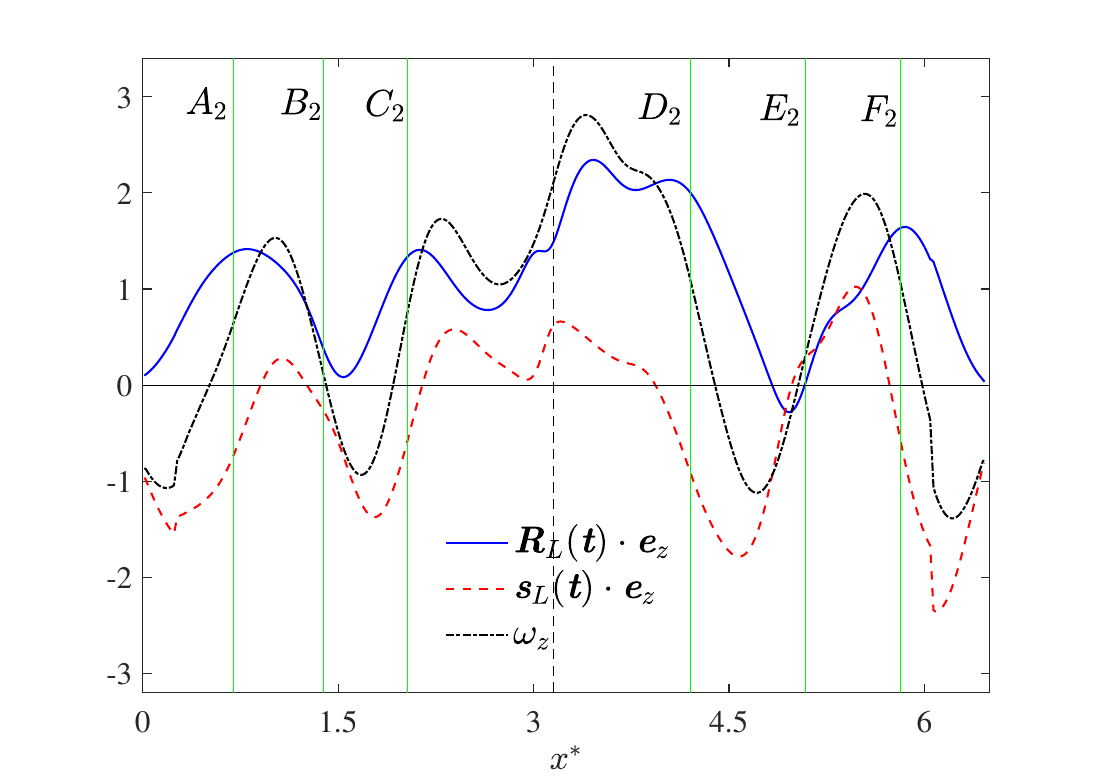}
			\label{Global_RSO1}
		\end{minipage}%
	}%
	\caption{ $(\bm{R}_{L}(\bm{t}),\bm{s}_{L}(\bm{t}),\bm{\omega})$ along (a) the vertical line $x^{*}=3.155$ and (b) the horizontal line $y^{*}=3.155$.} 
	\label{RSO12}
\end{figure}
\begin{figure}[t]
	\centering
	\subfigure[$\bm{R}_{L}(\bm{t})\bm{\cdot}\bm{e}_{z}$]{
		\begin{minipage}[t]{0.5\linewidth}
			\centering
			\includegraphics[width=1.0\columnwidth,trim={0cm 0.0cm 0.0cm 0.0cm},clip]{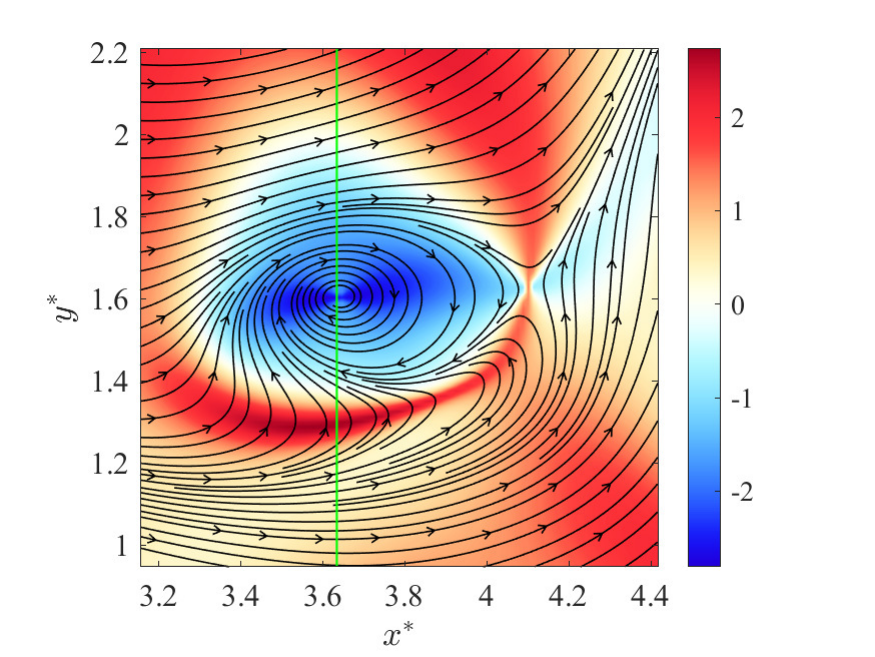}
			\label{Second_Rtz_contour_streamlines}
		\end{minipage}%
	}%
	\subfigure[$\bm{s}_{L}(\bm{t})\bm{\cdot}\bm{e}_{z}$]{
		\begin{minipage}[t]{0.5\linewidth}
			\centering
			\includegraphics[width=1.0\columnwidth,trim={0cm 0.0cm 0.0cm 0.0cm},clip]{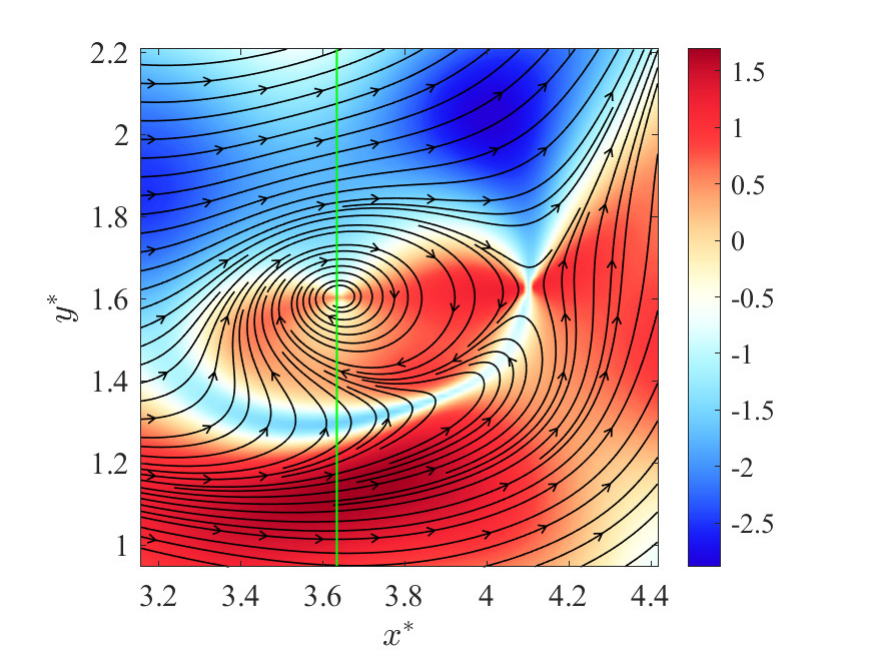}
			\label{Second_Stz_contour_streamlines}
		\end{minipage}%
	}%
	
	\subfigure[$\omega_{z}$]{
		\begin{minipage}[t]{0.5\linewidth}
			\centering
			\includegraphics[width=1.0\columnwidth,trim={0cm 0.0cm 0.0cm 0.0cm},clip]{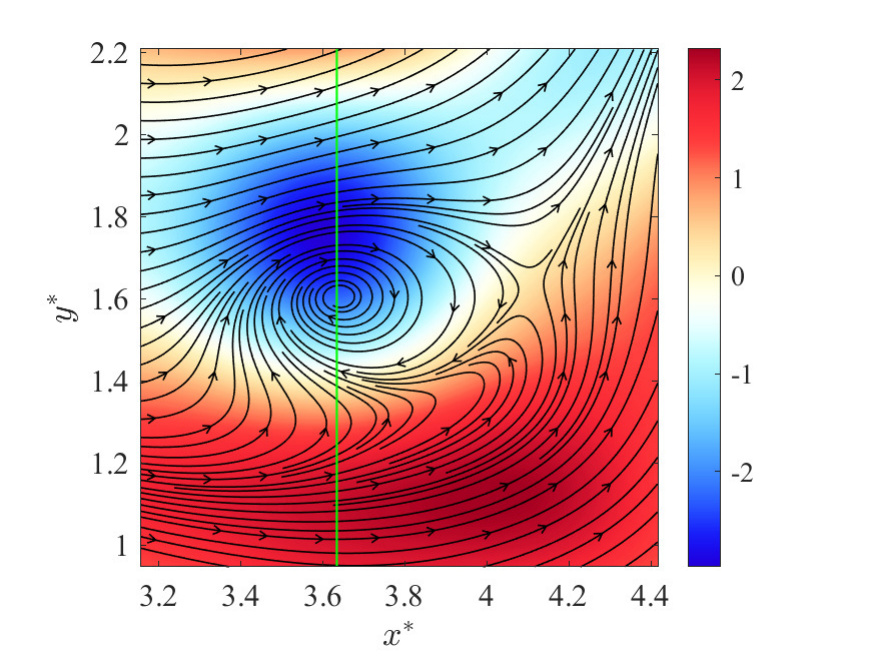}
			\label{Second_oz_contour_streamlines}
		\end{minipage}%
	}%
	\subfigure[$-Q$]{
		\begin{minipage}[t]{0.5\linewidth}
			\centering
			\includegraphics[width=1.0\columnwidth,trim={0cm 0.0cm 0.0cm 0.0cm},clip]{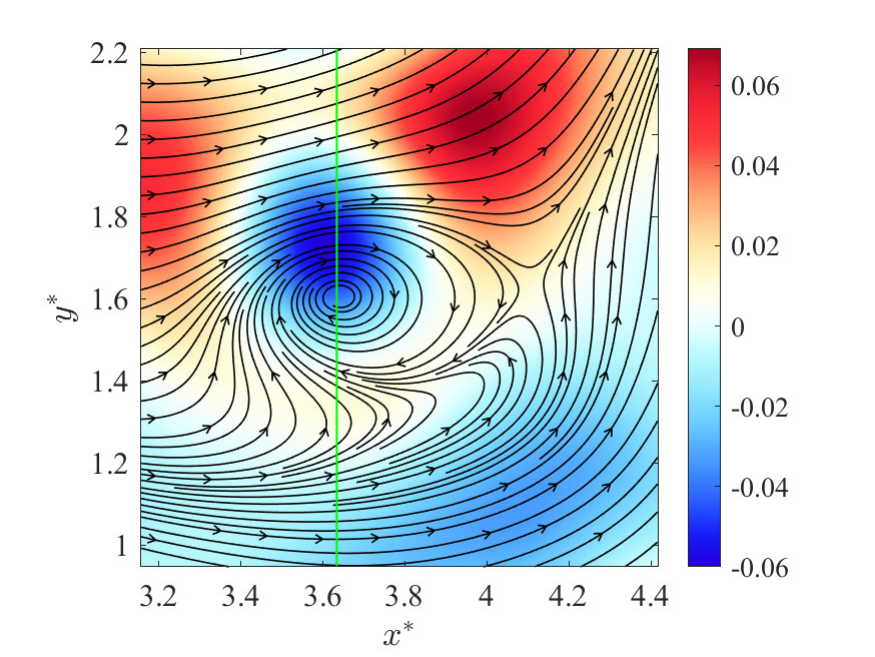}
			\label{Minus_Qc}
		\end{minipage}%
	}%
	\caption{Contour maps of vortex characteristics in the hurricane eye. (a) the orbital rotation $\bm{R}_{L}(\bm{t})\bm{\cdot}\bm{e}_{z}$; (b) the spin $\bm{s}_{L}(\bm{t})\bm{\cdot}\bm{e}_{z}$; (c) the vorticity $\omega_{z}$; (d) the second principal invariant of the VGT $Q$ (with a minus sign). The green solid line marks the cross-section $x^{*}=3.634$ across the center of the hurricane eye.} 
	\label{Second_vorticity_components}
\end{figure}

The velocity field and streamline topology are analyzed to  identify the primary flow structures in Saturn's north polar region. 
Figure~\ref{ux_y} presents the (normalized) velocity component $u_{x}$ along the $y$-direction (at $x^{*}=3.155$) whereas figure~\ref{Streamlines_ux_y} displays the corresponding global streamline pattern. Key features are marked in both figures, with peak and valleys points $(A_{1}-F_{1})$ of $u_{x}$ indicated by circles and dashed lines, respectively. The streamline analysis reveals a well-defined hexagonal structure in the outer region, containing both the counterclockwise-rotating NPV and a clearly distinguishable secondary vortex (hurricane eye) manifesting clockwise rotation.
Within the core region $(C_{1},D_{1})$ around the NPV, the flow inherently exhibits a quasi-linear velocity profile characteristic of rigid-body rotation, with velocity-magnitude maxima occuring at the endpoints $C_{1}$ and $D_{1}$. 
The Hexagon (i.e., the high-speed jet stream confined in the ring-belt region) consists of the radial segments $(A_{1},C_{1})$ and $(D_{1},F_{1})$, where the velocity profile exhibits concave curvature relative to the jet stream, with  local minima occurring at points $B_{1}$ and $E_{1}$, respectively.
The nonlinear variation of $u_{x}$ within these segments indicates that the shearing motion plays a crucial role in modulating the jet's rotational motion.
Viscous dissipation at small scales leads to velocity decay 
in the far field (beyond $A_{1}$ and $F_{1}$).
Complementary data in Figures~\ref{uy_x} and~\ref{Streamlines_uy_x} show similar behavior for the velocity component $u_{y}$ along the $x$-direction ($y^{*}=3.155$), further confirming the NPV and Hexagon as the dominant coherent structures governing the global flow pattern.

\begin{figure}[t]
	\centering
	\subfigure[$\bm{R}_{N}^{+}\bm{\cdot}\bm{e}_{z}$]{
		\begin{minipage}[t]{0.5\linewidth}
			\centering
			\includegraphics[width=1.0\columnwidth,trim={0cm 0.0cm 0.0cm 0.0cm},clip]{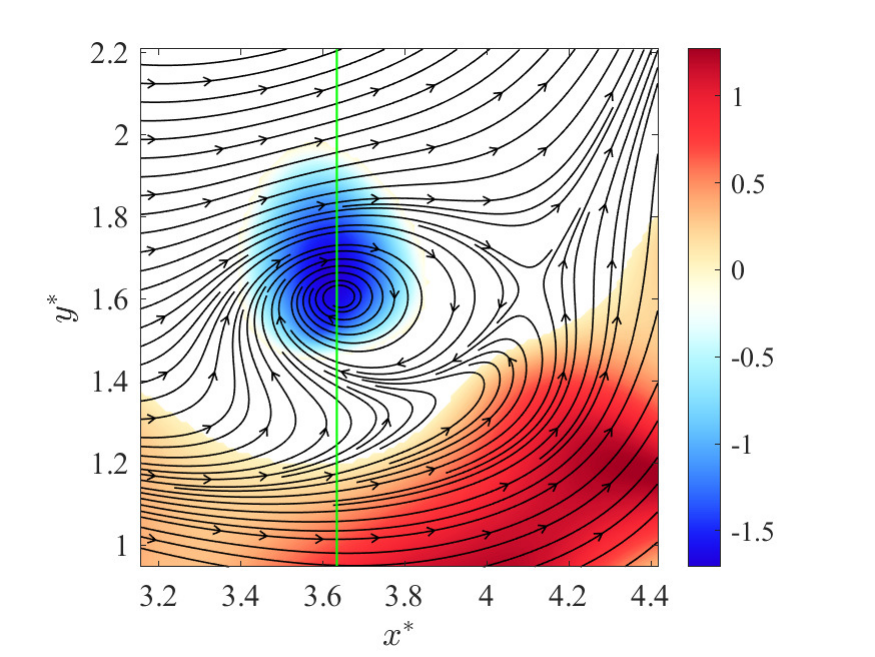}
			\label{Second_RNz_contour_streamlines}
		\end{minipage}%
	}%
	\subfigure[$\bm{R}_{N}^{-}\bm{\cdot}\bm{e}_{z}$]{
		\begin{minipage}[t]{0.5\linewidth}
			\centering
			\includegraphics[width=1.0\columnwidth,trim={0cm 0.0cm 0.0cm 0.0cm},clip]{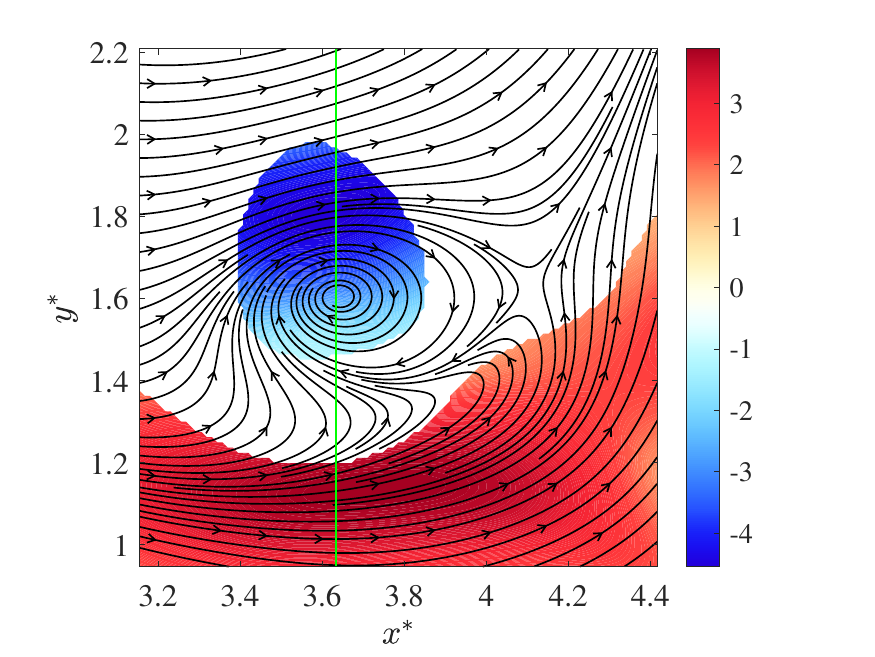}
			\label{Second_RNz1_contour_streamlines}
		\end{minipage}%
	}%
	\caption{Contour maps of the characteristic rigid rotation modes in the hurricane eye. (a) $\bm{R}_{N}^{+}\bm{\cdot}\bm{e}_{z}$ (Liutex) and (b) $\bm{R}_{N}^{-}\bm{\cdot}\bm{e}_{z}$. The green solid line indicates $x^{*}=3.634$ across the center of the hurricane eye.} 
	\label{Second_vorticity_components1}
\end{figure}

The streamline-based DVD in~\eqref{C8a12} is employed to decompose the total vorticity $\bm{\omega}$ into its constituent components: the orbital rotation mode $\bm{R}_{L}(\bm{t})$ and the spin mode $\bm{s}_{L}(\bm{t})$. Within this field description framework, 
the orbital rotation mode represents a natural extension of rigid-rotation concept to characterize the rotational behavior of a material line element instantaneously coinciding with the streamline.
Figure~\ref{Global_Rtz_contour_streamlines} presents the global contour distribution of the axial vorticity component $\bm{R}_{L}(\bm{t})$ (i.e., $\bm{R}_{L}(\bm{t})\bm{\cdot}\bm{e}_{z}$) associated with the orbital rotation mode while figure~\ref{Global_Stz_contour_streamlines} displays the corresponding axial component $\bm{s}_{L}(\bm{t})\bm{\cdot}\bm{e}_{z}$ for the spin mode.
In figure~\ref{Global_Rtz_contour_streamlines},
we observe six distinct petal-shaped regions of positive orbital rotation surrounding the NPV, though with significantly weaker magnitudes compared to the NPV core.
Most notably, the analysis clearly identifies a clockwise-rotating hurricane eye in the lower-right quadrant $(x^{*},y^{*})\in[3.2,4.4]\times[1,2.2]$, characterized by negative orbital-rotation component, which coexists with both the NPV and Hexagon. Figure~\ref{Global_Stz_contour_streamlines} demonstrates that while the spin mode shows relatively weak intensity within the NPV core region, the surrounding annular shear zone exhibits pronounced spin magnitude with alternating positive and negative vorticity bands aligned radially. This distinctive pattern results from strong velocity gradients at the transition region between the NPV and the hexagonal jet stream, in agreement with the velocity profiles previously shown in figures~\ref{ux_y} and~\ref{uy_x}.
The distribution of vorticity is displayed in figure~\ref{Global_oz_contour_streamlines}.
The combined analysis of figures~\ref{Global_Rtz_contour_streamlines},~\ref{Global_Stz_contour_streamlines}, and~\ref{Global_oz_contour_streamlines} confirms that the NPV is predominantly governed by the orbital rotation mode, whereas the Hexagon exhibits dominance of the spin mode. These findings are consistent with prior interpretation based on the velocity field and streamline patterns.
Furthermore, quantitative comparisons along the lines $x^{*}=3.155$ and $y^{*}=3.155$ (figures~\ref{Global_RSO2} and~\ref{Global_RSO1})
provide additional validation of these dynamical distinctions. 
Near the NPV, both the variation trend and magnitude of the vorticity $\omega_{z}$ closely resemble those of the orbital rotation $\bm{R}_{L}(\bm{t})$. In contrast, within the Hexagon, while the variation trend of vorticity aligns with the spin mode 
$\bm{s}_{L}(\bm{t})$, its magnitude exhibits additional modulation by the orbital rotation $\bm{R}_{L}(\bm{t})$.

For incompressible flows, the second principal invariant of $\bm{A}$ is expressed as $Q=\frac{1}{2}(\lVert\bm{\varOmega}\rVert_{F}^{2}-\lVert\bm{D}\rVert_{F}^{2})$, which characterizes the local competition between shear strain rate and vorticity magnitude~\citep{Hunt1988,Jeong1995}. Here, the Frobenius norm of an arbitrary second-order tensor $\bm{\Phi}$ is given by $\lVert\bm{\Phi}\rVert_{F}\equiv[{\rm tr}(\bm{\Phi}\bm{\Phi}^{\rm{T}})]^{1/2}$, with $\rm{tr}$ denoting the trace. It has been commonly used as a scalar criterion for vortex identification in complex flows. Physically, $Q$ serves as the source term in the pressure Poisson equation $\nabla^{2}p=2\rho{Q}$. The presence of a vortex core usually corresponds to a low pressure region which can be captured by a positive $Q$, although there is no explicit connection between $Q>0$ and local pressure minimum. Figure~\ref{Minus_Qc_global} displays the contour map of $-Q$ for comparative analysis. The results demonstrate that while the NPV and hurricane eye (represented as column vortex) are effectively captured by $-Q<0$, the Hexagon, predominantly dominated by the spin mode, exhibits less distinct differentiation.
This reduced sensitivity arises because both the strain-rate tensor $\bm{D}$ and rotation tensor $\bm{\varOmega}$ incorporate physical effects attributable to the spin mode, leading to partial cancellation of their respective contributions to $Q$. Consequently, the spin-dominated dynamics in the Hexagon region are not as clearly captured by $Q$-criterion.
\begin{figure}[t]
	\centering
	\subfigure[]{
		\begin{minipage}[t]{0.5\linewidth}
			\centering
			\includegraphics[width=1.0\columnwidth,trim={0cm 0.0cm 0.0cm 0.0cm},clip]{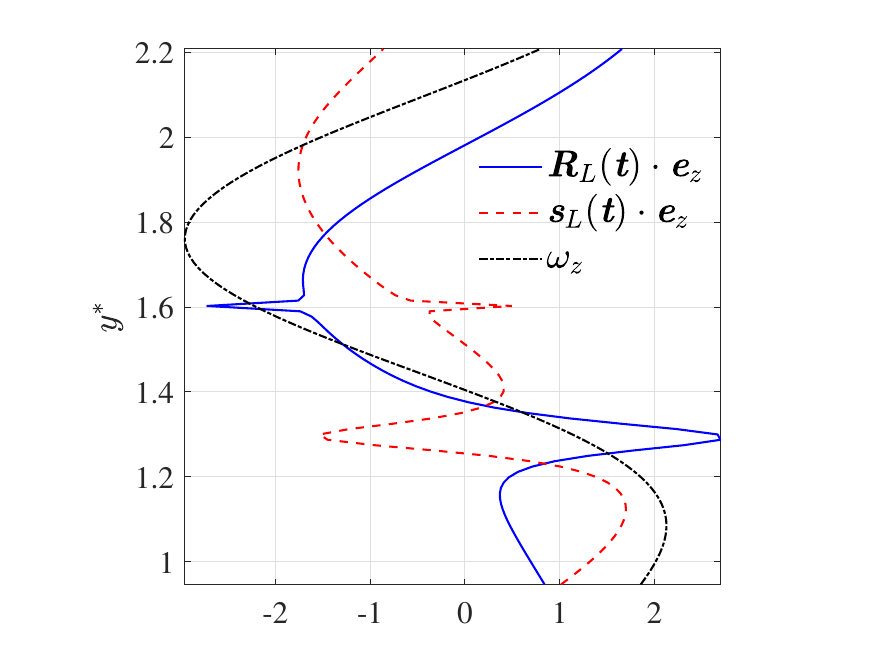}
			\label{Second_compare_RSO}
		\end{minipage}%
	}%
	\subfigure[]{
		\begin{minipage}[t]{0.5\linewidth}
			\centering
			\includegraphics[width=1.0\columnwidth,trim={0cm 0.0cm 0.0cm 0.0cm},clip]{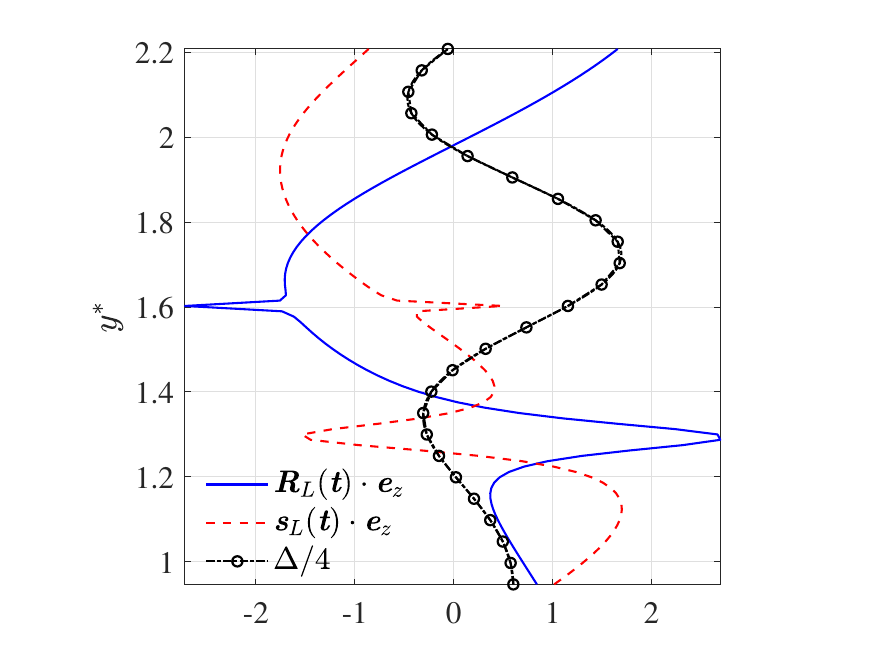}
			\label{Second_compare_RSDelta}
		\end{minipage}%
	}%
	
	\subfigure[]{
		\begin{minipage}[t]{0.5\linewidth}
			\centering
			\includegraphics[width=1.0\columnwidth,trim={0cm 0.0cm 0.0cm 0.0cm},clip]{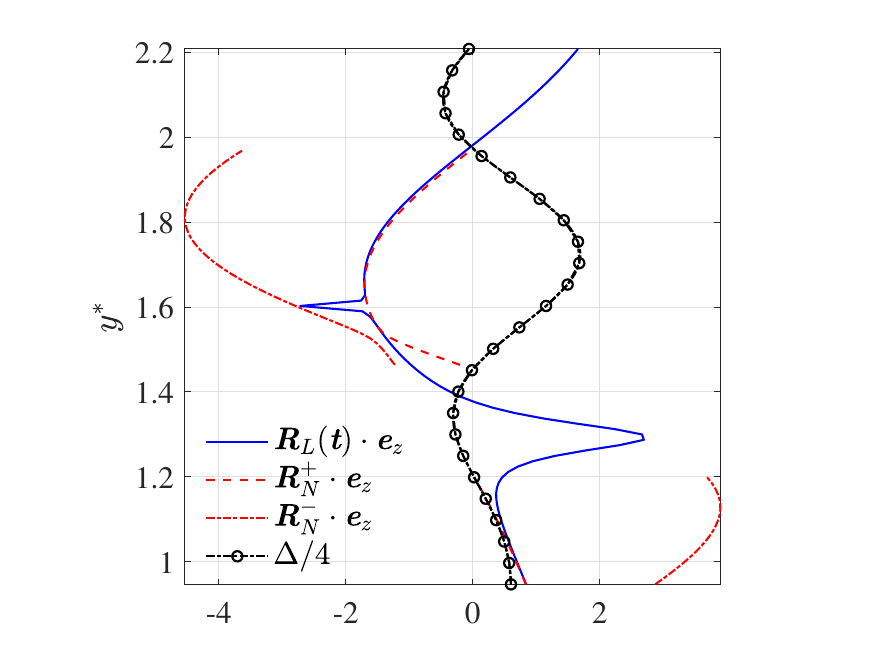}
			\label{Second_compare_RRNDelta}
		\end{minipage}%
	}%
	\subfigure[]{
		\begin{minipage}[t]{0.5\linewidth}
			\centering
			\includegraphics[width=1.0\columnwidth,trim={0cm 0.0cm 0.0cm 0.0cm},clip]{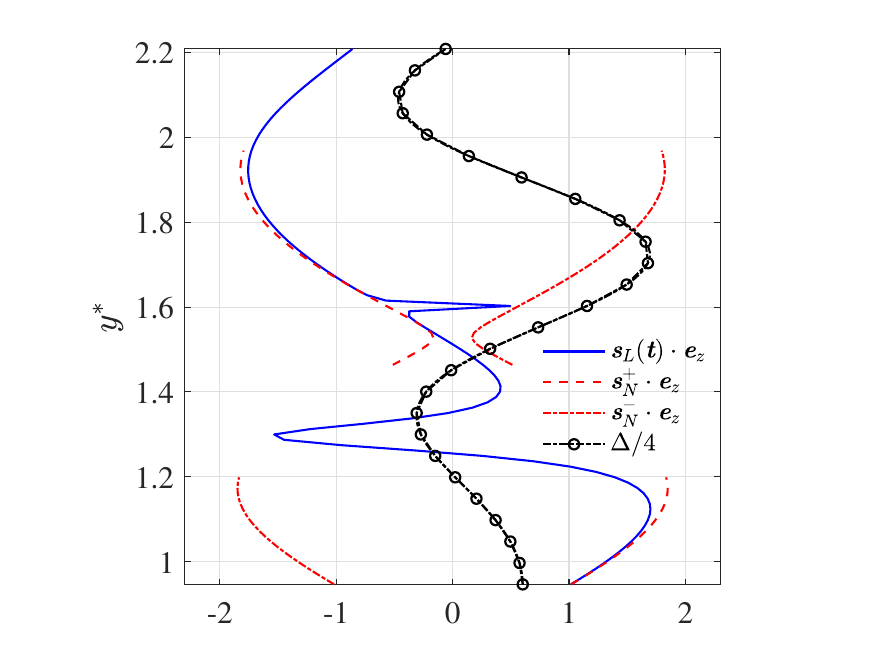}
			\label{Second_compare_ssNDelta}
		\end{minipage}%
	}%
	\caption{(a) $(\bm{R}_{L}(\bm{t}),\bm{s}_{L}(\bm{t}),\omega_{z})$, (b) $(\bm{R}_{L}(\bm{t}),\bm{s}_{L}(\bm{t}),\Delta)$, (c) $(\bm{R}_{L}(\bm{t}),\bm{R}_{N}^{\pm},\Delta)$, and (d) $(\bm{s}_{L}(\bm{t}),\bm{s}_{N}^{\pm},\Delta)$ along the vertical line $x^*=3.634$ in the hurricane eye. The normalized discriminant $\Delta^{*}$ has been scaled by a factor of $1/4$.} 
	\label{NEW20250430}
\end{figure}

Figure~\ref{Second_vorticity_components} provides zoom-in views of the quantities previously shown in figure~\ref{Global_vorticity_components}, focusing on the hurricane eye region with superimposed streamlines.
The primary structure of the hurricane eye exhibits a characteristic saddle-node pair configuration, which maintains the required topological balance between node and saddle points in the flow field  according to the topological rule~\citep{Tobak1982}.
The orbital-rotation component $\bm{R}_{L}(\bm{t})\bm{\cdot}\bm{e}_{z}$ (figure~\ref{Second_Rtz_contour_streamlines}) exhibits a coherent negative-valued region with high magnitude, whose spatial distribution precisely matches the characteristic eye-shaped pattern.
The superimposed streamlines reveal a distinct clockwise spiraling motion converging toward the rotation center, clearly demonstrating the hallmark features of an intense swirling vortex.
Along the lower periphery of the hurricane eye, intense shear interaction between the eastward jet current and westward vortex flow induces abrupt streamline deflection. This strong velocity gradient generates a banded region of high-magnitude orbital rotation that correlates with the zone of maximum streamline curvature. 
The close proximity near the saddle point displays both the positive and negative orbital rotation components due to the competing influences between the rotational flow and adjacent jet stream.
Notably, the spin component $\bm{s}_{L}(\bm{t})\bm{\cdot}\bm{e}_{z}$ (figure~\ref{Second_Stz_contour_streamlines})
displays a remarkably similar spatial distribution to the orbital rotation. Immediately beneath the eye structure, the negative spin is also a direct consequence of the high-curvature streamline geometry.
The superposition of these two modes yields the vorticity $\omega_{z}$
in figure~\ref{Second_oz_contour_streamlines}. A distinct blue-shaded region exhibiting negative vorticity is observed, with its centroid displaying significant spatial offset from the orbital rotation center due to the influence of the spin mode.

Moreover, the second invariant $Q>0$ in figure~\ref{Minus_Qc} reveals a flow pattern qualitatively similar to $\omega_{z}$, effectively capturing both the primary vortex core dominated by orbital rotation and portions of the Hexagon. According to the characteristic algebraic description, $Q$ can be equivalently expressed as $Q=\lambda_{\rm ci}^{2}-\frac{1}{3}\varepsilon^{2}$, where $\varepsilon\equiv\lambda_{r}-\chi$ indicates the difference between the relative stretching rates along the NND axes~\citep{Shrestha2021,LiuCQ2020,LiZhen2014}. This result implies that the region with $Q>0$ must satisfy $\lambda_{ci}\neq{0}$ and $\Delta>0$, but the converse does not hold. Therefore, the region identified by $Q>0$ is a subset of $\Delta>0$.

In figures~\ref{Second_RNz_contour_streamlines} and~\ref{Second_RNz1_contour_streamlines}, we show the contour maps of two physical quantities for extracting the characteristic rigid rotation modes: $\bm{R}_{N}^{+}\equiv2\psi^{+}\bm{e}_{3}$ (i.e., the Liutex vector) and $\bm{R}_{N}^{-}\equiv2\psi^{-}\bm{e}_{3}$, in the region $\Delta>0$.
While exhibiting distinct magnitude differences, they demonstrate qualitatively similar spatial distributions, effectively capturing the overall rigid-rotation features in both the hurricane eye (represented by a singly-connected blue region), and
the peripheral zone of the Hexagon (marked by a red region).
However, when compared to the orbital rotation $\bm{R}_{L}(\bm{t})$ (figure~\ref{Second_Rtz_contour_streamlines}) in the field description, these invariant physical measures demonstrate certain limitations in capturing the flow details such as high-curvature streamline region, and localized flow turning phenomena near the saddle point.

Figure~\ref{NEW20250430} presents quantitative comparisons of the DVD vorticity modes along the centerline of the hurricane eye $(x^{*}=3.634)$.
As the center of the hurricane eye is approached, the velocity magnitude attains its local minimum value $(q\rightarrow{0},\partial_{n}q\rightarrow{0})$, indicating the behavior of a stagnation point; and the streamline curvature becomes singular $(\kappa^{-1}\rightarrow{0}~\text{as}~\kappa\rightarrow\infty)$. 
Therefore, the vorticity modes should satisfy $\bm{R}_{L}(\bm{t})\bm{\cdot}\bm{e}_{z}\sim-2\kappa{q}$ and $\bm{s}_{L}(\bm{t})\bm{\cdot}\bm{e}_{z}\sim\kappa{q}$, which explain the negative and positive peaks observed in figures~\ref{Second_compare_RSO} and~\ref{Second_compare_RSDelta}, respectively, within the resolution limits of our dataset.
From the perspective of critical-point theory~\citep{Chong1990}, a well-defined vortex structure can only exist within regions satisfying $\Delta>0$, where the instantaneous streamlines exhibit a closed or swirling pattern in a reference frame co-moving with a fluid particle. For instance, the Liutex is only defined in the region $\Delta>0$ and becomes null for $\Delta<0$ where the vorticity reduces to the characteristic spin~\citep{GaoLiu2019,LiuCQ2020}. In contrast, the DVD vorticity modes $(\bm{R}_{L}(\bm{t}),\bm{s}_{L}(\bm{t}))$ are well defined throughout the entire flow domain (figure~\ref{Second_compare_RSDelta}), independent of the sign of $\Delta$ that determines the local flow topology.
For example, two additional peaks emerge within the region where $\Delta<0$ (between $y^{*}=1.2$ and $y^{*}=1.4$), coinciding spatially with zones of elevated streamline curvature. In this region without classical swirling regime, the Liutex $\bm{R}_{N}^{+}$ vanishes while the orbital rotation mode $\bm{R}_{L}(\bm{t})$ exhibits pronounced dominance.

In figures~\ref{Second_compare_RRNDelta} and~\ref{Second_compare_ssNDelta}, a comparison is made between the $\bm{e}_{z}$-components of the DVD vorticity modes $(\bm{R}_{L}(\bm{t}),\bm{s}_{L}(\bm{t}))$ (in~\eqref{C8a3} and~\eqref{C8a10}) and the two formulations of the IVD vorticity modes $(\bm{R}_{N}^{\pm},\bm{s}_{N}^{\pm})$ (in~\eqref{plus_expression1} and~\eqref{plus_expression2}), evaluated along the line $x^{*}=3.634$. 
As shown in figure~\ref{Second_compare_RRNDelta}, the orbital rotation $\bm{R}_{L}(\bm{t})$ is constrained between the upper bound $\bm{R}_{N}^{+}$ (Liutex) and the lower bound $\bm{R}_{N}^{-}$ in the region $\Delta>0$. Interestingly, $\bm{R}_{L}(\bm{t})$ exhibits close agreement with $\bm{R}_{N}^{+}$ (with only minor deviation) while remaining distinctly separate from $\bm{R}_{N}^{-}$, showing inverse characteristics compared to the Burgers vortex (\S\ref{Burgers vortex}).
Therefore, with reference to $\bm{R}_{N}^{-}$, the Liutex $\bm{R}_{N}^{+}$ represents a more suitable physical quantity for extracting the characteristic rigid rotation mode in this example.
Similarly, figure~\ref{Second_compare_ssNDelta} demonstrates that the DVD spin mode is bounded by the IVD modes $(\bm{s}_{N}^{+}\equiv\gamma^{+}\bm{e}_{3}~\text{and}~\bm{s}_{N}^{-}\equiv\gamma^{-}\bm{e}_{3})$, with $\bm{s}_{N}^{+}$ closely approximating $\bm{s}_{L}(\bm{t})$. The comparison further confirms the general inequalities presented in~\eqref{UD}.

\section{Conclusion and discussion}\label{Conclusions and discussions}
The fundamental concept of fluid rotation has evolved along two distinct but deeply interconnected paths (\S\ref{sec1p1}) since the classical Cauchy-Stokes theorem, revealing the rotational complexity of elementary fluid elements and vortex structures. Along the second research line, we present a unified theoretical framework for studying the kinematics of directed fluid elements and intrinsic vorticity decompositions, integrating the material, field, and characteristic algebraic descriptions. The main findings are summarized below, followed by further discussion.

1.~\textbf{Rotation of directed material line element and vorticity decomposition} 

We investigate the angular velocity of rigid rotation for a material line element $\delta\bm{r}\equiv\delta{r}\bm{e}$, where $\bm{e}$ denotes the unit orientation vector (\S\ref{Rotation of directed material line element and vorticity decomposition}). Theoretically, this angular velocity $\bm{W}_{L}(\bm{e})$ in~\eqref{pp6} consists of a classical contribution driven by the vorticity (i.e., $\frac{1}{2}\bm{\omega}$), and an additional term induced by the strain-rate tensor $\bm{D}$ (i.e., the specific angular velocity $\bm{W}_{D}(\bm{e})$ in~\eqref{pp3}). The surface average of $\bm{W}_{D}(\bm{e})$ over all possible orientations $\bm{e}$ vanishes so that the net rigid rotation of a fluid volume element (without preferred spatial directionality) reduces to $\bm{W}_{V}=\langle\bm{W}_{L}(\bm{e})\rangle=\frac{1}{2}\bm{\omega}$. Building on these results, we propose a line-element-based vorticity decomposition (\eqref{pp8} and~\eqref{eq14}), in which the vorticity constituents $(\bm{R}_{L}(\bm{e}),\bm{s}_{L}(\bm{e}),\bm{g}_{L}(\bm{e}))$ explicitly incorporate directional dependence.

2.~\textbf{Rotation of directed material surface element and vorticity decomposition} 

Beginning with the surface rate of deformation tensor $\bm{B}$ in~\eqref{ss1}, we discuss the angular velocity of rigid rotation of a directed material surface element $\delta\bm{\Sigma}\equiv\delta\Sigma\bm{n}_{\Sigma}$ (denoted as $\bm{W}_{\Sigma}(\bm{n}_{\Sigma})$), with $\bm{n}_{\Sigma}$ representing the surface unit normal vector (\S\ref{Rotation of directed material surface element and vorticity decomposition}). We demonstrate in~\eqref{mmm1} that $\bm{W}_{\Sigma}(\bm{n}_{\Sigma})$ arises from the combined influence of vorticity and strain rate, with the specific angular velocity identified as $-\bm{W}_{D}(\bm{n}_{\Sigma})$, a sign reversal compared to the line-element case. Consequently, $\bm{W}_{V}$ can equivalently be interpreted as the
mean angular velocity of its constituent surface elements, averaged over all possible spatial orientations. Building on this, we propose a surface-element-based vorticity decomposition (\eqref{ss9} and~\eqref{SEB}), where the constituent modes $(\bm{R}_{\Sigma}(\bm{n}_{\Sigma}),\bm{s}_{\Sigma}(\bm{n}_{\Sigma}),\bm{g}_{\Sigma}(\bm{n}_{\Sigma}))$ show directional dependence explicitly.

3.~\textbf{Intrinsic relations for orthogonal line and surface element)}

We establish the intrinsic relationships between the angular velocities $(\bm{W}_{L}(\bm{n}_{\Sigma}),\bm{W}_{\Sigma}(\bm{n}_{\Sigma}))$ of an orthogonal pair of line and surface elements $(\delta\bm{r},\delta\bm{\Sigma})$, revealing their complementary nature in both kinematic and geometric aspects (\S\ref{intrinsic}). Notably, the angular
velocity of rigid rotation for a fluid volume element can be interpreted as the arithmetic mean of those associated with the orthogonal line and
surface elements (see~\eqref{VVV2}).

4.~\textbf{Physical interpretation on spin mode of vorticity} 

Applying the generalized Caswell formula~\eqref{uuu1} (also known as Casewell-Wu formula) to the surface element $\delta\bm{\Sigma}$ yields an elegant identity~\eqref{uuu2} on the surface: $\bm{s}_{\Sigma}(\bm{n}_{\Sigma})=\bm{\omega}_{r}$ which reveals that the spin $\bm{s}_{\Sigma}(\bm{n}_{\Sigma})$ (in the surface-element-based vorticity decomposition) is identical to the relative vorticity $\bm{\omega}_{r}\equiv\bm{\omega}-2\bm{W}_{\Sigma}(\bm{n}_{\Sigma})$, accounting for surface shear stress $\bm{\tau}$ in Newtonian fluids (\S\ref{spin_mode_interpretation}). This fundamental relationship provides a rational foundation for the spin mode representation. Further, we analyze the decomposition of both spin and surface shear stress based on the intrinsic triple decomposition of the strain-rate tensor.
From the perspective of Helmholtz-Hodge decomposition, we show that both $\bm{\omega_{r}}$ and $\bm{\tau}$ are inherently affected by the potential flow component when considering arbitrarily moving and deformable surfaces.

5.~\textbf{Field description based on streamline and streamsurface} 

Within the field-theoretic framework and employing differential geometry, we propose a streamline-based triple vorticity decomposition~\eqref{C8a12}, where $\bm{t}$ denotes the unit tangent vector of a streamline. The corresponding vorticity constituents $(\bm{R}_{L}(\bm{t}),\bm{s}_{L}(\bm{t}),\bm{g}_{L}(\bm{t}))$ represent the rigid rotation, spin, and streaming modes, respectively.
These quantities can be quantitatively determined from velocity fields obtained through either experiments or numerical simulations. Using the surface-attached reference frame, we further introduce a streamsurface-based triple vorticity decomposition~\eqref{sb1} (with $\bm{n}_{S}$ as the unit normal vector of a streamsurface), characterized by the vorticity constituents $(\bm{R}_{\Sigma}(\bm{n}_{S}),\bm{s}_{\Sigma}(\bm{n}_{S}),\bm{g}_{\Sigma}(\bm{n}_{S}))$. Consistency between streamline- and streamsurface-based decompositions are validated by analyzing a cylindrical streamsurface (\S\ref{consistency}). The surface-element-based vorticity decomposition can also be extended to vorticity surface. The results remain equally valid for generic 2D flows.

6.~\textbf{Physical roles of NND rotational invariants and IVD vorticity modes} 

The normal-nilpotent decomposition (NND) of the VGT $\bm{A}$ originates from the Schur theorem in the linear algebra, where a set of six rotational invariants $(\chi,\lambda_{r},\alpha,\beta,\gamma,\psi)$ is introduced to characterize the local flow behaviors (\S\ref{NND_IVD}).
Combining NND and SAD yields the invariant vorticity decomposition (IVD) which reveals the characteristic vorticity modes, specifically the rigid rotation $\bm{R}_{N}$ and the spin $\bm{s}_{N}$ (\S\ref{NND_IVD}). $\bm{R}_{N}$ dominates the axial vortices while $\bm{s}_{N}$ governs the sheet vortices. Unifying NND/IVD and DVD enables us to elucidate the physical roles of these rotational invariants, through the generalized Caswell formula for the strain-rate tensor (\S\ref{NND_and_Caswell}) and Helmholtz decomposition (\S\ref{NND_Helmholtz}). 

For the 2D projected velocity field in the invariant plane (perpendicular to the swirling axis $\bm{e}_{3}$), exact transformation identities are derived to clarify the relationship between the NND variables and DVD vorticity modes (\S\ref{Transformation between characteristic and field descriptions}). Importantly, we find that the intrinsic structure of $\bm{A}$ under the Frenet-Serret frame attached to a streamline exhibits an irreducible real Schur form, which aligns perfectly with the matrix representation under the algebraic framework. In the region with positive discriminant $(\Delta>0)$, our analysis shows that the physically admissible DVD vorticity modes $(\bm{R}_{L}(\bm{t}),\bm{s}_{L}(\bm{t}))$ in the field description are strictly bounded by the IVD vorticity modes $(\bm{R}_{N}^{\pm},\bm{s}_{N}^{\pm})=(2\psi^{\pm},\gamma^{\pm})\bm{e}_{3}$ (\S\ref{sign_of_gamma}). These vorticity modes are constrained to lie along the principal diagonal of the square phase space (figure~\ref{RS_phase}). This finding leads to a minimization principle for determining the characteristic rigid rotation mode (i.e., Liutex) fundamentally embedded in the Liutex-shear decomposition.

7.~\textbf{Demonstrative examples} 

By introducing the specific angular velocity of a material line element, we resolve Joseph Bertrand’s puzzle for simple shear flow ($\Delta=0$), in agreement with IVD-NND analysis (\S\ref{Simple shear flow and Bertrand's puzzle}). The Liutex (in NND-I) disappears for potential flow outside a point vortex (where $\Delta<0$), whereas NND-II captures the underlying velocity gradient structure, aligning with the streamline-based description (\S\ref{point_vortex}). 
The Burgers vortex is adopted as an analytical example to examine the intrinsic connection and difference between algebraic and field descriptions (\S\ref{Burgers vortex}). In this case, $\bm{R}_{N}^{+}\equiv{2\psi^{+}}\bm{e}_{3}=\partial_{r}u_{\theta}\bm{e}_{3}$ (Liutex) lacks a direct kinematic link to orbital rotation and deviates more significantly from the DVD orbital rotation mode $\bm{R}_{L}(\bm{t})$ when compared to $\bm{R}_{N}^{-}\equiv2\psi^{-}\bm{e}_{3}=({u_{\theta}}/r)\bm{e}_{3}$. This comparative study highlights the complementary nature of IVD and DVD for physical insights in vortex motions.

We propose a novel vorticity-dynamical interpretation of vortex-sheet instability under small perturbation: the $\frac{1}{2}\pi$ phase difference between the perturbation-induced local strength density $\bm{\Gamma}^{\prime}$ and the disturbance
displacement $\eta$ is predominated by the disturbance spin mode $\bm{s}_{L}^{\prime}(\bm{t})$, with minor modulation from the orbital rotation mode (\S\ref{Vortex-sheet instability sec}).
This deepens the classical interpretation established by \citet{Batchelor1967}.
Additionally, we demonstrate the rotation patterns of material line elements in triaxial strain background field and preliminarily investigate the orientation instability of disturbed filaments (\S\ref{Moffatt-Kida-Ohkitani vortex solution}). Finally, leveraging the optical flow method in flow visualization, we examine the coherent structures on Saturn's north pole~--~including the Hexagon, NPV, and secondary hurricane-eye vortex~--~under both characteristic and field descriptions (\S\ref{Hexagon and North Polar Vortex on Saturn}). Our results show that the DVD vorticity modes $(\bm{R}_{L}(\bm{t}),\bm{s}_{L}(\bm{t}))$ are constrained by the IVD modes $(\bm{R}_{N}^{+},\bm{R}_{N}^{-})$ in the region $\Delta > 0$. In contrast to the Burgers vortex, $\bm{R}_{N}^{+}$ (Liutex) exhibits stronger alignment with the orbital rotation mode $\bm{R}_{L}(\bm{t})$ than $\bm{R}_{N}^{-}$.

8.~\textbf{Outlook and future work}

The present work establishes a general theory on kinematics of direct fluid elements and intrinsic vorticity decomposition in the context of vorticity and vortex dynamics. In future studies, we will apply this theory to analyze the kinematic and dynamic evolution of coherent structures in three-dimensional turbulent flows, both isotropic and wall-bounded, elucidating their formation and evolution mechanisms through interactions and transformations among distinct vorticity modes. Further extensions to sound source structures and sound generation mechanisms in complex flows present compelling avenues for research.
Since the rotation of a material line or surface element in a background flow involves more degrees of freedom than a single volume element, new types of hydrodynamic instabilities may emerge for vortex filaments and surfaces, extending beyond the scope of traditional vortex stability theory. This remains to be a fundamental subject worth further exploring by virtue of the existing visualization methods, for example, the theory of vortex-surface field linked with spherical Clebsch representation~\citep{Chern2017,Yangyue2023}.
\begin{bmhead}[Declarations of competing interests.]
The authors declare that they have no known competing financial
interests or personal relationships that could have appeared to influence
the work reported in this paper.
\end{bmhead}
\begin{bmhead}[Funding.]
The work was supported by the National Natural Science
Foundation of China (NSFC Award No. 12402262); the John O. Hallquist Endowed Professorship; and Presidential Innovation Professorship at Western Michigan University.  
\end{bmhead}

\begin{appen}
\section{Surface average of specific angular velocity}\label{AP0}
The dot product of the displacement vector $\delta\bm{r}\equiv\bm{x}^{\prime}-\bm{x}=\delta{r}\bm{e}$ and the strain-rate tensor $\bm{D}$ is given by
\begin{eqnarray}\label{cw1}
	\delta\bm{r}\bm{\cdot}\bm{D}=\bm{\nabla}^{\prime}\left(\frac{1}{2}\delta\bm{r}\bm{\cdot}\bm{D}\bm{\cdot}\delta\bm{r}\right).
\end{eqnarray}
where the operator $\bm{\nabla}^{\prime}=\bm{\nabla}_{\bm{x}^{\prime}}$ evaluates the spatial gradient with respect to $\bm{x}^{\prime}$.
By using~\eqref{cw1} and applying the chain rule of partial derivatives, we obtain
\begin{eqnarray}\label{cw2}
	\bm{e}\bm{\cdot}\bm{D}&=&\frac{1}{\delta r}\bm{\nabla}^{\prime}\left(\frac{1}{2}\delta\bm{r}\bm{\cdot}\bm{D}\bm{\cdot}\delta\bm{r}\right)\nonumber\\
	&=&\bm{\nabla}^{\prime}\left(\frac{1}{\delta r}\frac{1}{2}\delta\bm{r}\bm{\cdot}\bm{D}\bm{\cdot}\delta\bm{r}\right)-\frac{1}{2}\left(\delta\bm{r}\bm{\cdot}\bm{D}\bm{\cdot}\delta\bm{r}\right)\bm{\nabla}^{\prime}\left(\frac{1}{\delta r}\right).
\end{eqnarray}
On the right hand side of~\eqref{cw2}, the bracketed expression in the first term evaluates to
\begin{eqnarray}\label{cw3}
	\frac{1}{\delta r}\frac{1}{2}\delta\bm{r}\bm{\cdot}\bm{D}\bm{\cdot}\delta\bm{r}=\frac{1}{2}\delta{r}\left(\bm{e}\bm{\cdot}\bm{D}\bm{\cdot}\bm{e}\right)=\frac{1}{2}\frac{D\delta r}{Dt}.
\end{eqnarray}
We remark that through~\eqref{cw3}, equation~\eqref{cw1} admits the alternative form
\begin{eqnarray*}
\delta\bm{r}\bm{\cdot}\bm{D}=\bm{\nabla}^{\prime}\left(\frac{1}{4}\frac{D\delta r^2}{Dt}\right).
\end{eqnarray*}
Evaluating the spatial gradients of the squared displacement relation $\delta{r}^{2}=\delta\bm{r}\bm{\cdot}\delta\bm{r}$ gives
\begin{eqnarray}\label{cw4}
	2\delta{r}\bm{\nabla}^{\prime}\delta{r}=2\bm{\nabla}^{\prime}\delta\bm{r}\bm{\cdot}\delta\bm{r}=2\delta\bm{r},
\end{eqnarray}
and therefore, the spatial gradient of $\delta{r}$ with respect to $\bm{x}^{\prime}$ is
\begin{eqnarray}\label{cw5}
	\bm{\nabla}^{\prime}\delta{r}=\frac{\delta\bm{r}}{\delta r}=\bm{e}.
\end{eqnarray}
Consequently, using~\eqref{cw5}, the last term on the right hand side of~\eqref{cw2} is evaluated as
\begin{eqnarray}\label{cw6}
	-\frac{1}{2}\left(\delta\bm{r}\bm{\cdot}\bm{D}\bm{\cdot}\delta\bm{r}\right)\bm{\nabla}^{\prime}\left(\frac{1}{\delta r}\right)=\frac{1}{2}\left(\bm{e}\bm{\cdot}\bm{D}\bm{\cdot}\bm{e}\right)\bm{e}.
\end{eqnarray}
Substituting~\eqref{cw3} and~\eqref{cw6} into~\eqref{cw2} yields
\begin{eqnarray}\label{cw7}
	\bm{e}\bm{\cdot}\bm{D}=\bm{\nabla}^{\prime}\left(\frac{1}{2}\frac{D\delta r}{Dt}\right)+\frac{1}{2}\left(\bm{e}\bm{\cdot}\bm{D}\bm{\cdot}\bm{e}\right)\bm{e}.
\end{eqnarray}
Then, by taking the dot product of both sides of~\eqref{cw7} with $\bm{e}$, we obtain
\begin{eqnarray}\label{cw8}
	\bm{W}_{D}(\bm{e})\equiv\bm{e}\times(\bm{e}\bm{\cdot}\bm{D})=\bm{e}\times\bm{\nabla}^{\prime}\left(\frac{1}{2}\frac{D\delta r}{Dt}\right).
\end{eqnarray}
Finally, applying the Gauss's divergence theorem in conjunction with~\eqref{cw8}, the surface average of the specific angular velocity $\bm{W}_{D}(\bm{e})$ over the boundary surface $\bm{\Sigma}(\bm{e})=\bm{\Sigma}(\theta,\phi)$ of a unit sphere $\mathcal{B}(\bm{x})$ centered at $\bm{x}$ is evaluated as
\begin{eqnarray}\label{pp4}
	\langle\bm{W}_{D}(\bm{e})\rangle&\equiv&\frac{1}{4\pi}\int_{\bm{\Sigma}(\bm{e})}\bm{e}\times\left(\bm{e}\bm{\cdot}\bm{D}(\bm{x})\right)dS^{\prime}\nonumber\\
	&=&\frac{1}{4\pi}\int_{\mathcal{B}(\bm{x})}\bm{\nabla}^{\prime}\times\bm{\nabla}^{\prime}\left(\frac{1}{2}\frac{D\delta r}{Dt}\right)dV^{\prime}\nonumber\\
	&=&\bm{0}.
\end{eqnarray}
\section{Surface vorticity components under the surface-attached frame}\label{AP1}
The spatial gradient operator $\bm{\nabla}$ can be decomposed as $\bm{\nabla}=\bm{\nabla}_{\pi}+\bm{n}_{\Sigma}\partial_{3}$, where $\bm{\nabla}_{\pi}$ is the surface tangential gradient operator, and $\partial_{3}=\partial_{n}$ evaluates the surface-normal derivative.
Then, by virtue of the surface-attached orthonormal frame $\left\{\overline{\bm{e}}_{1},\overline{\bm{e}}_{1},\overline{\bm{e}}_{3}\right\}~(\overline{\bm{e}}_{3}=\bm{n}_{\Sigma})$, the vorticity on a generic surface element $\delta\bm{\Sigma}$ is decomposed as
\begin{eqnarray}\label{GG1}
	\bm{\omega}=\overline{\bm{e}}_{1}\times\partial_{1}\bm{u}+\overline{\bm{e}}_{2}\times\partial_{2}\bm{u}+\bm{n}_{\Sigma}\times\partial_{3}\bm{u},
\end{eqnarray}
where the partial derivatives $\partial_{i}~(i=1,2)$ follow the definition in~\eqref{partial_i}.

By applying techniques from differential geometry following \citep{ChenWH2002}, we evaluate the terms on the right-hand side of equation~\eqref{GG1} as
\begin{eqnarray*}\label{GG2}
	\overline{\bm{e}}_{1}\times\partial_{1}\bm{u}=(-\partial_{1}u_{3}-b_{11}u_{1}-b_{12}u_{2})\overline{\bm{e}}_{2}
	+(\partial_{1}u_{2}+\kappa_{g,1}u_{1}-b_{12}u_{3})\overline{\bm{e}}_{3}.
\end{eqnarray*}
\begin{eqnarray*}\label{GG3}
	\overline{\bm{e}}_{2}\times\partial_{2}\bm{u}=\left(\partial_{2}u_{3}+b_{12}u_{1}+b_{22}u_{2}\right)\overline{\bm{e}}_{1}+\left(-\partial_{2}u_{1}-\kappa_{g,2}u_{2}+b_{12}u_{3}\right)\overline{\bm{e}}_{3}.
\end{eqnarray*}
\begin{eqnarray}\label{GG4}
	\bm{n}_{\Sigma}\times\partial_{3}\bm{u}=-\partial_{3}u_{2}\overline{\bm{e}}_{1}
	+\partial_{3}u_{1}\overline{\bm{e}}_{2},
\end{eqnarray}
where the geodesic curvatures of the coordinate curves are determined by
\begin{eqnarray*}
	\kappa_{g,1}=-\frac{1}{\sqrt{g_{22}}}\frac{\partial\ln\sqrt{g_{11}}}{\partial{x}_{2}}=-\partial_{2}\ln\sqrt{g_{11}},
\end{eqnarray*}
\begin{eqnarray}\label{geodesic}
	\kappa_{g,2}=-\frac{1}{\sqrt{g_{11}}}\frac{\partial\ln\sqrt{g_{22}}}{\partial{x}_{1}}=-\partial_{1}\ln\sqrt{g_{22}}.
\end{eqnarray}
Substituting~\eqref{GG4} into~\eqref{GG1} yields
\begin{eqnarray*}\label{C9X1}
	\omega_{1}&=&\partial_{2}u_{3}-\partial_{3}u_{2}+b_{12}u_{1}+b_{22}u_{2}\nonumber\\
	&=&\frac{1}{\sqrt{g_{22}}}\frac{\partial u_3}{\partial x_2}-\frac{\partial u_2}{\partial x_3}+b_{12}u_{1}+b_{22}u_{2},
\end{eqnarray*}
\begin{eqnarray*}\label{C9X2}
	\omega_{2}&=&\partial_{3}u_{1}-\partial_{1}u_{3}-b_{11}u_{1}-b_{12}u_{2}\nonumber\\
	&=&\frac{\partial u_1}{\partial x_3}-\frac{1}{\sqrt{g_{11}}}\frac{\partial u_3}{\partial x_1}-b_{11}u_{1}-b_{12}u_{2},
\end{eqnarray*}
\begin{eqnarray}\label{C9X3}
	\omega_{3}&=&\partial_{1}u_{2}-\partial_{2}u_{1}+\kappa_{g,1}u_{1}-\kappa_{g,2}u_{2}\nonumber\\
	&=&\frac{1}{\sqrt{g_{11}}}\frac{\partial u_2}{\partial x_1}
	-\frac{1}{\sqrt{g_{22}}}\frac{\partial u_1}{\partial x_2}+\kappa_{g,1}u_{1}-\kappa_{g,2}u_{2}.
\end{eqnarray}
We remark that the surface-normal vorticity component $\omega_{3}$ in~\eqref{C9X3} is entirely determined by the surface-parallel motion itself (including the tangential velocity field $\bm{u}_{\pi}=(u_{1},u_{2})$ and its surface-parallel derivatives) and the local geometric properties of the surface, independent of the off-surface velocity information and associated gradients. By using~\eqref{geodesic}, $\omega_{3}$ admits a more compact representation as
\begin{eqnarray}
\omega_{3}=\frac{1}{\sqrt{g_{11}}\sqrt{g_{22}}}\left[\frac{\partial}{\partial x_{1}}\left(\sqrt{g_{22}}u_{2}\right)-\frac{\partial}{\partial x_{2}}\left(\sqrt{g_{11}}u_{1}\right)\right].
\end{eqnarray}
Moreover, ${\omega}_{3}$ completely determines the last term in~\eqref{uuuu1} via
\begin{eqnarray}\label{C100}
	2\mathscr{A}[{\left(\bm{\nabla}_{\pi}\bm{u}\right)_{\pi}}]=\omega_{3}\overline{\bm{e}}_{1}\overline{\bm{e}}_{2}-\omega_{3}\overline{\bm{e}}_{2}\overline{\bm{e}}_{1}=\omega_{3}\overline{\bm{e}}_{1}\wedge\overline{\bm{e}}_{2},
\end{eqnarray}
where $\wedge$ is the wedge product describing the exterior form~\citep{ChenWH2002}. By acting the Hodge star operator $(*)$ on both sides of~\eqref{C100}, we obtain
\begin{eqnarray}\label{C101}
	2*\mathscr{A}[{\left(\bm{\nabla}_{\pi}\bm{u}\right)_{\pi}}]=\omega_{3}\bm{n}_{\Sigma}=\bm{g}_{\Sigma}(\bm{n}_{\Sigma}).
\end{eqnarray}

\section{Helmholtz decomposition of Burgers vortex solution}\label{AP3}
From the velocity representation of the Burgers vortex under the basis vectors $\left\{\bm{e}_{r},\bm{e}_{\theta},\bm{e}_{z}\right\}$ of the cylindrical coordinate system, we obtain the following Helmholtz decomposition:
\begin{eqnarray}\label{C1}
	\bm{u}=\bm{u}_{\Phi}+\bm{v}.
\end{eqnarray}
The potential velocity $\bm{u}_{\phi}$ is expressed as
\begin{eqnarray}\label{C2}
	\bm{u}_{\Phi}=\bm{\nabla}\Phi=-\frac{1}{2}ar\bm{e}_{r}+az\bm{e}_{z}=-\frac{1}{2}a{x}\bm{e}_{x}-\frac{1}{2}ay\bm{e}_{y}+az\bm{e}_{z},
\end{eqnarray}
where the scalar potential is given by
\begin{eqnarray}\label{C3}
	\Phi(r,z)=-\frac{1}{4}ar^{2}+\frac{1}{2}az^2.
\end{eqnarray}
The vortical velocity $\bm{v}$ is expressed as
\begin{eqnarray}\label{C4}
	\bm{v}=\bm{\nabla}\times\bm{\Psi}=u_{\theta}(r)\bm{e}_{\theta},~\bm{\Psi}={\Psi}_{z}(r)\bm{e}_{z},
\end{eqnarray}
where the vectorial potential is given by
\begin{eqnarray}\label{C5}
	\Psi_{z}(r)=\frac{\Gamma}{4\pi}{\rm Ei}\left(-\frac{r^2a}{4\nu}\right)-\frac{\Gamma}{2\pi}\ln{r}.
\end{eqnarray}
Here, ${\rm Ei}(\eta)$ is the exponential integral function, being defined as
\begin{eqnarray}\label{C6}
	{\rm Ei}(\eta)=-\int_{-\eta}^{+\infty}\frac{\exp(-\xi)}{\xi}d\xi,~\text{for}~\eta<0.
\end{eqnarray}
From~\eqref{C2} and~\eqref{C4}, the gradients of $\bm{u}_{\Phi}$ and $\bm{v}$ are determined as
\begin{eqnarray*}
	\bm{\nabla}\bm{u}_{\Phi}=\bm{\nabla\nabla}\Phi
	=	\begin{bmatrix}
		-\frac{1}{2}a &0& 0\\
		0 &-\frac{1}{2}a&0\\
		0&0& a
	\end{bmatrix}
	=	\begin{bmatrix}
		\chi &0& 0\\
		0 &\chi&0\\
		0&0& \lambda_{r}
	\end{bmatrix},
\end{eqnarray*}
\begin{eqnarray}
	\bm{\nabla}\bm{v}=\begin{bmatrix}
		0&\partial_{r}u_{\theta}& 0\\
		-\frac{1}{r}u_{\theta} &0&0\\
		0&0& 0
	\end{bmatrix}
	=\begin{bmatrix}
		0 &\psi^{-}+\gamma^{-}& 0\\
		-\psi^{-} &0&0\\
		0&0& 0
	\end{bmatrix}.
\end{eqnarray}
\end{appen}

\bibliography{jfm}

\end{document}